\documentclass[aps,twocolumn,showpacs,amsmath,amssymb,prc]{revtex4-2}
\usepackage{graphicx,here}
\usepackage{multirow}
\usepackage{tikz}
\usepackage{epstopdf}
\usepackage[percent]{overpic}
\usepackage{scrextend}
\usepackage{xcolor,xspace}
\usepackage[ulem=normalem]{changes}
\usepackage{hyperref}
\hypersetup{colorlinks=True,urlcolor=blue,linkcolor=blue,citecolor=blue,filecolor=black}

\colorlet{Changes@Color}{red}
\usepackage{dcolumn,color,footnote,bm,braket}
\usepackage{url,longtable,tabularx}
\usepackage{threeparttable}

\usepackage{fancybox}
\usetikzlibrary{matrix}

\newcommand\+{\dagger}
\newcommand\jr{j_{\rho}}
\newcommand\mr{m_{\rho}}

\newcommand\mgt{M^{\mathrm{GT}}_{2\nu}}
\newcommand\mf{M^{\mathrm{F}}_{2\nu}}
\newcommand\mgtbeta{M({\mathrm{GT}})}
\newcommand\mfbeta{M({\mathrm{F}})}

\newcommand\mbb{M_{2\nu}}
\newcommand\mbbe{M_{2\nu}^{\mathrm{eff}}}

\newcommand\mbbqp{M_{2\nu}^{\mathrm{(I)}}}
\newcommand\mbbqe{M_{2\nu}^{\mathrm{(II)}}}

\newcommand\hb{\hat{H}_{\mathrm{B}}}
\newcommand\hf{\hat{H}_{\mathrm{F}}}
\newcommand\hbf{\hat{V}_{\mathrm{BF}}}
\newcommand\hff{\hat{V}_{\nu\pi}}

\newcommand\db{\beta\beta}
\newcommand\tnbb{2\nu\beta\beta}
\newcommand\znbb{0\nu\beta\beta}
\newcommand\gae{g_{\mathrm{A,eff}}}
\newcommand\gaeb{g_{\mathrm{A,eff,\beta}}}
\newcommand\gaep{g_{\mathrm{A,eff}}^{\mathrm{(I)}}}
\newcommand\gaee{g_{\mathrm{A,eff}}^{\mathrm{(II)}}}
\newcommand\ga{g_{\mathrm{A}}}
\newcommand\gv{g_{\mathrm{V}}}

\newcommand\trgg{0^+_1\to0^+_1}
\newcommand\trge{0^+_1\to0^+_2}

\newcommand\ft{\log{ft}}
\newcommand\btm{\beta^{-}}

\newcommand\egs{E_{\mathrm{gs}}}

\newcommand\qbb{Q_{\beta\beta}}
\newcommand\qbbt{Q_{\beta\beta,\mathrm{th}}}
\newcommand\qbbe{Q_{\beta\beta,\mathrm{ex}}}

\newcommand\taubb{\tau_{1/2}^{(2\nu)}}
\newcommand\vd{v_{\mathrm{d}}}
\newcommand\vssd{v_{\mathrm{ssd}}}
\newcommand\vsss{v_{\mathrm{ss}}}
\newcommand\vt{v_{\mathrm{t}}}

\begin{document}

\title{Two-neutrino double-$\beta$ decay in the mapped interacting boson model}

\author{Kosuke Nomura}
\email{knomura@phy.hr}
\affiliation{Department of Physics, Faculty of Science, 
University of Zagreb, HR-10000 Zagreb, Croatia}

\date{\today}

\begin{abstract}
A calculation of 
two-neutrino double-$\beta$ ($\tnbb$) 
decay matrix elements 
within the interacting boson model (IBM) 
that is based on the nuclear density functional theory 
is presented. 
The constrained self-consistent mean-field 
(SCMF) calculation using a universal 
energy density functional (EDF) 
and a pairing interaction provides 
potential energy surfaces 
with triaxial quadrupole degrees of freedom  
for even-even nuclei corresponding to 
the initial and final states 
of the $\tnbb$ decays of interest. 
The SCMF energy surface is then mapped 
onto the bosonic one, and this procedure 
determines the IBM Hamiltonian for 
the even-even nuclei. 
The same SCMF calculation provides the essential 
ingredients of the interacting boson fermion-fermion 
model (IBFFM) for the intermediate odd-odd nuclei 
and the Gamow-Teller and Fermi transition operators. 
The EDF-based IBM and IBFFM provide a 
simultaneous description of excitation 
spectra and electromagnetic transition rates 
for each nucleus, and single-$\beta$ and $\db$ decay properties. 
The calculated $\tnbb$-decay nuclear matrix elements 
are compared with experiment and with those from 
earlier theoretical calculations. 
\end{abstract}

\maketitle

\section{Introduction}

Double-$\beta$ ($\db$) decay is a fundamental nuclear 
process that is expected to occur between the atomic 
nuclei with $(A,Z)$ and $(A,Z\pm2)$. 
The process can be classified into basically two types of the decay  
mode according to whether or not it emits neutrinos/antineutrinos, that is, 
whether or not the process conserves the number of leptons. 
Neutrinoless $\db$ ($\znbb$) decay is a 
lepton-number-violating nuclear process that does 
not emit neutrinos and is
allowed to occur 
if neutrinos have mass and are Majorana particles, i.e., 
they are their own antiparticles. 
The $\znbb$ decay has a number of implications for 
the various symmetry properties of the standard model of the 
electroweak interactions and, in that context, experiments 
have been performed at a number of facilities 
around the world, that are aimed at the direct 
observations of the decay and determination 
of its half-life (see, e.g., Refs.~\cite{avignone2008,ejiri2019}, 
and references are therein). 
There have also been extensive theoretical investigations 
on the $\db$ decay, with much emphasis on the zero-neutrino modes 
\cite{primakoff1959,haxton1984,doi1985,tomoda1991,suhonen1998,faessler1998,vogel2012,vergados2012,engel2017}. 
In particular, 
the reliable prediction of the $\znbb$-decay half-life requires inputs 
from nuclear-structure theory 
through the accurate calculation of the nuclear matrix element (NME), 
and numerous attempts have been made 
for the calculation of the $\znbb$-decay NME from a number of 
theoretical approaches at various levels of sophistication 
(see Ref.~\cite{engel2017} for a recent review). 

Another type of the $\db$ decay 
is the two-neutrino $\db$ ($\tnbb$) decay. 
This decay process involves two electrons (or positrons) 
and two electron antineutrinos (or electron neutrinos), 
and occurs regardless of whether or not the neutrinos are 
their antiparticles. 
Since this nuclear process conserves lepton number, 
it is allowed by the standard model, 
and indeed has been observed in a number of experiments, 
a recent compilation of the $\tnbb$-decay data 
being provided, e.g., in Ref.~\cite{barabash2020}. 
Given a wealth of the experimental data, 
theoretical calculation of the $\tnbb$-decay properties 
serves as a meaningful benchmark for a given 
theoretical model. Furthermore, most of the theoretical 
tools that are needed for calculating the 
$\tnbb$-decay NME, including the model assumptions and parameters, 
and the calculated wave functions for the initial 
and final even-even nuclei can be commonly used for 
the prediction of the $\znbb$-decay NME.

In the calculations of 
$\znbb$-decay NME, on one hand, the closure approximation has been 
frequently used, since the virtual neutrinos exchanged in the process 
have momenta much larger ($\approx100$ MeV) 
than the typical energy scale of nuclear excitations
($\approx10$ MeV). 
In the case of the $\tnbb$ decay, on the other hand, the 
neutrino momenta are about the same order of magnitude 
as the nuclear excitation energies, hence the closure 
approximation is not expected to be very good. 
One would then have to explicitly  calculate 
the intermediate states of the neighboring odd-odd nuclei, 
and their single-$\beta$ decay matrix elements. 
Calculations for the $\db$ decay without 
using the closure approximations have been 
performed, e.g., within the quasiparticle 
random-phase approximation (QRPA) 
\cite{suhonen1998,pirinen2015,simkovic2018}, 
the large-scale shell model (LSSM) 
\cite{caurier2007,yoshinaga2018,caurier1990,caurier2012,senkov2016,coraggio2019}, 
and the interacting boson model (IBM) 
\cite{yoshida2013}.

This paper presents a calculation 
of the $\tnbb$-decay NMEs within the framework of the 
IBM that is based on the nuclear density functional theory. 
The starting point shall be a set of the 
constrained self-consistent 
mean-field (SCMF) calculations \cite{RS} using 
a universal EDF and a pairing interaction 
\cite{RS,bender2003,vretenar2005,niksic2011,robledo2019} 
to yield potential energy surfaces  
as functions of the triaxial quadrupole deformations 
for the even-even nuclei involved in the $\db$ decays of  
$^{48}$Ca, $^{76}$Ge, $^{82}$Se, $^{96}$Zr, $^{100}$Mo, 
$^{110}$Pd, $^{116}$Cd, $^{124}$Sn, $^{128}$Te,
$^{130}$Te, $^{136}$Xe, $^{150}$Nd, and $^{198}$Pt. 
The SCMF energy surface is then mapped onto 
the corresponding bosonic one, 
and the strength parameters of the IBM Hamiltonian 
for describing spectroscopic properties of the 
even-even nuclei are determined by this mapping 
procedure \cite{nomura2008}. 
Here the closure approximation is not made, but 
the intermediate states of the odd-odd 
neighbors are explicitly calculated in the 
interacting boson-fermion-fermion model (IBFFM) 
\cite{brant1984,IBFM}. 
The essential ingredients of the IBFFM 
and the Gamow-Teller and Fermi transition 
operators are provided by the 
same SCMF method \cite{nomura2016odd,nomura2019dodd}. 
The resulting IBM and IBFFM wave functions 
are then used to compute single-$\beta$ 
and $\tnbb$ decay properties.

The aforementioned SCMF-to-IBM mapping procedure 
\cite{nomura2008}, and its extensions to odd-mass 
and odd-odd nuclear systems 
\cite{nomura2016odd,nomura2019dodd}, 
have so far been applied to and shown to be valid in 
the studies of various nuclear structure phenomena 
in medium-heavy and heavy nuclei in a wide range 
of the nuclear chart, 
including the shape evolution and coexistence 
\cite{nomura2012sc,nomura2016sc,nomura2016zr}, the onset of 
octupole deformation and octupole collective excitations  
\cite{nomura2013oct,nomura2014}, the quantum 
phase transitions in odd-mass nuclear systems 
\cite{nomura2016qpt,nomura2017odd-1}, and the 
$\beta$ decays of the odd-$A$ \cite{nomura2020beta-1} 
and even-$A$ \cite{nomura2020beta-2} 
nuclei in the mass $A\approx130$ region. 
It is then another purpose 
of this paper to investigate whether the EDF-based 
IBM and IBFFM approaches are able to give a consistent 
description of the spectroscopic properties of 
low-lying states for each 
relevant nucleus and single-$\beta$ and $\tnbb$-decay 
matrix elements.

At this point, it is worth noting that 
for the last decade the IBM framework has been 
extensively applied to the studies of $\db$ decays 
\cite{barea2009,barea2013,barea2015,barea2015b,deppisch2020,kotila2021}. 
In these references, 
wave functions of the IBM Hamiltonian that 
gives an excellent phenomenological description 
of the relevant even-even nuclei were used for 
computing the $\db$ decay NMEs, 
while the $\db$-decay operators were derived 
from multi-nucleon systems 
within the generalized seniority scheme \cite{OAI}. 
A calculation of the $\tnbb$ decays of $^{128}$Te and 
$^{130}$Te without the closure 
approximation was also made 
within the IBM and IBFFM with the ingredients 
determined mostly on phenomenological grounds 
\cite{yoshida2013}. 
In Ref.~\cite{vanisacker2017}, an isospin invariant 
version of the IBM that is derived from a realistic 
shell-model interaction has been employed to 
identify the importance of neutron-proton isoscalar pairing 
in the $\znbb$ decay in the $^{48}$Ca region.

The paper is organized as follows. 
In Sec.~\ref{sec:theory}, 
the theoretical procedures to construct 
the IBM and IBFFM Hamiltonians and the $\db$-decay operators 
are outlined. 
In Sec.~\ref{sec:energy}, calculated 
excitation spectra of the 
even-even initial and final nuclei, 
and odd-odd intermediate nuclei 
are briefly discussed. 
Section~\ref{sec:beta} presents the calculated 
single-$\beta$-decay properties between 
the even-even and odd-odd nuclei. 
In Sec.~\ref{sec:dbeta}, 
results for the $\tnbb$-decay NMEs in comparison 
with the experimental and earlier 
theoretical values, and effective values of 
the axial vector coupling constant are shown. 
In the same section the sensitivity of the 
results to several model assumptions is investigated. 
Finally, Sec.~\ref{sec:summary} gives 
a summary of the main results.

\section{Theoretical framework\label{sec:theory}}

\subsection{Self-consistent mean-field calculations}

The constrained SCMF calculations for 
the even-even nuclei are performed within 
the relativistic Hartree-Bogoliubov (RHB) framework
\cite{vretenar2005,niksic2011,DIRHB} 
with the density-dependent point-coupling (DD-PC1) 
\cite{DDPC1} functional for the particle-hole channel, 
and a separable pairing force of 
finite range \cite{tian2009} 
for the particle-particle channel. 
The constraints are imposed on the mass 
quadrupole moments, which are related to the 
axially-symmetric deformation $\beta$ 
and triaxiality $\gamma$ \cite{BM}. 
The constrained calculations produce 
the potential energy surface 
as a function of the $\beta$ and $\gamma$ deformations, 
which is denoted as 
$E_{\mathrm{SCMF}}(\beta,\gamma)$.

\begin{figure*}[ht]
\begin{center}
\includegraphics[width=\linewidth]{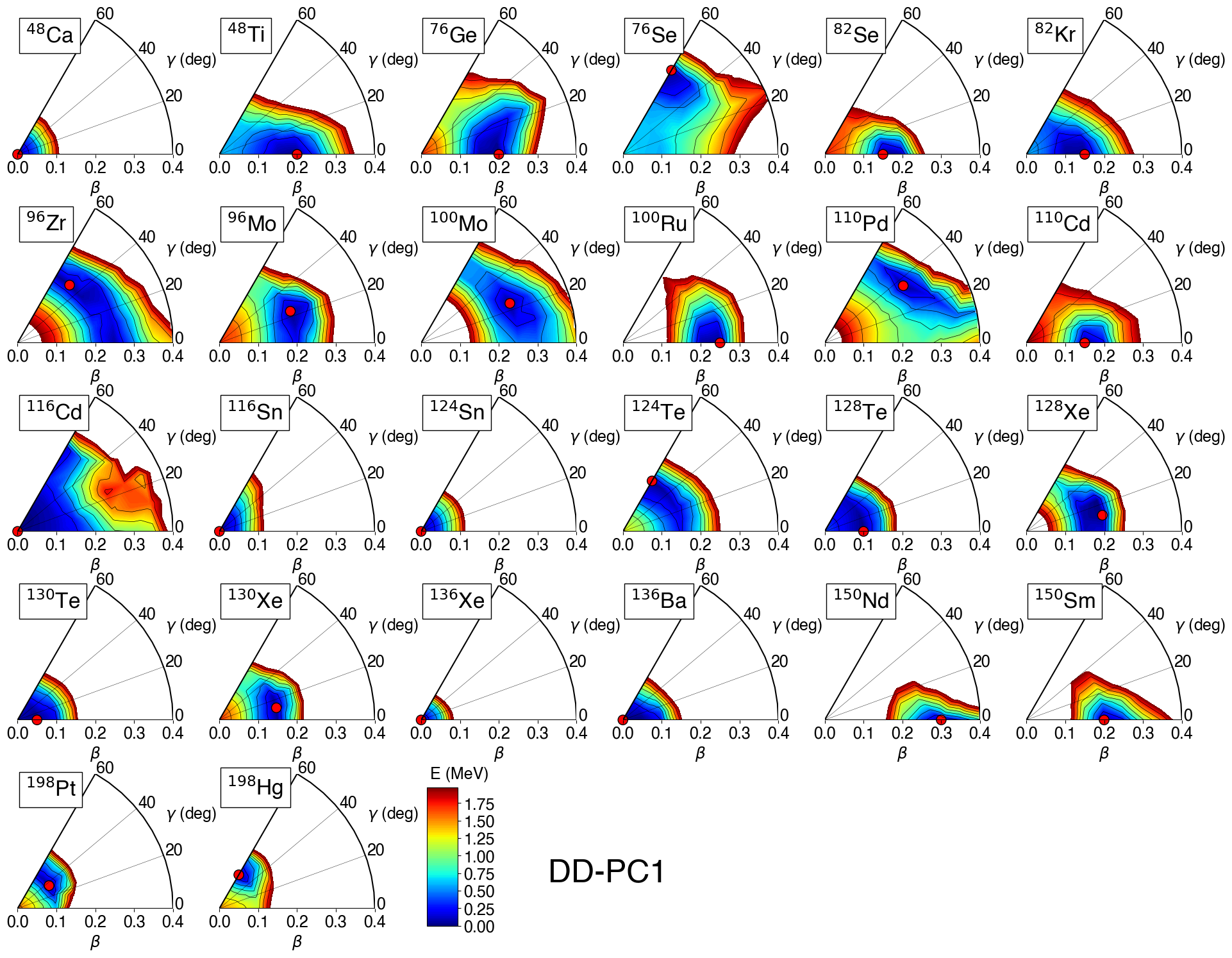}
\caption{SCMF quadrupole triaxial 
$(\beta,\gamma)$ deformation energy surfaces 
for the initial and final even-even nuclei in the 
$\tnbb$ decays under study. 
The energy difference between 
neighboring contours is 250 keV. The global minimum 
is identified by the solid circle. 
}
\label{fig:pesdft}
\end{center}
\end{figure*}

The contour plots of the SCMF $(\beta,\gamma)$-deformation 
energy surfaces are shown in Fig.~\ref{fig:pesdft}. 
For many of the nuclei, either a nearly spherical 
minimum with  $\beta\approx0$, or an 
axially-symmetric prolate $(\beta\neq0,\gamma=0^{\circ})$, 
or oblate $(\beta\neq0,\gamma=60^{\circ})$ minimum 
occurs in their energy surfaces. 
For the mass $A\approx100$ nuclei, 
the energy surfaces are notably 
soft especially in $\gamma$ direction, 
with a shallow triaxial 
minimum within the range $20^{\circ}<\gamma<50^{\circ}$. 
The softness implies that a substantial degree of 
shape mixing is expected to occur near the ground state. 
For the rare-earth nuclei $^{150}$Nd and $^{150}$Sm, 
a well-developed axially-deformed prolate 
deformation is suggested, with the minimum 
at $(\beta,\gamma)\approx(0.3,0^{\circ})$. 
There are also considerable differences between 
the topology of the energy surfaces for 
initial and final states of a given $\tnbb$ decay. 
For instance, the energy surface for the nucleus $^{76}$Ge
exhibits a prolate deformed minimum, while the oblate 
ground state is predicted for the corresponding 
final-state nucleus $^{76}$Se.

In addition to the DD-PC1 EDF, 
the same constrained RHB calculations have been carried out 
for a couple of cases, based on the density-dependent 
meson-exchange (DD-ME2) interaction \cite{DDME2}, 
another representative relativistic functional. 
It is shown, however, that essentially 
no notable qualitative or quantitative difference 
emerges between the SCMF results obtained 
with the DD-PC1 and DD-ME2 EDFs. 

\begin{figure*}[ht]
\begin{center}
\includegraphics[width=\linewidth]{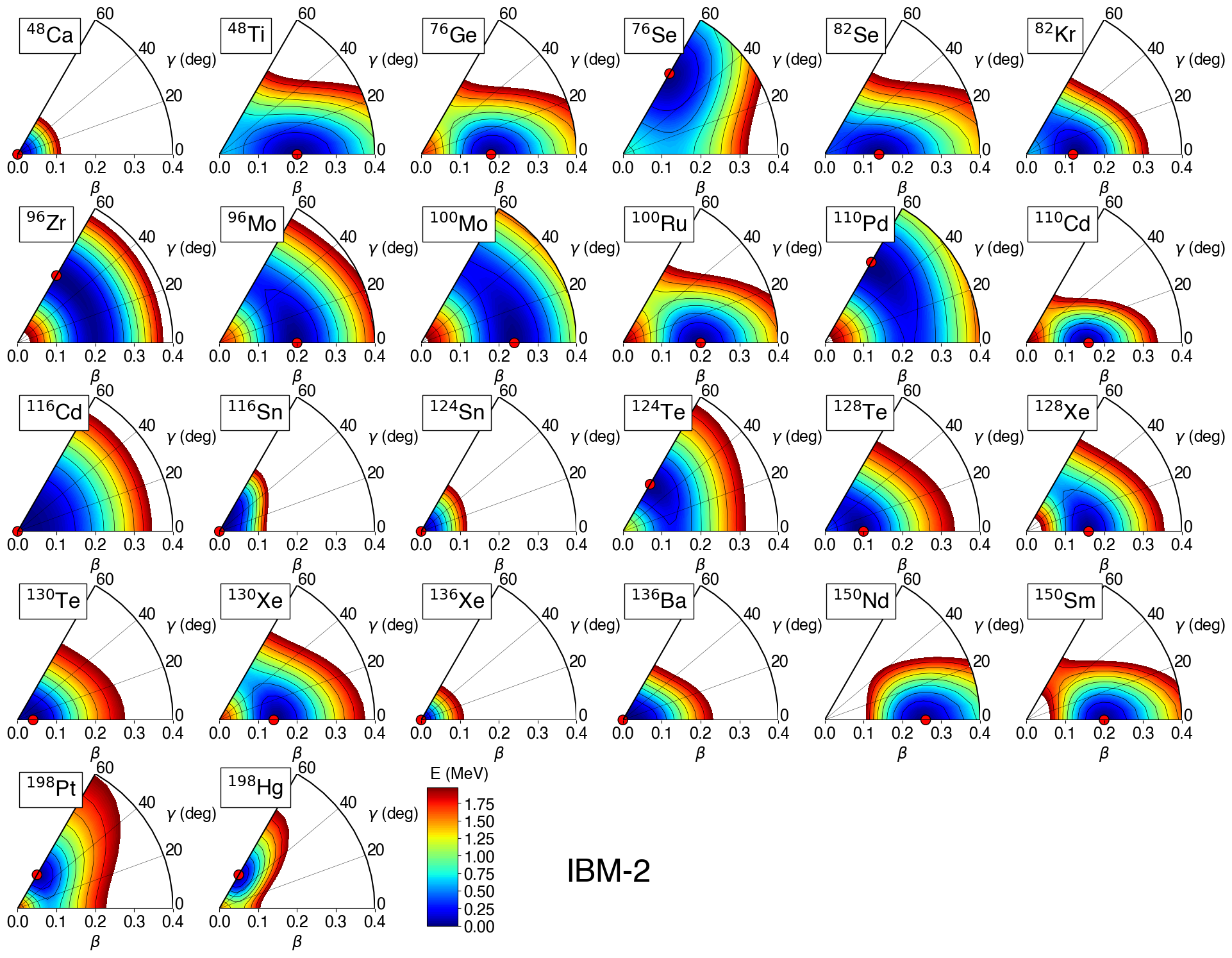}
\caption{Same as Fig.~\ref{fig:pesdft}, 
but for the IBM-2 energy surfaces.}
\label{fig:pesibm}
\end{center}
\end{figure*}

\subsection{IBM-2 for initial and final nuclei}

In the present study the neutron-proton 
IBM (IBM-2) \cite{OAI} is considered, since its connection to 
a microscopic picture is clearer than the original 
version of the model (IBM-1), which does not make a distinction 
between the neutron and proton degrees of freedom. 
The IBM-2 is built on the neutron and proton 
monopole ($s_{\nu}$ and $s_{\pi}$), and 
quadrupole ($d_{\nu}$ and $d_{\pi}$) bosons. 
From a microscopic point of view \cite{OAIT,OAI}, 
the $s_{\nu}$ ($s_\pi$) and $d_\nu$ ($d_\pi$) bosons 
are associated with the 
collective $S_\nu$ ($S_\pi$) and $D_\nu$ ($D_\pi$) 
pairs of valence neutrons (protons) 
with angular momenta $J=0^{+}$ and $J=2^{+}$, 
respectively.

The IBM-2 Hamiltonian employed in this study takes the form
\begin{align}
\label{eq:hb}
 \hb = 
&\epsilon_{d}(\hat{n}_{d_{\nu}}+\hat{n}_{d_{\pi}})
+\kappa\hat{Q}_{\nu}\cdot\hat{Q}_{\pi}
\nonumber\\
&+ \kappa_{\nu}\hat{Q}_{\nu}\cdot\hat{Q}_{\nu}
+ \kappa_{\pi}\hat{Q}_{\pi}\cdot\hat{Q}_{\pi}
+ \kappa'\hat{L}\cdot\hat{L},
\end{align}
where in the first term, 
$\hat{n}_{d_\rho}=d^\+_\rho\cdot\tilde d_{\rho}$ 
($\rho=\nu,\pi$) is the $d$-boson number operator, 
with $\epsilon_{d}$ the single $d$-boson
energy relative to the $s$-boson one, and 
$\tilde d_{\rho\mu}=(-1)^\mu d_{\rho-\mu}$. 
The second term is the quadrupole-quadrupole 
interaction between neutron and proton bosons, 
with 
$\hat Q_{\rho}=d_{\rho}^\+ s_{\rho} + s_{\rho}^\+\tilde d_{\rho}^\+ + \chi_{\rho}(d^\+_{\rho}\times\tilde{d}_{\rho})^{(2)}$ the 
bosonic quadrupole operator. 
The third and fourth terms are the 
quadrupole-quadrupole interactions between 
identical bosons. 
From the microscopic considerations \cite{OAIT}, 
in medium-heavy and heavy nuclei, 
the quadrupole-quadrupole interaction between 
non-identical bosons $\hat{Q}_{\nu}\cdot\hat{Q}_{\pi}$ 
is shown to be more important than 
the ones between identical bosons. 
In most cases, therefore, the interaction 
terms between identical bosons 
$\hat{Q}_{\rho}\cdot\hat{Q}_{\rho}$ are omitted here, 
but are considered only for those nuclei with 
$N_{\pi}=0$ or $N_{\nu}=0$. 
The last term in Eq.~(\ref{eq:hb}) is a rotational 
term with 
$\hat{L}=\sqrt{10}\sum_{\rho}(d^{\+}_{\rho}\times\tilde{d}_{\rho})^{(1)}$ 
being the bosonic angular momentum operator.

The geometrical structure of 
a given IBM-2 Hamiltonian can be formulated in terms of 
the boson coherent state \cite{dieperink1980,ginocchio1980}, 
which is defined by
\begin{align}
\label{eq:coherent}
 \ket{\Phi}=\prod_{\rho=\nu,\pi}
\left[
s_{\rho}^{\+}+\sum_{\mu=-2}^{+2}\alpha_{\rho\mu}d_{\rho\mu}^{\+}
\right]^{N_{\rho}}\ket{0},
\end{align}
up to a normalization factor. 
The amplitudes $\alpha_{\rho\mu}$ are given 
as $\alpha_{\rho0}=\beta_{\rho}\cos{\gamma_{\rho}}$, 
$\alpha_{\rho\pm1}=0$, and 
$\alpha_{\rho\pm2}=\beta_{\rho}\sin{\gamma_{\rho}}/\sqrt{2}$, 
where $\beta_{\rho}$ and $\gamma_{\gamma}$ are 
boson analogs of the deformation variables. 
$\ket{0}$ represents the boson vacuum, i.e., the inert core. 
$N_{\nu}$ ($N_{\pi}$) is the number of neutron 
(proton) bosons, and is counted as half the number 
of valence neutron (proton) particles/holes 
\cite{OAIT,OAI}. 
For a given even-even nucleus under study, the nearest 
doubly-magic nucleus is here taken as the inert core. 
Specifically for $^{48}$Ca and $^{48}$Ti, however, 
the nucleus $^{40}$Ca is taken as the inert core, in order to 
have finite number of active bosons enough to 
produce energy spectra. 
It is assumed that the deformations for neutron and 
proton bosons are equal to each other, 
$\beta_{\nu}=\beta_{\pi}$ and 
$\gamma_{\nu}=\gamma_{\pi}$. 
It is further assumed that the fermionic and bosonic 
deformation variables can be related to each other 
in such a way that $\beta_{\nu}=\beta_{\pi}\propto\beta$ 
and $\gamma_{\nu}=\gamma_{\pi}\equiv\gamma$ 
\cite{ginocchio1980,nomura2008}. 
Under these assumptions, 
the energy surface for the boson system is given by taking 
the energy expectation value in the coherent state 
$E_{\mathrm{IBM}}(\beta,\gamma)\equiv
\braket{\Phi|\hb|\Phi}/\braket{\Phi|\Phi}$.

The IBM-2 Hamiltonian is built by following the SCMF-to-IBM mapping 
procedure of Ref.~\cite{nomura2008}. 
In this procedure, the parameters 
$\epsilon_d$, $\kappa$, $\kappa_{\nu}$, 
$\kappa_{\pi}$, $\chi_{\nu}$, and $\chi_{\pi}$ in the 
Hamiltonian (\ref{eq:hb}) are determined, for each 
nucleus, so that the approximate equality
\begin{align}
\label{eq:map}
 E_{\mathrm{IBM}}(\beta,\gamma)
\approx
E_{\mathrm{SCMF}}(\beta,\gamma)
\end{align}
should be satisfied in the vicinity of the global minimum. 
To be more specific, the IBM-2 parameters are calibrated 
so that the fermionic and bosonic 
energy surfaces become similar to each other, 
within the excitation energy of a few MeV 
with respect to the global minimum. 
The reason for restricting the region for the mapping (\ref{eq:map})
to that energy range is that, in the SCMF model the low-energy 
collective states are supposed to be dominated by the configurations 
in the vicinity of the global minimum, while, 
at higher-excitation energy corresponding to large $\beta$ 
deformations, noncollective (or quasiparticle) excitations 
come to play a role, but are not 
included in the present IBM-2 model space by construction. 
The remaining parameter $\kappa'$ 
of the $\hat{L}\cdot\hat{L}$ 
term (\ref{eq:hb}) is fixed separately, 
in such a way \cite{nomura2011rot} 
that the cranking moment of inertia 
calculated in the intrinsic frame of the boson system 
at the global minimum should be equal to the Inglis-Belyaev 
\cite{inglis1956,belyaev1961} value calculated by the 
constrained RHB method. 
In these procedures, the IBM-2 parameters are determined 
based only on the EDF calculations, that is, 
no phenomenological adjustment of 
the parameters to the experimental 
data is made. The derived IBM-2 parameters 
are listed in Table~\ref{tab:parab} of Appendix~\ref{sec:parab}. 

The mapped IBM-2 energy surfaces, 
shown in Fig.~\ref{fig:pesibm}, are basically 
similar to the SCMF ones (see Fig.~\ref{fig:pesdft}) 
in the vicinity of the global minimum. 
There are, however, several discrepancies that 
are worth some remarks.  
First, a shallow triaxial minimum 
is found in the SCMF energy surfaces 
for $^{96}$Zr, $^{96}$Mo, $^{100}$Mo, 
$^{110}$Pd, $^{128}$Xe, $^{130}$Xe, and $^{198}$Pt, 
but not in the IBM counterparts. 
It was suggested in Ref.~\cite{nomura2012tri} that, 
within the IBM-2 framework the triaxial minimum 
could be produced by introducing a specific 
three-body boson term in the IBM-2 Hamiltonian.  
The higher-order term was also shown to play an important role 
to reproduce correctly the energy-level systematics 
of quasi-$\gamma$ bands in medium-heavy and heavy nuclei. 
However, the triaxial minimum suggested in most of the SCMF 
energy surfaces of the considered even-even nuclei 
is so shallow that the inclusion 
of the higher-order term in the boson Hamiltonian $\hb$ 
is assumed to play a minor role in the final results 
of the $\tnbb$-decay NME, 
for which only the lowest-lying states in the yrast band 
are relevant. 
Second, in the region with 
larger $\beta$ deformation, the SCMF energy surface 
becomes steeper, while the bosonic one becomes rather flat. 
This reflects the fact that, while the SCMF model comprises  
all the constituent nucleons, the IBM-2 is built 
on the more restricted space consisting of the finite number 
of the valence $S$ and $D$ nucleon pairs only. 
The adopted boson Hamiltonian (\ref{eq:hb}) might 
also be of a too simplified form of the most general 
IBM-2 Hamiltonian to reproduce the original SCMF 
surface very precisely.

\subsection{IBFFM-2 for intermediate nuclei}

To describe the intermediate odd-odd nuclei, in addition 
to the collective (boson) degrees of freedom, 
an unpaired neutron and an unpaired proton degrees 
of freedom, as well as the boson-fermion and fermion-fermion 
couplings, are considered within the neutron-proton 
IBFFM (IBFFM-2). 
The Hamiltonian of IBFFM-2 is given by
\begin{align}
\label{eq:ham-ibffm2}
 \hat{H}=\hb + \hf^{\nu} + \hf^{\pi} + \hbf^{\nu} + \hbf^{\pi} + \hff,
\end{align}
where $\hb$ is the same IBM-2 Hamiltonian
as in (\ref{eq:hb}) that represents the 
bosonic even-even core, $\hf^{\nu}$ ($\hf^{\pi}$) 
is the single-neutron (proton) 
Hamiltonian, $\hbf^{\nu}$ ($\hbf^{\pi}$) represents 
the interaction between the odd neutron (proton) and 
the even-even IBM-2 core, 
and the last term $\hff$ is the residual 
neutron-proton interaction. 
Table~\ref{tab:odda} summarizes the intermediate 
odd-odd nuclei considered in this study, 
described as a system of the even-even core 
nucleus with one neutron and one proton added or subtracted, 
and the neighboring odd-$N$ and odd-$Z$ nuclei. 

\begin{table}
\caption{\label{tab:odda}
Intermediate odd-odd nuclei, the corresponding 
even-even core nuclei, and the neighboring 
odd-$N$ and odd-$Z$ nuclei. 
}
 \begin{center}
 \begin{ruledtabular}
  \begin{tabular}{lcc}
Odd-odd nucleus & Odd-$N$ nucleus & Odd-$Z$ nucleus \\
\hline
$^{48}_{21}$Sc$_{27}=^{46}_{20}$Ca$_{26}+n+p$ & 
$^{47}_{20}$Ca$_{27}$ & $^{47}_{21}$Sc$_{26}$ \\
$^{76}_{33}$As$_{43}=^{76}_{32}$Ge$_{44}-n+p$ & 
$^{75}_{32}$Ge$_{43}$ & $^{77}_{33}$As$_{44}$ \\
$^{82}_{35}$Br$_{47}=^{82}_{34}$Se$_{48}-n+p$ & 
$^{81}_{34}$Se$_{47}$ & $^{83}_{35}$Br$_{48}$ \\
$^{96}_{41}$Nb$_{55}=^{96}_{42}$Mo$_{54}+n-p$ & 
$^{97}_{42}$Mo$_{55}$ & $^{95}_{41}$Nb$_{54}$ \\
$^{100}_{43}$Tc$_{57}=^{100}_{44}$Ru$_{56}+n-p$ & 
$^{101}_{44}$Ru$_{57}$ & $^{99}_{43}$Tc$_{56}$ \\
$^{110}_{47}$Ag$_{63}=^{110}_{48}$Cd$_{62}+n-p$ & 
$^{111}_{48}$Cd$_{63}$ & $^{109}_{47}$Ag$_{62}$ \\
$^{116}_{49}$In$_{67}=^{118}_{50}$Sn$_{68}-n-p$ & 
$^{117}_{50}$Sn$_{67}$ & $^{117}_{49}$In$_{68}$ \\
$^{124}_{51}$Sb$_{73}=^{124}_{50}$Sn$_{74}-n+p$ & 
$^{123}_{50}$Sn$_{73}$ & $^{125}_{51}$Sb$_{74}$ \\
$^{128}_{53}$I$_{75}=^{128}_{52}$Te$_{76}-n+p$ & 
$^{127}_{52}$Te$_{75}$ & $^{129}_{53}$I$_{76}$ \\
$^{130}_{53}$I$_{77}=^{130}_{52}$Te$_{78}-n+p$ & 
$^{129}_{52}$Te$_{77}$ & $^{131}_{53}$I$_{78}$ \\
$^{136}_{55}$Cs$_{81}=^{136}_{54}$Xe$_{82}-n+p$ & 
$^{135}_{54}$Xe$_{81}$ & $^{137}_{55}$Cs$_{82}$ \\
$^{150}_{61}$Pm$_{89}=^{148}_{60}$Nd$_{88}+n+p$ & 
$^{149}_{60}$Nd$_{89}$ & $^{149}_{61}$Pm$_{88}$ \\
$^{198}_{79}$Au$_{119}=^{200}_{80}$Hg$_{120}-n-p$ & 
$^{199}_{80}$Hg$_{119}$ & $^{199}_{79}$Au$_{120}$ \\
  \end{tabular}
 \end{ruledtabular}
 \end{center}
\end{table}

The single-nucleon Hamiltonian 
$\hf^{\rho}$ takes the form 
\begin{align}
\label{eq:hf}
 \hf^{\rho} = -\sum_{\jr}\epsilon_{\jr}\sqrt{2\jr+1}
  (a_{\jr}^\+\times\tilde a_{\jr})^{(0)}
\equiv
\sum_{\jr}\epsilon_{\jr}\hat{n}_{\jr},
\end{align}
where $\epsilon_{\jr}$ stands for the 
single-particle energy of the odd neutron 
or proton orbital $\jr$, with $\rho=\nu$ or $\pi$. 
$a_{\jr}^{(\+)}$ represents 
a particle annihilation (or creation) operator, 
with $\tilde{a}_{\jr}$ defined by 
$\tilde{a}_{\jr\mr}=(-1)^{\jr -\mr}a_{\jr-\mr}$. 
The operator $\hat{n}_{\jr}$ stands for the number operator 
for the unpaired particle.

The boson-fermion interaction 
$\hbf^{\rho}$ here has a specific form
\begin{equation}
\label{eq:hbf}
 \hbf^{\rho}
=\Gamma_{\rho}\hat{V}_{\mathrm{dyn}}^{\rho}
+\Lambda_{\rho}\hat{V}_{\mathrm{exc}}^{\rho}
+A_{\rho}\hat{V}_{\mathrm{mon}}^{\rho}. 
\end{equation}
The first, second, and third 
terms are dynamical quadrupole, 
exchange, and monopole interactions, respectively, 
and are given as
\begin{align}
\label{eq:dyn}
&\hat{V}_{\mathrm{dyn}}^{\rho}
=\sum_{\jr\jr'}\gamma_{\jr\jr'}
(a^{\+}_{\jr}\times{a}_{\jr'})^{(2)}
\cdot\hat{Q}_{\rho'},\\
\label{eq:exc}
&\hat{V}^{\rho}_{\mathrm{exc}}
=-\left(
s_{\rho'}^\+\times\tilde{d}_{\rho'}
\right)^{(2)}
\cdot
\sum_{\jr\jr'\jr''}
\sqrt{\frac{10}{N_{\rho}(2\jr+1)}}
\beta_{\jr\jr'}\beta_{\jr''\jr} \nonumber \\
&{\quad}\times:\left(
(d_{\rho}^{\+}\times\tilde{a}_{\jr''})^{(\jr)}\times
(a_{\jr'}^{\+}\times\tilde{s}_{\rho})^{(\jr')}
\right)^{(2)}:
+ (H.c.),\\
\label{eq:mon}
&\hat{V}_{\mathrm{mon}}^{\rho}
=\hat{n}_{d_{\rho}}\hat{n}_{\rho}. 
\end{align} 
Here the factors 
$\gamma_{\jr\jr'}=(u_{\jr}u_{\jr'}-v_{\jr}v_{\jr'})Q_{\jr\jr'}$, 
and $\beta_{\jr\jr'}=(u_{\jr}v_{\jr'}+v_{\jr}u_{\jr'})Q_{\jr\jr'}$, 
with 
$Q_{\jr\jr'}=\braket{\ell_{\rho}\frac{1}{2}\jr\|Y^{(2)}\|\ell'_\rho\frac{1}{2}\jr'}$ 
the matrix element of the fermion 
quadrupole operator in the single-particle basis. 
$\hat{Q}_{\rho'}$ in (\ref{eq:dyn}) is the same boson 
quadrupole operator as in the boson 
Hamiltonian (\ref{eq:hb}). 
The notation $:(\cdots):$ in (\ref{eq:exc}) 
means normal ordering. 

The choice of the boson-fermion interactions 
$\hbf^{\rho}$ of the specific forms 
(\ref{eq:dyn}), (\ref{eq:exc}), and (\ref{eq:mon}) 
follows the earlier microscopic 
considerations, based on the generalized seniority scheme 
\cite{scholten1985,IBFM}, that the dynamical 
and exchange terms 
are dominated by the interactions between 
unlike particles, while the monopole term between like particles. 
In addition, within this scheme 
the single-particle energy $\epsilon_{\jr}$ 
in Eq.~(\ref{eq:hf}) is replaced with the quasiparticle energy 
denoted by $\tilde \epsilon_{\jr}$. 

Special attention is paid to the treatment of 
those nuclei with $N_{\nu}=0$ and/or $N_{\pi}=0$. 
If the boson core has 
$N_{\nu}=0$ and $N_{\pi}\neq0$ ($N_{\pi}=0$ and $N_{\nu}\neq0$) 
bosons, and the odd particle is a neutron (proton), 
then any like-particle interaction such as the 
monopole interaction of the form (\ref{eq:mon}) 
vanishes, and only the dynamical term (\ref{eq:dyn}) 
is considered. 
There is also no exchange interaction in this case, 
as it includes like-particle couplings [see the second 
line of Eq.~(\ref{eq:exc})]. 
If, on the other hand, the boson core has   
$N_{\nu}\neq0$ and $N_{\pi}=0$ ($N_{\pi}\neq0$ and $N_{\nu}=0$) 
bosons, and the odd particle is a neutron (proton), then 
there is no unlike particle coupling and 
the dynamical and exchange terms in Eqs.~(\ref{eq:dyn}) and 
(\ref{eq:exc}) are replaced with the like-particle
interactions of the forms
\begin{align}
\label{eq:dyn-semimagic}
\hat{V}^{\rho}_{\mathrm{dyn}}
=&\sum_{\jr\jr'}\gamma_{\jr\jr'}(a^{\+}_{\jr}\times{a}_{\jr'})^{(2)}
\cdot\hat{Q}_{\rho}, \\
\label{eq:exc-semimagic}
\hat{V}^{\rho}_{\mathrm{exc}}=
&-\sum_{j_{\rho}j_{\rho}^{\prime}j_{\rho}^{\prime\prime}}
2\sqrt{\frac{5}{2\jr''+1}}\beta_{\jr\jr''}
\beta_{j_{\rho}'j_{\rho}''} \nonumber \\
&\times:\left(
(d_{\rho}^{\+}\times\tilde a_{j_{\rho}})^{(\jr'')}
\times
(a^{\+}_{j_{\rho}^{\prime}}\times\tilde{d}_{\rho})^{(\jr'')}
\right)^{(0)}:.
\end{align}

As for the residual neutron-proton interaction $\hff$ 
in (\ref{eq:ham-ibffm2}), 
the following form \cite{brant1988} is used:
\begin{align}
\label{eq:hff}
\hff
&=4\pi
\left[{\vd}+\vssd
{\bm\sigma}_{\nu}\cdot{\bm\sigma}_{\pi}
\right]
\delta(\bm{r})
\delta(\bm{r}_{\nu}-r_0)
\delta(\bm{r}_{\pi}-r_0)
\nonumber\\
&-\frac{1}{\sqrt{3}}\vsss
{\bm\sigma}_{\nu}\cdot{\bm\sigma}_{\pi}
+\vt
\left[
\frac{3({\bm\sigma}_{\nu}\cdot{\bf r})
({\bm\sigma}_{\pi}\cdot{\bf r})}{r^2}
-{\bm{\sigma}}_{\nu}
\cdot{\bm{\sigma}}_{\pi}
\right], 
\end{align}
where the first term comprises delta  
and spin-spin delta interactions, and the second, 
and third terms denote spin-spin, and tensor 
interactions, respectively. 
$\vd$, $\vssd$, $\vsss$, and $\vt$
are parameters, while 
$\bm{r}=\bm{r}_{\nu}-\bm{r}_{\pi}$ 
and $r_0=1.2A^{1/3}$ fm.

With the IBM-2 Hamiltonian determined by the procedure 
described in the previous section, 
the full IBFFM-2 Hamiltonian (\ref{eq:ham-ibffm2}) is 
constructed by following the procedures developed in 
Refs.~\cite{nomura2016odd,nomura2019dodd}: 
\begin{enumerate}
 \item[(i)] 
The quasiparticle energies 
$\tilde\epsilon_{\jr}$ and occupation probabilities 
$v^{2}_{\jr}$ of the odd nucleons 
are calculated self-consistently 
by the RHB method constrained to zero 
deformation $\beta=0$. 
These quantities are then 
input to the single-nucleon Hamiltonian $\hf^\rho$ 
(\ref{eq:hf}) 
and the boson-fermion interactions $\hbf^\rho$, 
defined in 
Eqs.~(\ref{eq:hbf}), (\ref{eq:dyn}), (\ref{eq:exc}), 
(\ref{eq:dyn-semimagic}), and (\ref{eq:exc-semimagic}). 
The single-particle energies $\epsilon_{\jr}$ 
and the $v^2_{\jr}$ values 
are shown in Tables~\ref{tab:spe1}, \ref{tab:spe2}, 
\ref{tab:spe3}, and \ref{tab:spe4} of Appendix~\ref{sec:spe}. 
\item[(ii)]
The strength parameters $\Gamma_{\rho}$, $\Lambda_{\rho}$, 
and $A_{\rho}$ are determined so as to reproduce 
the experimental data on the low-energy excitation spectra 
of neighboring odd-$N$ and odd-$Z$ nuclei within 
the neutron-proton 
interacting boson-fermion model (IBFM-2) 
\cite{iachello1979,scholten1985,IBFM}, separately for 
positive- and negative-parity states. 
The relevant odd-mass nuclei, based on 
which the boson-fermion interaction strengths for the IBFFM-2 
Hamiltonian are determined, are summarized in 
Table~\ref{tab:odda}. 
\item[(iii)] 
The parameters for the residual interaction $\hff$, 
i.e., $\vd$, $\vssd$, $\vsss$, and $\vt$, are 
determined to reproduce excitation spectra of low-lying 
positive-parity states for each odd-odd nucleus. 
The spin-spin term in Eq.~(\ref{eq:hff}) 
is found to make a minor contribution to the low-lying states, 
hence is neglected. 
The adopted strength parameters for the interaction term 
$\hff$, as well as the ones for $\hbf^\rho$, 
are listed in Table~\ref{tab:paraff} of Appendix~\ref{sec:paraff}.
\end{enumerate}
The IBFFM-2 Hamiltonian thus determined is diagonalized 
using the computer program TWBOS \cite{TWBOS} 
in the basis 
$\ket{\left[(L_{\nu}L_{\pi})^{(L)}(j_{\nu}j_{\pi})^{(J)}\right]^{(I)}}$, 
where $L_{\nu}$ ($L_{\pi}$) and $L$ are angular momentum 
for neutron (proton) boson system, and the total angular 
momentum of the even-even boson core, respectively. 
$I$ stands for the total angular momentum of the 
coupled system.

\subsection{$\tnbb$ decay within the IBM}

To a good approximation, 
the $\tnbb$-decay half-life $\tau_{1/2}^{(2\nu)}$ 
can be expressed by the factorized 
form (see, e.g., \cite{tomoda1991})
\begin{align}
\label{eq:taubb}
 \left[\tau_{1/2}^{(2\nu)}\right]^{-1}
=G_{2\nu}|M_{2\nu}|^{2}
\end{align}
where $G_{2\nu}$ is the phase-space factor  
in year$^{-1}$ for the $0^+\to0^+$ $\tnbb$ decay, 
and $M_{2\nu}$ represents the NME given by
\begin{align}
\label{eq:mbb}
 M_{2\nu}=\ga^{2}\cdot{m_{e}c^{2}}
\left[
\mgt - \left(\frac{g_\mathrm{V}}{\ga}\right)^{2}\mf
\right],
\end{align}
with $\gv=1$ and $\ga=1.269$ the vector and 
axial vector coupling constants, respectively. 
$\mgt$ and $\mf$ are Gamow-Teller (GT) and Fermi (F) 
matrix elements, respectively, and are given by
\begin{align}
\label{eq:mgt}
 &\mgt=\sum_{N}
\frac{\braket{0^{+}_{F}\|t^{+}\sigma\|1^{+}_{N}}\braket{1^{+}_{N}\|t^{+}\sigma\|0^{+}_{1,I}}}
{E_{N}-E_{I}+\frac{1}{2}(Q_{\beta\beta}+2m_{e}c^{2})}
\\
\label{eq:mf}
&\mf=\sum_{N}
\frac{\braket{0^{+}_{F}\|t^{+}\|0^{+}_{N}}\braket{0^{+}_{N}\|t^{+}\|0^{+}_{1,I}}}
{E_{N}-E_{I}+\frac{1}{2}(Q_{\beta\beta}+2m_{e}c^{2})},
\end{align}
where $t^{\pm}$ represents the isospin raising or 
lowering operator, $\bm{\sigma}$ is the spin operator, 
$\qbb$ is the $\db$ 
decay $Q$ value, and $E_{I}$ ($E_{N}$) stands for the energy 
of the initial (intermediate) state. 
The sums in Eqs.~(\ref{eq:mgt}) and (\ref{eq:mf}) are 
taken over all the intermediate states $1^{+}_{N}$ and 
$0^{+}_{N}$ with the excitation energies $E_x$ 
below 10 MeV. 
The validity of this energy cutoff is discussed 
in Sec.~\ref{sec:comp-ex}. 
The IBFFM-2 code used in this calculation generates, but 
is limited to, a maximum of $N=200$ eigenvalues 
for a given angular momentum $I$.

The boson images of the Fermi ($t^\pm$) and Gamow-Teller 
($t^\pm\sigma$) transition operators, denoted respectively 
by $\hat{T}^\mathrm{F}$ and 
$\hat{T}^\mathrm{GT}$, take the forms
\begin{align}
\label{eq:ofe}
&t^\pm\longmapsto\hat{T}^{\rm F}
=\sum_{j_{\nu}j_{\pi}}
\eta_{j_{\nu}j_{\pi}}^{\mathrm{F}}
\left(\hat P_{j_{\nu}}\times\hat P_{j_{\pi}}\right)^{(0)}, \\
\label{eq:ogt}
&t^{\pm}\sigma\longmapsto\hat{T}^{\rm GT}
=\sum_{j_{\nu}j_{\pi}}
\eta_{j_{\nu}j_{\pi}}^{\mathrm{GT}}
\left(\hat P_{j_{\nu}}\times\hat P_{j_{\pi}}\right)^{(1)}, 
\end{align}
where the coefficients 
$\eta_{j_{\nu}j_{\pi}}^{\mathrm{F}}$ and 
$\eta_{j_{\nu}j_{\pi}}^{\mathrm{GT}}$ are, 
to the lowest order,  
\begin{align}
\label{eq:eta}
\eta_{j_{\nu}j_{\pi}}^{\mathrm{F}}
&=-\sqrt{2j_{\nu}+1}
\delta_{j_{\nu}j_{\pi}},
\\
\eta_{j_{\nu}j_{\pi}}^{\mathrm{GT}}
&= - \frac{1}{\sqrt{3}}
\left\langle
\ell_{\nu}\frac{1}{2};j_{\nu}
\bigg\|{\bm\sigma}\bigg\|
\ell_{\pi}\frac{1}{2};j_{\pi}
\right\rangle
\delta_{\ell_{\nu}\ell_{\pi}}.
\end{align}
$\hat P_{\jr}$ is here given by one of the 
one-particle creation operators
\begin{subequations}
 \begin{align}
\label{eq:creation1}
&A^{\+}_{\jr\mr} = \zeta_{\jr} a_{{\jr}\mr}^{\+}
 + \sum_{\jr'} \zeta_{\jr\jr'} s^{\+}_\rho (\tilde{d}_{\rho}\times a_{\jr'}^{\+})^{(\jr)}_{\mr}
\\
\label{eq:creation2}
&B^{\+}_{\jr\mr}
=\theta_{\jr} s^{\+}_\rho\tilde{a}_{\jr\mr}
 + \sum_{\jr'} \theta_{\jr\jr'} (d^{\+}_{\rho}\times\tilde{a}_{\jr'})^{(\jr)}_{\mr},
\end{align}
and annihilation operators
\begin{align}
\label{eq:annihilation1}
&\tilde{A}_{\jr\mr}=(-1)^{\jr-\mr}A_{\jr-\mr}\\ 
\label{eq:annihilation2}
&\tilde{B}_{\jr\mr}=(-1)^{\jr-\mr}B_{\jr-\mr}.  
\end{align}
\end{subequations}
The operators in Eqs.~(\ref{eq:creation1}) 
and (\ref{eq:annihilation1}) 
conserve the boson number, whereas those 
in Eqs.~(\ref{eq:creation2}) and (\ref{eq:annihilation2}) 
do not. 
The GT (\ref{eq:ogt}) and Fermi (\ref{eq:ofe}) 
operators are formed as a pair of the above operators, 
depending on the type of the $\beta$ decay 
under study (i.e., $\beta^+$ or $\btm$) and 
on the particle or hole nature of bosons in 
the even-even IBM-2 core. 
It is also noted that the expressions in 
(\ref{eq:creation1}), (\ref{eq:creation2}), 
(\ref{eq:annihilation1}), and (\ref{eq:annihilation2}) 
are of simplified forms of 
the most general one-particle transfer operators 
in the IBFM-2 \cite{IBFM}. 
Coefficients $\zeta_{j}$, $\zeta_{jj'}$, 
$\theta_{j}$, and $\theta_{jj'}$ 
are determined by using the same $v_{\jr}^2$ values 
used in the IBFFM-2 calculations for 
the odd-odd nuclei. 
The expressions for these coefficients are 
given in Appendix~\ref{sec:opt}. 
Note that no phenomenological 
parameter is introduced in the formulas in 
(\ref{eq:creation1})--(\ref{eq:annihilation2}). 

%
\begin{figure*}
\begin{center}
\includegraphics[width=.8\linewidth]{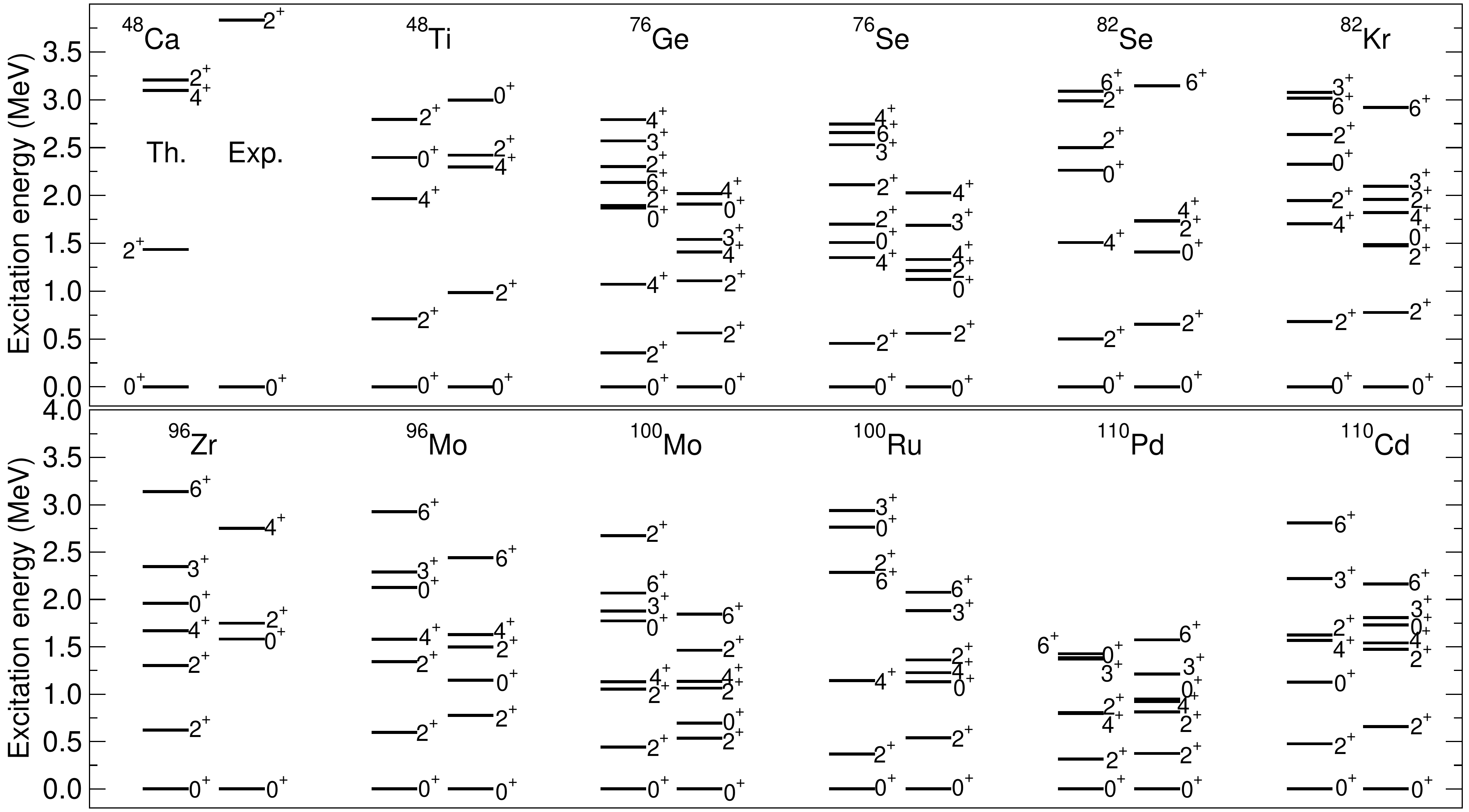}
\includegraphics[width=.8\linewidth]{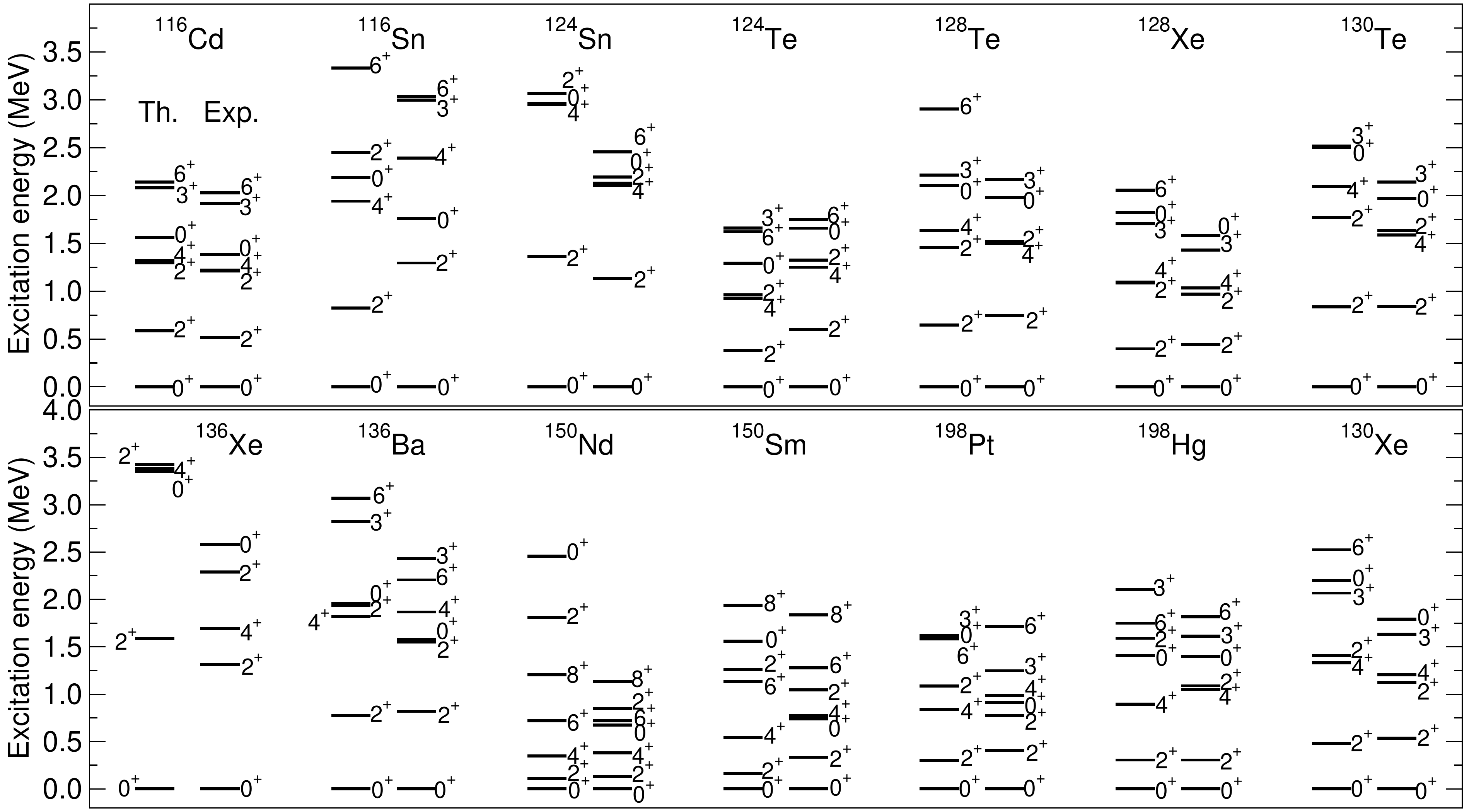}
\caption{Comparison of theoretical and experimental 
low-energy spectra of the initial and final even-even nuclei 
of the $\tnbb$ decays with mass number $A=48$ 
($^{48}$Ca and $^{48}$Ti) to 
$A=198$ ($^{198}$Pt and $^{198}$Hg). 
For each nucleus, the calculated energy spectrum
is shown on the left-hand side, and 
the experimental one on the right. 
The experimental data are taken from Ref.~\cite{data}.}
\label{fig:ee}
\end{center}
\end{figure*}

%
\begin{figure*}
\begin{center}
\includegraphics[width=.8\linewidth]{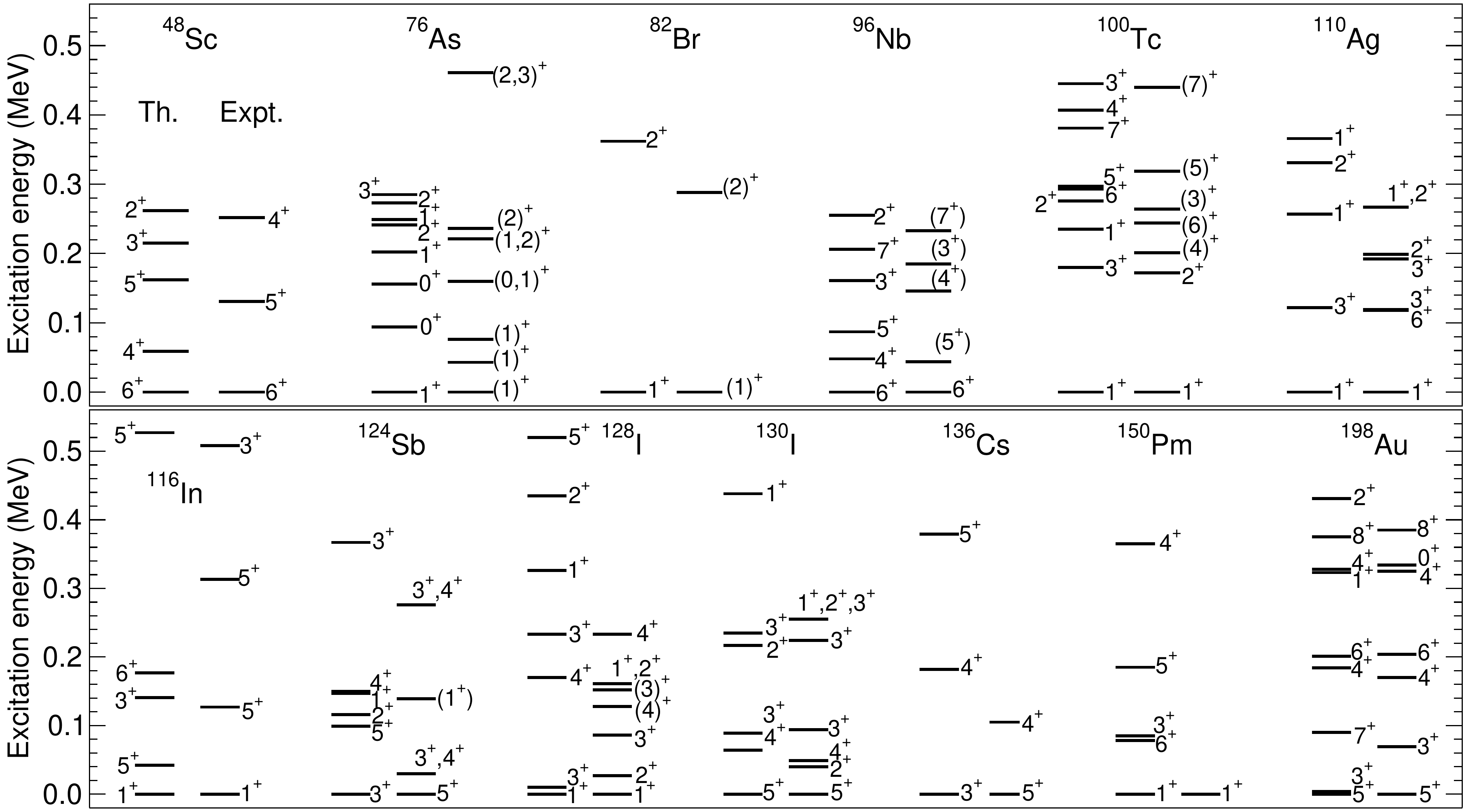}
\caption{Same as Fig.~\ref{fig:ee}, but for the intermediate 
odd-odd nuclei.}
\label{fig:oo}
\end{center}
\end{figure*}

\section{Excitation spectra\label{sec:energy}}

\subsection{Initial and final even-even nuclei }

Figure~\ref{fig:ee} 
shows the results for low-energy excitation 
spectra of the even-even initial and 
final nuclei, obtained by the diagonalization \cite{NPBOS} of the mapped 
IBM-2 Hamiltonian. 
In general, the excitation spectra calculated 
within the mapped IBM-2 are in reasonable 
agreement with the experimental ones.

It is worth mentioning that the energies of the 
excited $0^{+}$ states 
are systematically overestimated. 
This can be explained in part by the too large 
quadrupole-quadrupole boson interaction strength 
$\kappa$ [cf. Eq.~(\ref{eq:hb})], as compared to those 
values used in the standard IBM fitting calculations 
(see, e.g., Table XXIII of \cite{barea2013}). 
The unusually large strength parameter might also 
imply a certain deficiency of the underlying EDF. 
The present SCMF calculation within the RHB method 
with DD-PC1 functional generally 
yields a too steep valley around the global minimum. 
To reproduce this topology, an unexpectedly large 
quadrupole-quadrupole boson interaction 
strength has to be chosen. 
The same problem is encountered when other types of the 
relativistic and non-relativistic EDFs 
are used as the microscopic input to the IBM-2 
(see, e.g., \cite{nomura2010,nomura2011sys}). 
The discrepancy in the $0^+_2$ energy levels 
could also be attributed to 
the specific form of the IBM-2 Hamiltonian. 
In principle, a more general IBM-2 Hamiltonian should include 
some other terms such as the so-called 
Majorana terms \cite{OAIT}, while the boson model space 
needs to be extended to include some additional degrees 
of freedom such as the intruder states and the 
subsequent configuration mixing \cite{duval1981}. 
As shown in Refs.~\cite{nomura2012sc,nomura2016sc}, 
these extensions indeed improve 
the description of the excited $0^{+}$ states, 
but also involve more parameters. 
In order not to complicate the problem, 
this point will not be pursued further in the 
present paper.

The structure of $^{96}$Zr is of particular interest. 
In Fig.~\ref{fig:ee}, the experimental data show 
the high-lying $2^{+}_{1}$ state, indicating 
the proton $Z=40$ and neutron $N=56$ doubly subshell closure, 
and the presence of the low-lying $0^{+}_{2}$ level 
below the $2^{+}_{1}$ one, 
a signature of shape coexistence \cite{kremer2016}. 
The present IBM-2 produces a more rotational-like 
spectrum for $^{96}$Zr, giving much lower $2^+_1$ energy 
level than in the experimental spectrum. 
Note that the corresponding 
SCMF energy surface is soft in $\gamma$, 
but is rigid in $\beta$ 
deformations, while there is no spherical-deformed 
shape coexistence as suggested empirically 
(see Fig.~\ref{fig:pesdft}). 

The energy spectra of the doubly and semi-magic 
nuclei $^{48}$Ca, $^{116}$Sn, 
$^{124}$Sn, and $^{136}$Xe are also not satisfactorily 
reproduced. This is obviously because the IBM-2 comprises 
the collective degrees of freedom only, and 
is not able to reproduce the excitation spectra of 
these nuclei, which are supposed to be determined mainly 
by non-collective (or single-particle) excitations. 

The resultant IBM-2 wave functions give rise to the 
electromagnetic transition properties for the even-even 
initial and final nuclei of the $\tnbb$ decays under 
investigation. The relevant results 
are shown in Figs.~\ref{fig:ee1-em} and \ref{fig:ee2-em} 
of Appendix~\ref{sec:doo-em}, where some details about the 
results are also given.

\subsection{Intermediate odd-odd nuclei}

Figure~\ref{fig:oo} compares the IBFFM-2 and experimental 
\cite{data,bucurescu2012} energy spectra for the 
low-lying positive-parity states of the 
intermediate odd-odd nuclei. 
The agreement between the calculated and experimental 
spectra is generally good. 
Especially the fact that the correct ground-state 
spin is obtained for most of the odd-odd nuclei 
is satisfactory, since for the calculation 
of $\tnbb$ decay, the excitation energies and wave functions 
of the $1^{+}$ and $0^{+}$ states are particularly important. 
For all the odd-odd nuclei except for 
$^{76}$As, the $0^{+}$ states are, in general, 
predicted to be higher in energy than the $1^+$ states, 
with the lowest $0^+$ level calculated at the excitation 
energy $E_{x}$ higher than 0.5 MeV. 

By using the resulting IBFFM-2 wave functions, 
the electromagnetic properties of the odd-odd nuclei, 
including the 
electric quadrupole $Q(I)$ and magnetic dipole $\mu(I)$ moments, 
and the $B(E2)$ and $B(M1)$ transition rates, 
for lowest states are readily calculated. 
Especially, the calculation provides a reasonable description 
of the observed $Q(I)$ and $\mu(I)$ moments 
for a majority of the considered odd-odd nuclei, 
when typical effective 
$g$ factors and boson charges for the $M1$ and $E2$ operators 
are used. These electromagnetic properties are useful to 
judge if the wave functions for the odd-odd nuclei are 
reliable. Nevertheless, the primary goal of this work is the 
calculation of $\tnbb$ decay, hence some details about the 
results for the electromagnetic properties are 
given in Appendix~\ref{sec:doo-em}. 

\begin{figure}[ht]
\begin{center}
\includegraphics[width=.8\linewidth]{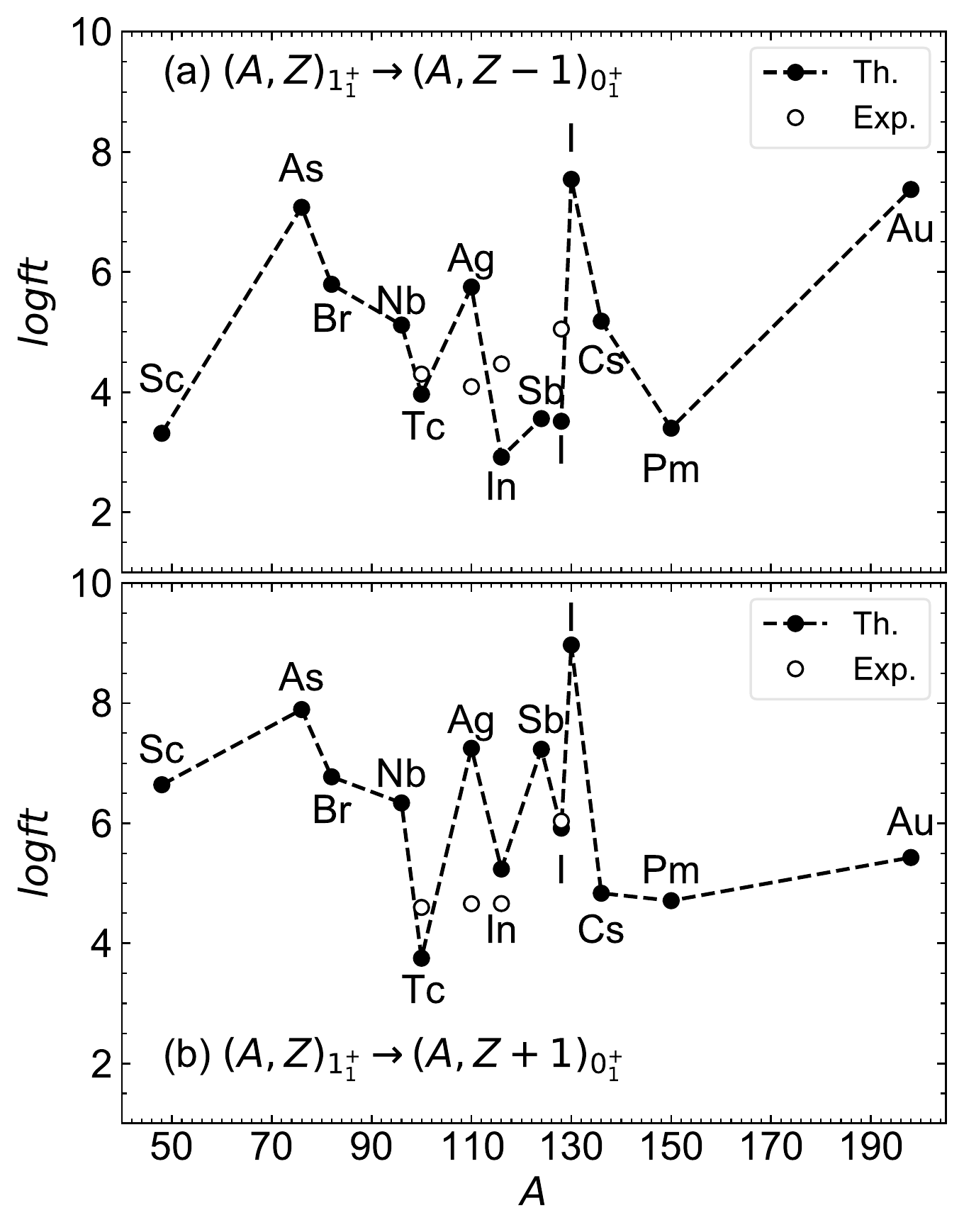}
\caption{Calculated and experimental 
\cite{data} $\log{ft}$ values 
for (a) electron-capture (EC) and 
(b) $\beta^{-}$ decays from the $1^{+}_{1}$ 
state of odd-odd nuclei to the ground state $0^{+}_{1}$ 
of initial- and final-state even-even nuclei, respectively. 
For the experimental data, error bars are not shown 
because they are smaller than the marker size.}
\label{fig:logft10}
\end{center}
\end{figure}

\begin{figure*}[ht]
\begin{center}
\begin{tabular}{rl}
\includegraphics[width=.49\linewidth]{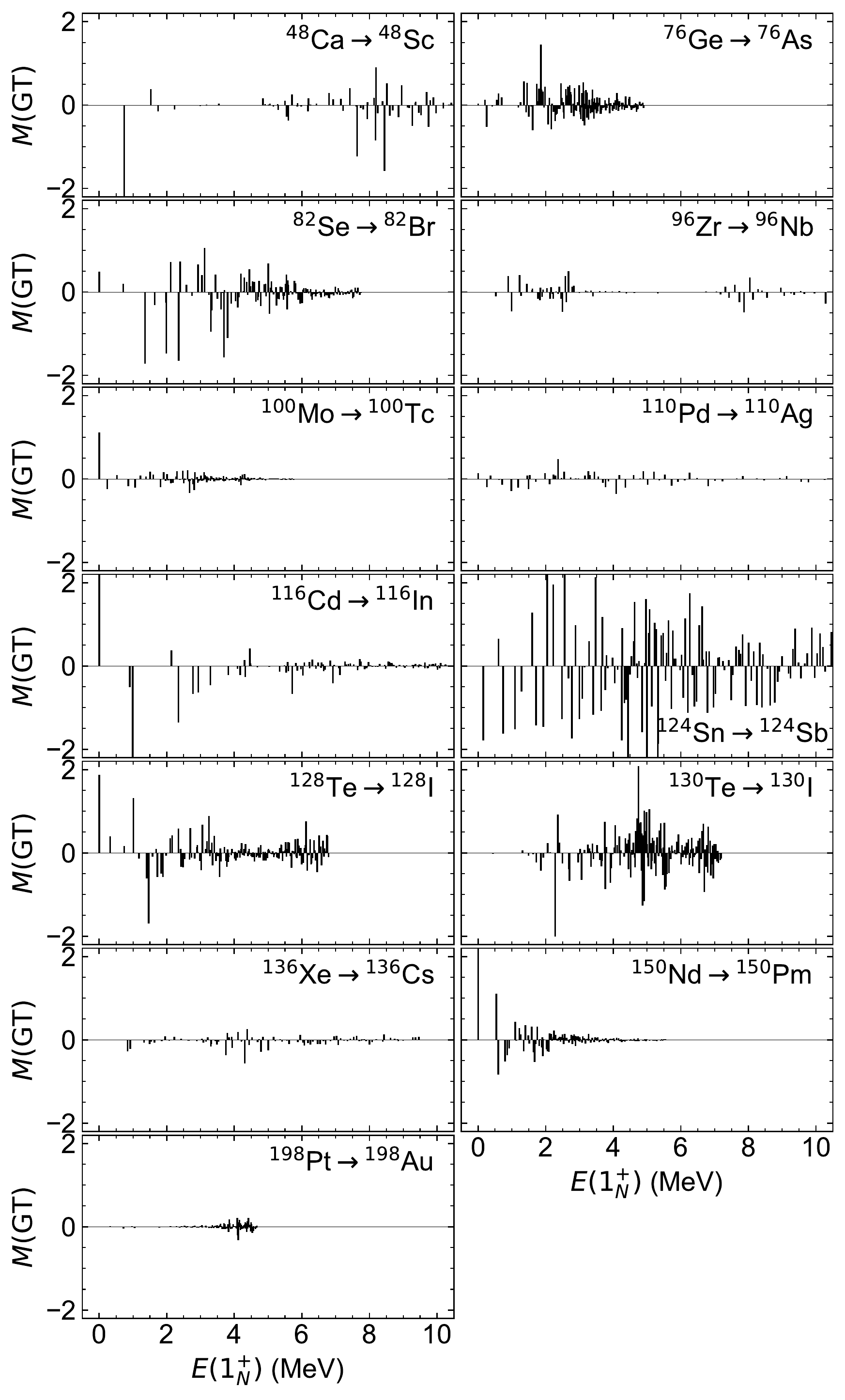} & 
\includegraphics[width=.49\linewidth]{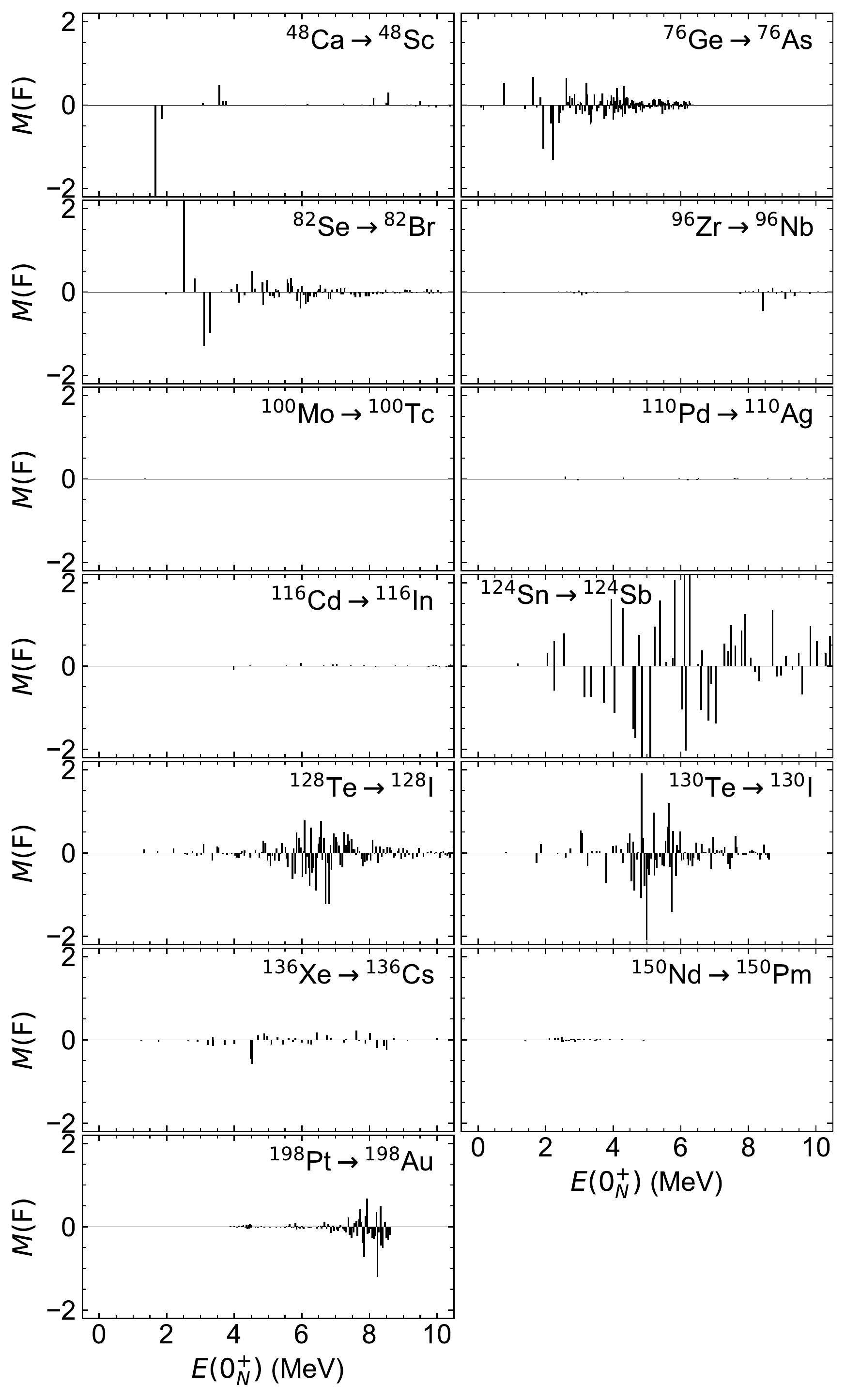}\\
\end{tabular}
\caption{Gamow-Teller $M(\mathrm{GT}; 0^+_{1,I}\to1^+_N)$
and Fermi $M(\mathrm{F}; 0^+_{1,I}\to0^+_N)$
matrix elements between the initial and 
the $N$th intermediate states, plotted against their 
excitation energies 
$E(1^{+}_{N})$ and $E(0^{+}_{N})$ (in MeV), respectively.}
\label{fig:dist_eo}
\end{center}
\end{figure*}

\begin{figure*}[ht]
\begin{center}
\begin{tabular}{rl}
\includegraphics[width=.49\linewidth]{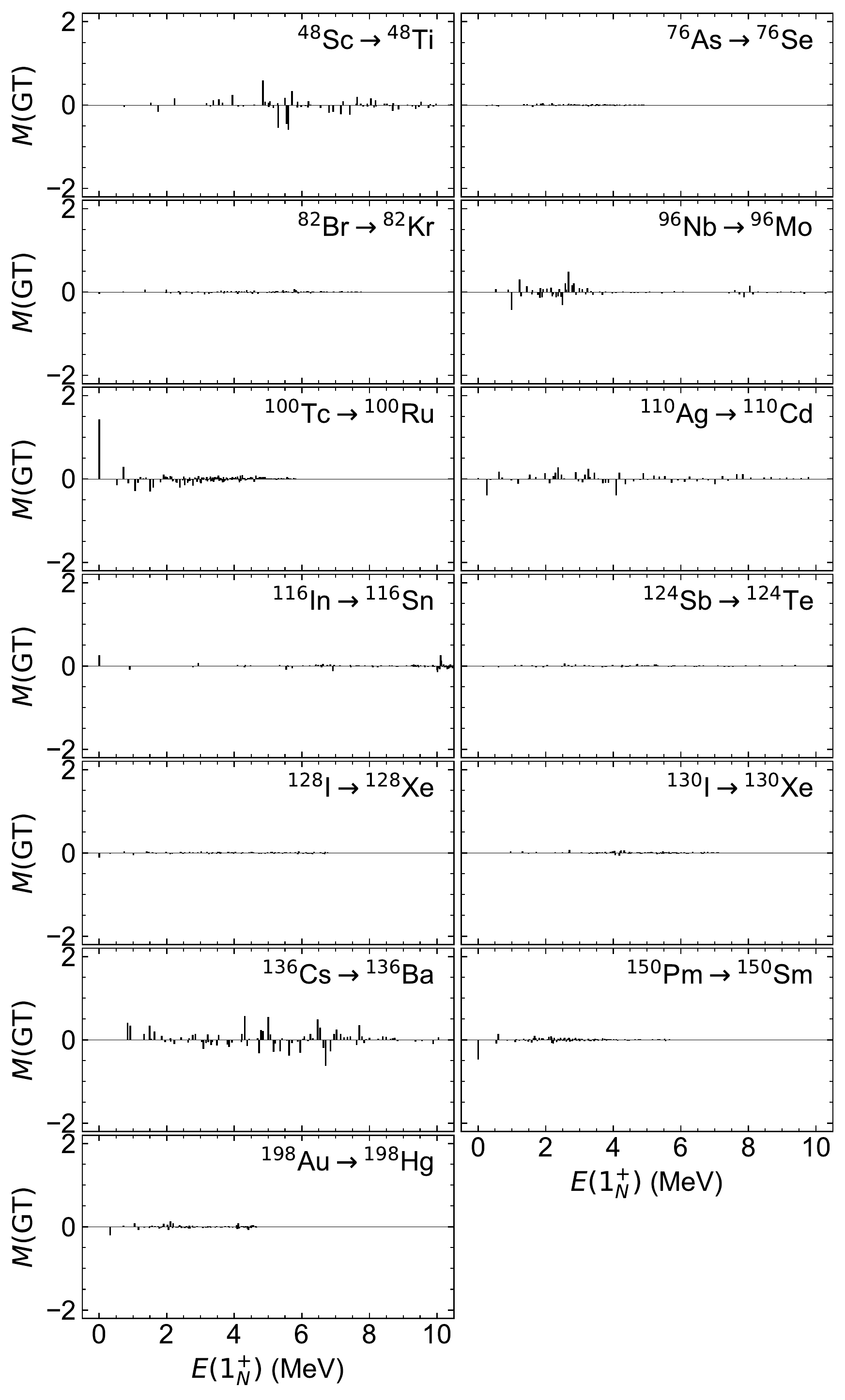} & 
\includegraphics[width=.49\linewidth]{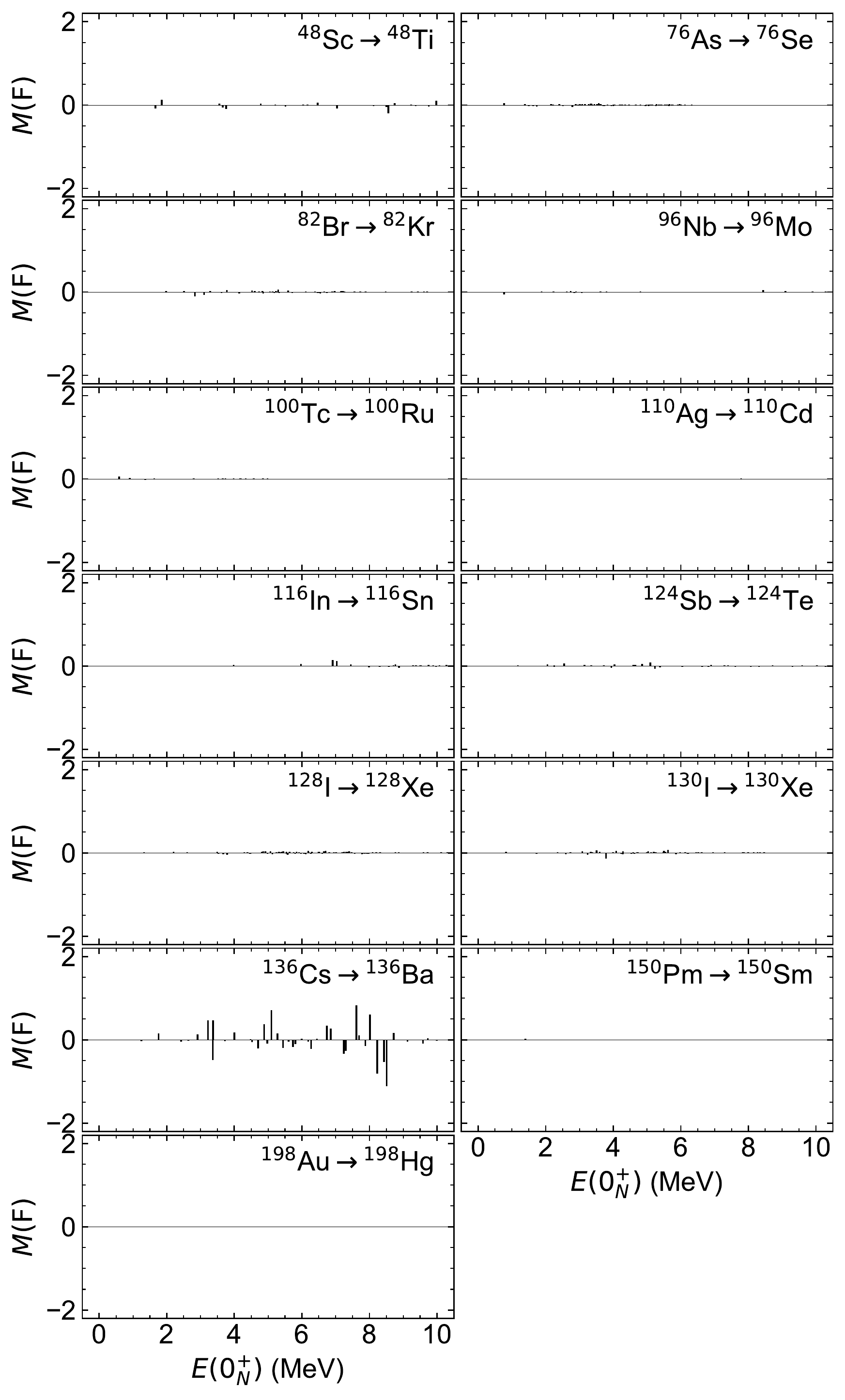}\\
\end{tabular}
\caption{Same as Fig.~\ref{fig:dist_eo}, but for 
$M(\mathrm{GT}; 1^+_{N}\to0^+_{1,F})$ and 
$M(\mathrm{F}; 0^+_{N}\to0^+_{1,F})$ 
between the $N$th intermediate 
and the final states.}
\label{fig:dist_oe}
\end{center}
\end{figure*}

\section{Single $\beta$ decay\label{sec:beta}}

The $ft$ value for the $\beta$ decay is given by
\begin{align}
\label{eq:ft}
ft=\frac{K}{|\mfbeta|^2+\left(\frac{\ga}{\gv}\right)^2|\mgtbeta|^2}. 
\end{align}
The constant $K=6163$ s, and 
$\mfbeta$ and $\mgtbeta$ are the reduced matrix elements 
of the Fermi $\hat{T}^{\mathrm{F}}$ (\ref{eq:ofe}) 
and Gamow-Teller $\hat{T}^{\mathrm{GT}}$ (\ref{eq:ogt}) 
operators, respectively. 
For the calculations of single-$\beta$ decays, 
the free value of the axial vector 
constant $\ga=1.269$ is used, i.e., no quenching 
is introduced.

Figure~\ref{fig:logft10} shows the $\ft$ values 
for (a) $\beta^{+}$ or 
electron-capture (EC) decays, and (b) $\beta^{-}$ 
decays of the $1^{+}_{1}$ state of the intermediate odd-odd nuclei 
into the ground states $0^{+}_{1}$ of the initial 
and final nuclei, respectively. 
The obtained $\ft$ values for the EC 
decay are typically within the range $3\lesssim\ft\lesssim7$, 
while the calculated values for the $\btm$ decays of most 
of the nuclei are larger than 
the ones for the EC decays, i.e., $4\lesssim\ft\lesssim9$. 
Both the EC and $\btm$ decay $\ft$ values strongly 
depend on the mass number $A$. 

On closer inspection, 
the $\ft$ value for the EC decay of the state 
$^{116}$In$(1^{+}_{1})$ is underestimated by a factor of 2. 
For the $^{116}$In nucleus, the neutron-proton pair 
configuration $[\nu{g_{7/2}}\otimes\pi{g_{9/2}}]^{(J=1)}$
accounts for $84\,\%$ of the IBFFM-2 wave function of 
the $1^{+}_{1}$ state. 
This configuration makes a large contribution 
to the $\pi{g_{9/2}}\leftrightarrow\nu{g_{7/2}}$ transitions 
in the GT matrix elements, 
hence the resultant $\ft$ values becomes 
small, particularly for the EC decay 
(Fig.~\ref{fig:logft10}(a)). 
The same observation applies to the single-$\beta$ 
decays of $^{100}$Tc, in which case the predicted $1^+_1$ wave 
function turns out to be similar in structure to the 
one for $^{116}$In, i.e., 
most (86 \%) of the $1^+_1$ wave function for $^{100}$Tc 
is built on the configuration 
$[\nu{g_{7/2}}\otimes\pi{g_{9/2}}]^{(J=1)}$. 
Therefore, the resultant $\ft$ values for the EC and $\btm$ decays 
of $^{100}$Tc are relatively small. 
On the other hand, the $\ft$ values for both the EC and $\btm$ 
decays of $^{110}$Ag are quite large, as compared to 
the neighboring nuclei $^{100}$Tc and $^{116}$In, and also 
overestimate the experimental values. 
In $^{110}$Ag, the neutron-proton pair 
component $[\nu{g_{7/2}}\otimes\pi{g_{9/2}}]^{(J=1)}$ 
plays a less dominant role than in 
$^{100}$Tc and $^{116}$In, as it 
constitutes 64 \% of the $1^+_1$ wave function. 
Consequently, the coupling 
$\pi{g_{9/2}}\leftrightarrow\nu{g_{7/2}}$ 
in the GT strengths plays a much less significant role 
for both the EC and $\btm$ decays 
than in the case of the $^{116}$In and $^{100}$Tc ones, 
yielding the large $\ft$ values. 
The nature of the wave function also appears to be 
sensitive to the parameters for the IBFFM-2. 
For $^{100}$Ag, particularly large strength parameter 
is chosen for the tensor term (see Table~\ref{tab:paraff}). 
It is also worth noticing that the $\ft$ values for 
both the EC and $\btm$ decays of $^{130}$I are quite large. 
In the corresponding IBFFM-2 wave function 
of the $1^{+}_{1}$ state, pair components are 
largely fragmented, with the dominant contributions 
coming from the 
$[\nu{s_{1/2}}\otimes\pi{g_{7/2}}]^{(J=3)}$ ($67\,\%$)
and 
$[\nu{d_{3/2}}\otimes\pi{s_{1/2}}]^{(J=2)}$ ($14\,\%$). 
These configurations do not make any contribution to the 
GT transition, since the only allowed GT transition 
in the present case is the one between states with 
$\ell_\nu=\ell_\pi$ [see Eq.~(\ref{eq:eta})].  
The GT matrix element of the $^{130}$I 
decay is also composed of numerous other terms with the 
amplitudes of small orders of magnitude, 
which cancel each other, leading to the large $\ft$ 
values. 
Cancellations of this kind are supposed to be sensitive 
to the wave functions of the IBM-2 and IBFFM-2, or the 
adopted single-particle energies.

In Fig.~\ref{fig:dist_eo}, the distributions 
of the $\mgtbeta$ and $\mfbeta$ 
matrix elements between the initial 
and the $N$th intermediate states $1^{+}_{N}$ and $0^{+}_{N}$ 
are shown 
as functions of their excitation energies below $E_{x}=10$ MeV. 
In most of the considered decay chains, the 
$M({\mathrm{GT}}; 0^{+}_{1,I}\to1^{+}_{N})$ matrix elements 
appear to be evenly distributed. 
The Fermi matrix elements 
$M({\mathrm{F}}; 0^{+}_{1,I}\to0^{+}_{N})$ are also 
fragmented to a large extent, with generally smaller 
magnitudes than $\mgtbeta$. 
In some of the decay chains, the $\mfbeta$ values 
almost vanish. 
For the $^{100}$Mo$\to^{100}$Tc decay, for instance, 
the only possible transition that contributes to 
$\mfbeta$ is the 
one between the neutron and proton ${1g_{7/2}}$ orbitals. 
But the corresponding pair configurations 
$[\nu{g_{7/2}}\otimes\pi{g_{7/2}}]^{(J)}$ in the IBFFM-2 
wave function of the low-lying $0^{+}$ states 
of $^{100}$Tc are negligibly small, since the proton 
$1g_{7/2}$ orbital basically belongs to the 
next major shell to the 
considered fermion space and its coupling to the 
neutron $1g_{7/2}$ orbital is weak.  
(see Table~\ref{tab:spe2}).

Figure~\ref{fig:dist_oe} shows the 
$M({\mathrm{GT}}; 1^{+}_{N}\to0^{+}_{1,F})$ and 
$M({\mathrm{F}}; 0^{+}_{N}\to0^{+}_{1,F})$ matrix 
elements between the $1^{+}_{N}$ and $0^{+}_{N}$ 
intermediate states to the final state $0^{+}_{1,F}$. 
Both the $\mgtbeta$ and $\mfbeta$ values 
for these decays are much smaller than in 
the case of the $\beta^{+}$/EC ones 
(see Fig.~\ref{fig:dist_eo}). 
It is remarkable that for the $^{100}$Tc$\to^{100}$Mo 
and $^{150}$Pm$\to^{150}$Sm $\btm$ decays, 
the largest $\mgtbeta$ value is obtained at 
the lowest-lying $1^+$ intermediate state.

\section{$\tnbb$ decay\label{sec:dbeta}}

\begin{figure*}[ht]
\begin{center}
\begin{tabular}{rl}
\includegraphics[width=.49\linewidth]{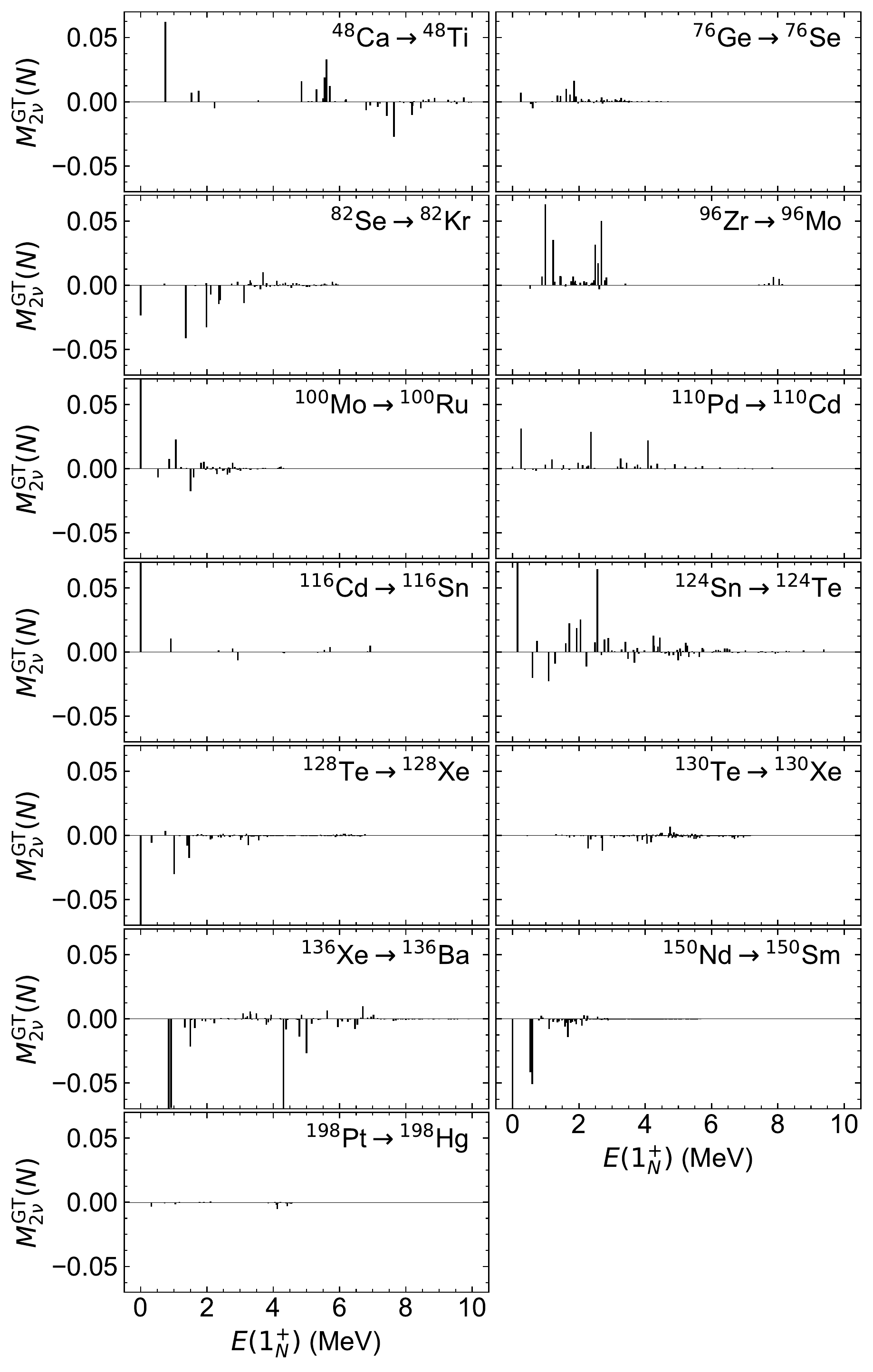} & 
\includegraphics[width=.49\linewidth]{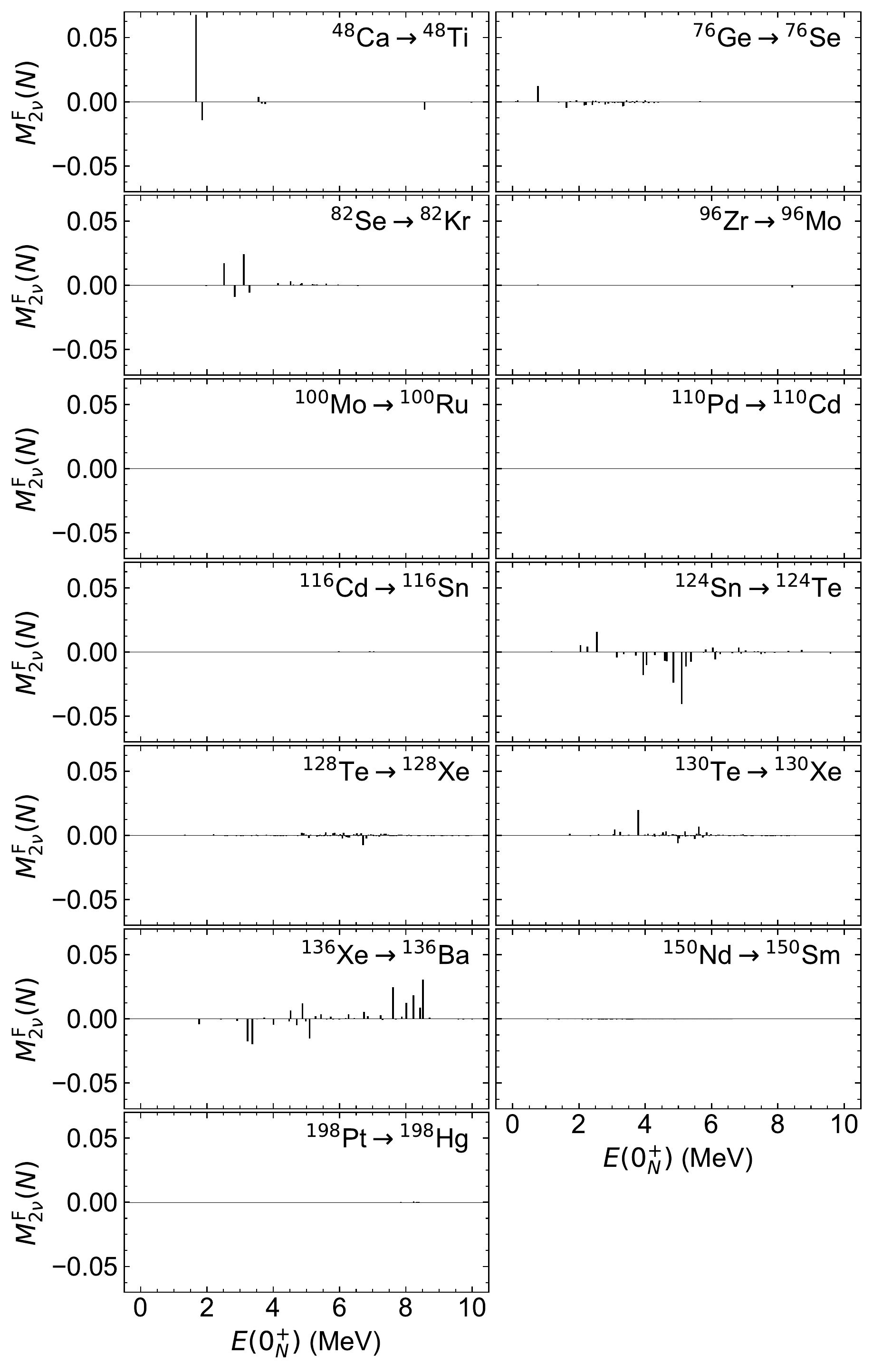}\\
\end{tabular}
\caption{Gamow-Teller 
$\mgt(N)$ (\ref{eq:mgt_dist}) (first and second columns) 
and Fermi $\mf(N)$ (\ref{eq:mf_dist}) (third and fourth columns) 
matrix elements of the $\tnbb$ decays, plotted against 
$E(1^+_N)$ and $E(0^+_N)$, respectively.}
\label{fig:dist_ee}
\end{center}
\end{figure*}

\subsection{Estimation of $\qbb$ values}

As shown in the formulas (\ref{eq:mgt}) 
and (\ref{eq:mf}), the $\qbb$ value is needed for the 
calculation of the $\tnbb$-decay NME. 
The $\qbb$ values is here calculated by using the formula
\begin{align}
\label{eq:qbb}
 \qbb=2(m_{n}-m_{p}-m_{e})c^{2}+\egs^{I}-\egs^{F}
\end{align}
where $m_{n}$, $m_{p}$, and $m_e$ are neutron, proton, 
and electron masses, respectively. 
$\egs^{I}$ (or $\egs^{F}$) stands for the 
ground-state energy of the IBM-2 for the 
initial (or final) nucleus, and is given by 
\begin{align}
\label{eq:egs}
 \egs = E_{\mathrm{IBM}}(0^{+}_{1}) + E_{0},
\end{align}
where the first term stands for the IBM-2 
eigenenergy of the 
$0^{+}_{1}$ ground state and $E_0$ is a constant term 
that depends on nucleon numbers but does not affect 
excitation energies \cite{IBM}.
For each of the initial 
and final nuclei, the constant $E_{0}$ is equated 
to the total mean-field energy 
at the spherical configuration, 
$E_\mathrm{SCMF}(0,\gamma)$  
(see Ref.~\cite{nomura2010} for details). 
One further needs the $Q$ value of the $\btm$ decay 
of the initial nucleus. It could also be determined 
by using the eigenenergies of the IBM-2 and IBFFM-2, 
and the total SCMF energies. 
However, 
the calculation of the $Q_\beta$ value of odd-odd nucleus 
within the present framework would be complicated, 
since it is strongly influenced by the combinations of 
a number of parameters for the IBFFM-2. 
In order to simplify the discussion, 
here the experimental $Q_{\beta}$ values taken from 
\cite{data} are employed.

\begin{table}
\caption{\label{tab:qbb}
Calculated and experimental $\qbb$ values 
(in MeV) of initial even-even nuclei, denoted by $\qbbt$ 
and $\qbbe$, respectively. The $\qbbe$ values are 
taken from Ref.~\cite{data}. 
The $Q_{\beta}$ values 
of the initial nuclei, adopted from 
the data \cite{data}, are also shown.}
 \begin{center}
 \begin{ruledtabular}
  \begin{tabular}{lccc}
Nucleus & $\qbbt$ (MeV)
& $\qbbe$ (MeV)
& $Q_{\beta}$ (MeV) \\
\hline
$^{48}$Ca & $1.8479$ & $4.2681$ & $0.2790$ \\
$^{76}$Ge & $0.8831$ & $2.0391$ & $0.9215$ \\
$^{82}$Se & $1.6356$ & $2.9979$ & $0.0952$ \\
$^{96}$Zr & $4.1285$ & $3.3560$ & $0.1640$ \\
$^{100}$Mo & $2.8338$ & $3.0344$ & $0.1721$ \\
$^{110}$Pd & $2.9081$ & $2.0171$ & $0.8736$ \\
$^{116}$Cd & $6.1166$ & $2.8135$ & $0.4627$ \\
$^{124}$Sn & $-0.3795$ & $2.2927$ & $0.6124$ \\
$^{128}$Te & $-0.1784$ & $0.8680$ & $1.2518$ \\
$^{130}$Te & $1.4466$ & $2.5290$ & $0.4200$ \\
$^{136}$Xe & $0.0989$ & $2.4579$ & $0.0903$ \\
$^{150}$Nd & $3.3123$ & $3.3714$ & $0.0830$ \\
$^{198}$Pt & $1.2895$ & $1.0503$ & $0.3232$ \\
  \end{tabular}
 \end{ruledtabular}
 \end{center}
\end{table}

Table~\ref{tab:qbb} shows the calculated and 
experimental \cite{data} $\qbb$ values, 
denoted by $\qbbt$ and $\qbbe$, respectively.
The $\qbbt$ values for many of the nuclei 
are close to the experimental ones, whereas 
negative $\qbb$ values are obtained for $^{124}$Sn 
and $^{128}$Te. 
The calculation of the $\qbb$ value depends 
on a subtle balance between the SCMF total energies 
for the initial and final nuclei, as well as between the 
eigenenergies of the IBM-2. 
Note that, as for $^{128}$Te, the deformed QRPA 
calculation for $\db$ decay 
with a Skyrme interaction also 
obtained the negative $\qbb$ value \cite{alvarez2004}. 
In what follows, the results obtained from both $\qbbt$ 
and $\qbbe$ values are discussed.

\subsection{GT and F matrix elements\label{sec:gtf}}

One can study strength distributions 
of each terms in the sums in the $\mgt$ (\ref{eq:mgt}) and 
$\mf$ (\ref{eq:mf}) matrix elements, i.e., 
\begin{align}
\label{eq:mgt_dist}
&\mgt(N)\equiv
\frac{\braket{0^{+}_{F}\|t^{+}\sigma\|1^{+}_{N}}\braket{1^{+}_{N}\|t^{+}\sigma\|0^{+}_{1,I}}}
{E_{N}-E_{I}+\frac{1}{2}(Q_{\beta\beta}+2m_{e}c^{2})}
\\
\label{eq:mf_dist}
&\mf(N)\equiv
\frac{\braket{0^{+}_{F}\|t^{+}\|0^{+}_{N}}\braket{0^{+}_{N}\|t^{+}\|0^{+}_{1,I}}}
{E_{N}-E_{I}+\frac{1}{2}(Q_{\beta\beta}+2m_{e}c^{2})}. 
\end{align}
The corresponding results for the ground-state-to-ground-state 
($\trgg$) transitions 
are shown in Fig.~\ref{fig:dist_ee} 
as functions of the excitation energies of the 
$N$th intermediate states $1^{+}_{N}$ and $0^{+}_{N}$. 
In the figure, only the results that employ 
the theoretical $\qbb$ values, i.e., $\qbbt$, calculated by 
using the formulas (\ref{eq:qbb}) and (\ref{eq:egs}), 
are considered, since 
there is no striking difference between the results 
using the $\qbbt$ and $\qbbe$ values at the qualitative level. 

From Fig.~\ref{fig:dist_ee}, 
for most of the decays, 
the low-lying $1^{+}$ states appear to make dominant 
contributions to the GT transitions. 
This finding more or less conforms to the single-state 
dominance (SSD) \cite{griffiths1992,civitarese1998} 
or low-lying state dominance (LLSD) \cite{moreno2008} 
hypotheses. 
In the earlier IBFFM-2 calculation of 
Ref.~\cite{yoshida2013}, the authors found 
more distinct SSD nature in  
the GT strength distributions  
of the $^{128}$Te and $^{130}$Te $\tnbb$ decays. 
One sees in Fig.~\ref{fig:dist_ee}  that 
the Fermi contributions are generally weak, except 
for the $^{48}$Ca, $^{124}$Sn, and $^{136}$Xe decays, 
and their strength distributions do not exhibit 
a characteristic trend as in the case of the GT ones $\mgt$.

Table~\ref{tab:gtf} gives the results 
for Gamow-Teller $\mgt$ 
and Fermi $\mf$ matrix elements for the $\trgg$ transition 
and the ground-state-to-first-excited-state ($\trge$) transition. 
For the $\trgg$ transitions, in most cases 
the calculated values $|\mgt(\trgg)|\approx0.05-0.15$, while 
particularly large $\mgt(\trgg)$ values are obtained 
for the $^{100}$Mo and $^{150}$Nd decays. 
There is also no significant difference 
in the results for the $\mgt$ and $\mf$ matrix elements 
between the calculations with the $\qbbt$ and $\qbbe$ values. 
As for the $\trge$ decays, on the other hand, 
the prediction of $\mgt$ and $\mf$ depends strongly on the 
choice of the $\qbb$ values, particularly  
for the $^{48}$Ca, $^{124}$Sn, and $^{136}$Xe decays. 
Indeed, the corresponding $\qbbt$ values for these processes 
are quite different from the experimental ones $\qbbe$ 
(see Table~\ref{tab:qbb}). 
These nuclei are also all doubly or semimagic nuclei, 
for which the IBM framework is not considered a very 
good approach to give a reasonable $\qbb$ value. 
In addition, both the calculated $\mgt$ and $\mf$ 
values for certain $\trge$ 
transitions, i.e., $^{48}$Ca$\to^{48}$Ti, 
$^{76}$Ge$\to^{76}$Se, and 
$^{82}$Se$\to^{82}$Kr, are much larger in 
magnitude than those for the $\trgg$ transitions, 
especially when the $\qbbt$ values are used. 

\begin{table*}
\caption{\label{tab:gtf}
Calculated Gamow-Teller $\mgt$ (\ref{eq:mgt}) 
and Fermi $\mf$ (\ref{eq:mf}) 
matrix elements for the $\trgg$ and $\trge$ 
$\tnbb$ decays. 
Each matrix element is presented for the two cases 
in which the experimental \cite{data} and theoretical 
$\qbb$ values, represented by $\qbbe$ and $\qbbt$, respectively, 
are used [see Table~\ref{tab:qbb} and Eq.~(\ref{eq:egs})]. 
}
 \begin{center}
 \begin{ruledtabular}
  \begin{tabular}{lcccccccc}
\multirow{3}{*}{Decay}
&\multicolumn{4}{c}{$0^{+}_{1}$}
&\multicolumn{4}{c}{$0^{+}_{2}$}\\
\cline{2-5}\cline{6-9}
&\multicolumn{2}{c}{$\mgt$}
&\multicolumn{2}{c}{$\mf$}
&\multicolumn{2}{c}{$\mgt$}
&\multicolumn{2}{c}{$\mf$}\\
\cline{2-3}\cline{4-5}\cline{6-7}\cline{8-9}
&  $\qbbt$ & $\qbbe$ & $\qbbt$ & $\qbbe$ 
&  $\qbbt$ & $\qbbe$ & $\qbbt$ & $\qbbe$ 
\\
\hline
$^{48}$Ca$\to^{48}$Ti & $0.060$ & $0.042$ & $0.024$ & $0.016$ & $0.325$ & $0.066$ & $-0.142$ & $-0.075$ \\ 
$^{76}$Ge$\to^{76}$Se & $0.040$ & $0.034$ & $-0.007$ & $-0.007$ & $0.097$ & $0.078$ & $-0.085$ & $-0.069$ \\ 
$^{82}$Se$\to^{82}$Kr & $-0.060$ & $-0.045$ & $0.017$ & $0.015$ & $0.124$ & $0.070$ & $-0.081$ & $-0.064$ \\ 
$^{96}$Zr$\to^{96}$Mo & $0.139$ & $0.154$ & $-0.001$ & $-0.001$ & $0.053$ & $0.063$ & $-0.000$ & $-0.000$ \\ 
$^{100}$Mo$\to^{100}$Ru & $0.513$ & $0.483$ & $-0.000$ & $-0.000$ & $-0.007$ & $-0.020$ & $0.000$ & $0.000$ \\ 
$^{110}$Pd$\to^{110}$Cd & $0.071$ & $0.080$ & $0.000$ & $0.000$ & $-0.052$ & $-0.062$ & $0.000$ & $0.000$ \\ 
$^{116}$Cd$\to^{116}$Sn & $0.148$ & $0.275$ & $0.001$ & $0.001$ & $0.032$ & $-0.037$ & $-0.001$ & $-0.001$ \\ 
$^{124}$Sn$\to^{124}$Te & $0.123$ & $0.074$ & $-0.054$ & $-0.045$ & $0.345$ & $-0.066$ & $0.016$ & $0.012$ \\ 
$^{128}$Te$\to^{128}$Xe & $-0.139$ & $-0.102$ & $0.006$ & $0.005$ & $0.108$ & $0.032$ & $0.002$ & $0.002$ \\ 
$^{130}$Te$\to^{130}$Xe & $-0.041$ & $-0.037$ & $0.025$ & $0.022$ & $0.043$ & $0.037$ & $-0.019$ & $-0.017$ \\ 
$^{136}$Xe$\to^{136}$Ba & $-0.173$ & $-0.102$ & $0.028$ & $0.028$ & $-2.807$ & $0.010$ & $0.009$ & $0.001$ \\ 
$^{150}$Nd$\to^{150}$Sm & $-0.375$ & $-0.369$ & $0.000$ & $0.000$ & $-0.414$ & $-0.390$ & $-0.000$ & $-0.000$ \\ 
$^{198}$Pt$\to^{198}$Hg & $-0.016$ & $-0.016$ & $0.001$ & $0.001$ & $-0.008$ & $-0.010$ & $-0.000$ & $-0.000$ \\
  \end{tabular}
 \end{ruledtabular}
 \end{center}
\end{table*}

\begin{table}
\caption{\label{tab:chif}
Ratio of the calculated $\mf$ to $\mgt$ matrix elements 
$\chi_F(0^+)=\mf/\mgt$ for the 
$\trgg$ and $\trge$ $\tnbb$ decays of the even-even nuclei 
under investigation in the cases where the two different $\qbb$ values, 
$\qbbt$ and $\qbbe$, are used. 
}
 \begin{center}
 \begin{ruledtabular}
  \begin{tabular}{lcccc}
\multirow{2}{*}{Nucleus}
&\multicolumn{2}{c}{$\qbbt$}
&\multicolumn{2}{c}{$\qbbe$}\\
\cline{2-3}\cline{4-5}
&  $\chi_F(0^+_1)$ & $\chi_F(0^+_2)$ & $\chi_F(0^+_1)$ & $\chi_F(0^+_2)$
\\
\hline
$^{48}$Ca & $0.401$ & $-0.435$ & $0.390$ & $-1.147$ \\ 
$^{76}$Ge & $-0.179$ & $-0.874$ & $-0.204$ & $-0.888$ \\ 
$^{82}$Se & $-0.287$ & $-0.650$ & $-0.328$ & $-0.923$ \\ 
$^{96}$Zr & $-0.006$ & $-0.003$ & $-0.005$ & $-0.002$ \\ 
$^{100}$Mo & $-0.000$ & $-0.005$ & $-0.000$ & $-0.002$ \\ 
$^{110}$Pd & $0.000$ & $-0.000$ & $0.000$ & $-0.000$ \\ 
$^{116}$Cd & $0.004$ & $-0.021$ & $0.002$ & $0.022$ \\ 
$^{124}$Sn & $-0.443$ & $0.046$ & $-0.608$ & $-0.182$ \\ 
$^{128}$Te & $-0.040$ & $0.016$ & $-0.051$ & $0.050$ \\ 
$^{130}$Te & $-0.104$ & $-0.093$ & $-0.601$ & $-0.460$ \\ 
$^{136}$Xe & $-0.163$ & $-0.003$ & $-0.278$ & $0.061$ \\ 
$^{150}$Nd & $-0.001$ & $0.000$ & $-0.001$ & $0.000$ \\ 
$^{198}$Pt & $-0.060$ & $0.036$ & $-0.059$ & $0.030$ \\
  \end{tabular}
 \end{ruledtabular}
 \end{center}
\end{table}

As one can see in Table~\ref{tab:gtf}, 
for both the $\trgg$ and $\trge$ decays, 
the Fermi NMEs $\mf$ are, as a whole, calculated to be 
smaller in magnitude than the GT ones, $\mgt$. 
However, for those even-even nuclei where protons and neutrons 
can occupy the same major shells, i.e., $^{48}$Ca, $^{76}$Ge, 
$^{82}$Se, $^{124}$Sn, $^{128,130}$Te, and $^{136}$Xe, 
the corresponding $\mf$ matrix elements are of the same 
order of magnitude as the $\mgt$ ones. In particular, for 
$^{48}$Ca the calculated $|\mf(\trge)|$ value is even larger than 
the $|\mgt(\trge)|$ one. 
This finding is further corroborated by the 
Fermi strength distribution systematics shown in Fig.~\ref{fig:dist_ee}, 
in which non-negligible contributions of the Fermi transition 
are already apparent  for the $^{48}$Ca, $^{124}$Sn, and $^{136}$Xe decays. 

To examine more quantitatively the contribution of the Fermi 
transition relative to the GT one, 
the ratio $\chi_F\equiv\mf/\mgt$ \cite{tomoda1991} is calculated 
for the $\tnbb$ decays of interest. 
The results are listed in Table~\ref{tab:chif}. 
If the isospin is a good symmetry quantum number, 
the ratio $\chi_F$ should be equal to zero. 
For the $\tnbb$ decay mode, the Fermi matrix element should vanish, 
while in the $\znbb$ decay it is expected to be non-vanishing but 
also quite small. 
The large $\mf$ matrix element or $\chi_F$ ratio 
therefore implies that there is a spurious isospin-symmetry 
breaking in the wave functions of the initial and 
final even-even nuclei generated by the employed model. 
For the aforementioned even-even nuclei $^{48}$Ca, $^{76}$Ge, 
$^{82}$Se, $^{124}$Sn, $^{128,130}$Te, and $^{136}$Xe, the ratio 
$|\chi_F|$ is here calculated to be substantially large, 
which is typically $|\chi_F|\gtrsim0.1$. 
One further observes in Table~\ref{tab:chif} that, in general, 
the calculation with the experimental $\qbb$ 
values, $\qbbe$, leads to a larger degree of the isospin 
symmetry breaking than with the theoretical $\qbb$ values, $\qbbt$. 
A similar degree of isospin symmetry breaking to the present 
case was reported in the earlier IBM-2 calculations in 
Refs.~\cite{barea2013,yoshida2013}. 
The problem was addressed in great detail in 
Refs.~\cite{barea2013,barea2015}. 
In Ref.~\cite{barea2015}, in particular, 
a method to restore the broken isospin symmetry 
in the calculation of the Fermi matrix element for 
the $\tnbb$ decay was proposed, that is to modify 
the Fermi transition operator so that the 
Fermi matrix element within the closure approximation 
should vanish. 

In the present theoretical framework, provided that the 
Fermi transition is actually spurious, then either 
the Fermi contribution $\mf$ would have to be simply neglected, 
or some modification should be 
made to the Fermi transition operator of (\ref{eq:ofe}) 
in a similar spirit to Ref.~\cite{barea2015}. 
The latter prescription points to an interesting 
question as to how the broken isospin symmetry can be treated 
in the mapped IBM-2 framework, but is at the same time 
beyond the scope of the present investigation. 
In the following, the Fermi matrix element is retained in the 
calculation of the $\tnbb$-decay NME, while it should be 
kept in mind that especially for those nuclei 
with approximately equal proton and neutron numbers 
a spurious isospin 
symmetry breaking may be present in the large Fermi 
matrix element. 
%

\subsection{$\tnbb$ NMEs}

In the second and fifth columns of 
Table~\ref{tab:nme}, the results for 
the $\tnbb$ NMEs are given. 
In the last column, the effective NMEs 
denoted by $\mbbe$, which are extracted 
from the observed $\tnbb$-decay half-lives \cite{barabash2020}, 
are also shown. 
The calculated NMEs are typically within the range 
$0.05\lesssim|\mbb|\lesssim0.3$, which are in most cases 
larger than the experimental values $|\mbbe|$. 
For the $\trgg$ decays of $^{100}$Mo, 
and both the $\trgg$ and $\trge$ decays of $^{150}$Nd, the 
corresponding $\mbb$ are particularly large. 

\begin{table*}
\caption{\label{tab:nme}
Calculated $\tnbb$-decay NMEs 
$\mbb$ (\ref{eq:mbb}) 
(second and fifth columns), and 
$\mbbqp$ (third and sixth columns) 
and $\mbbqe$ (fourth and seventh columns), 
quenched with the effective $\ga$ factors 
$\gaep$ (\ref{eq:gaep}) 
and $\gaee$ (\ref{eq:gaee}), respectively. 
The results obtained with the theoretical 
and experimental  $\qbb$ values are compared. 
The effective NMEs $\mbbe$ extracted from the $\taubb$ data 
\cite{barabash2020} are shown in the eighth column. 
}
 \begin{center}
 \begin{ruledtabular}
  \begin{tabular}{lccccccc}
\multirow{2}{*}{Decay}
&\multicolumn{3}{c}{$\qbbt$}
&\multicolumn{3}{c}{$\qbbe$}&
\multirow{2}{*}{$|\mbbe|$ \cite{barabash2020}}\\
\cline{2-4}\cline{5-7}
&{$|\mbb|$}&{$|\mbbqp|$}&{$|\mbbqe|$}
&{$|\mbb|$}&{$|\mbbqp|$}&{$|\mbbqe|$}&\\
\hline
$^{48}$Ca$\to^{48}$Ti & 0.073 & 0.020 & 0.034 & 0.051 & 0.014 & 0.024 & $0.035\pm0.003$ \\ 
$^{76}$Ge$\to^{76}$Se & 0.072 & 0.017 & 0.021 & 0.062 & 0.014 & 0.018 & $0.106\pm0.004$ \\ 
$^{82}$Se$\to^{82}$Kr & 0.115 & 0.026 & 0.031 & 0.087 & 0.020 & 0.024 & $0.085\pm0.001$ \\ 
$^{96}$Zr$\to^{96}$Mo & 0.225 & 0.048 & 0.048 & 0.249 & 0.053 & 0.054 & $0.088\pm0.004$ \\ 
$^{100}$Mo$\to^{100}$Ru & 0.827 & 0.174 & 0.167 & 0.778 & 0.164 & 0.157 & $0.185\pm0.002$ \\ 
$^{100}$Mo$\to^{100}$Ru$(0^+_2)$ & 0.011 & 0.002 & 0.002 & 0.032 & 0.007 & 0.007 & $0.151\pm0.004$ \\ 
$^{110}$Pd$\to^{110}$Cd & 0.115 & 0.023 & 0.020 & 0.128 & 0.026 & 0.022 & ${}$ \\ 
$^{116}$Cd$\to^{116}$Sn & 0.238 & 0.048 & 0.037 & 0.443 & 0.089 & 0.069 & $0.108\pm0.003$ \\ 
$^{124}$Sn$\to^{124}$Te & 0.253 & 0.050 & 0.035 & 0.164 & 0.032 & 0.022 & ${}$ \\ 
$^{128}$Te$\to^{128}$Xe & 0.229 & 0.044 & 0.030 & 0.169 & 0.033 & 0.022 & $0.043\pm0.003$ \\ 
$^{130}$Te$\to^{130}$Xe & 0.091 & 0.017 & 0.011 & 0.081 & 0.016 & 0.010 & $0.0293\pm0.0009$ \\ 
$^{136}$Xe$\to^{136}$Ba & 0.307 & 0.058 & 0.035 & 0.194 & 0.037 & 0.022 & $0.0181\pm0.0006$ \\ 
$^{150}$Nd$\to^{150}$Sm & 0.604 & 0.111 & 0.055 & 0.594 & 0.109 & 0.054 & $0.055\pm0.003$ \\ 
$^{150}$Nd$\to^{150}$Sm$(0^+_2)$ & 0.666 & 0.122 & 0.060 & 0.629 & 0.116 & 0.057 & $0.044\pm0.005$ \\ 
$^{198}$Pt$\to^{198}$Hg & 0.026 & 0.004 & 0.001 & 0.027 & 0.005 & 0.001 & ${}$ \\
  \end{tabular}
 \end{ruledtabular}
 \end{center}
\end{table*}

\begin{figure}[ht]
\begin{center}
\includegraphics[width=\linewidth]{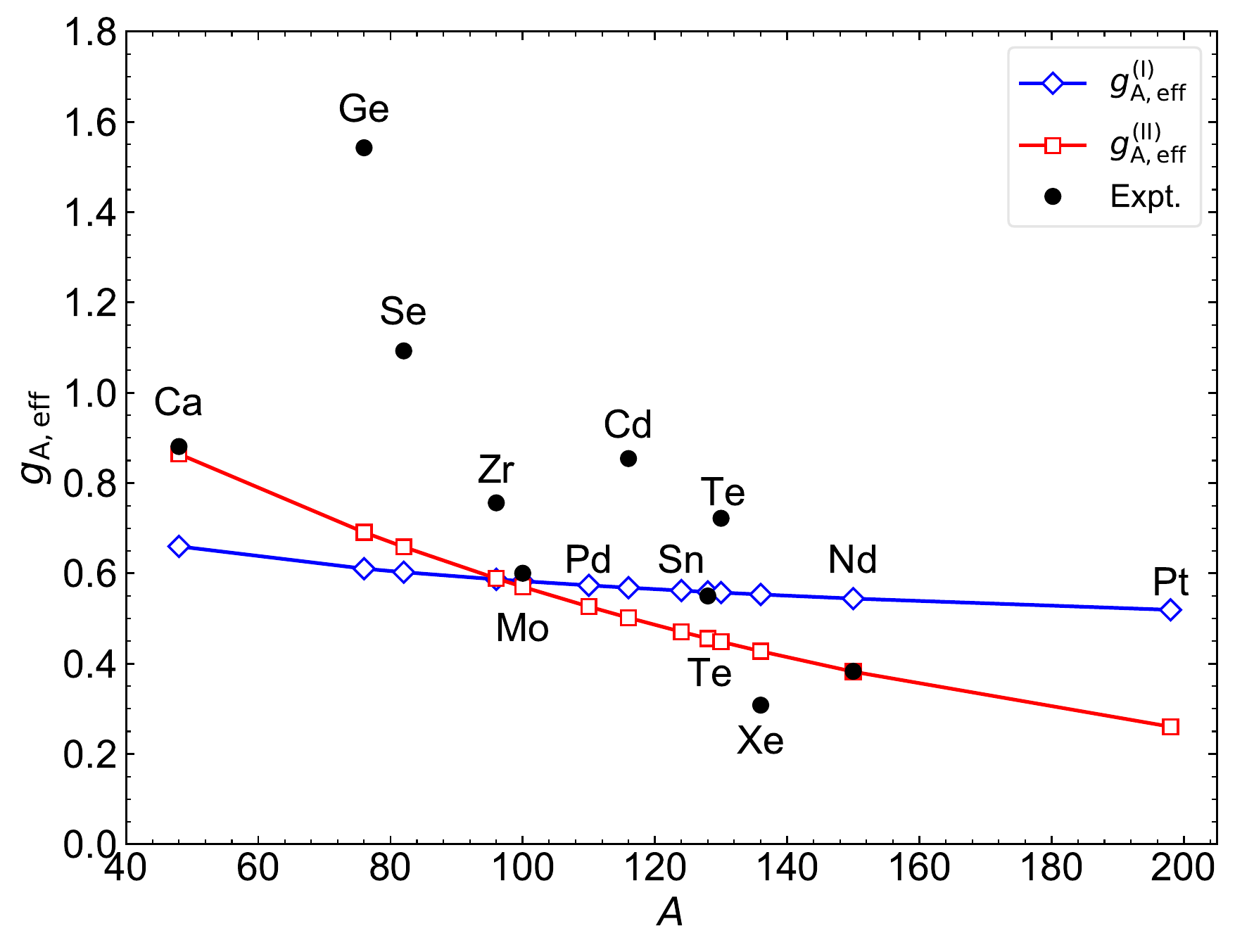}
\caption{Effective $\ga$ factors 
$\gaep=\ga{A}^{-0.169}$ (\ref{eq:gaep}) (diamonds) and 
$\gaee=\ga e^{-0.008A}$ (\ref{eq:gaee}) (squares). 
The solid circles denoted by ``Expt.'' 
represent $\gae$ values 
that would be required to reproduce 
the experimental NMEs $\mbbe$.}
\label{fig:ga}
\end{center}
\end{figure}

To make a reasonable comparison with experiment, 
one can consider quenching of the obtained NME. 
Here it is assumed that in Eq.~(\ref{eq:mbb}) 
only the $\ga$ factor is quenched, 
whereas the ratio $\gv/\ga$, as well as $\gv$, is not. 
The quenched NME can be then 
obtained by simply replacing the free value 
$\ga$ in (\ref{eq:mbb}) with the effective 
$\ga$ factor denoted by $\gae$, i.e.,
\begin{align}
\label{eq:mbbq}
 |\mbb|\rightarrow\left(\frac{\gae}{\ga}\right)^2 |\mbb|. 
\end{align}
Also the quenching factor $q$ is given by $q=\gae/\ga$. 
It is further assumed that the $\gae$ factor 
is a smooth function of the mass number $A$, and 
specifically the two parametrizations are here considered: 
\begin{align}
\label{eq:gaep}
 &\gaep=\ga\cdot{A}^{-a},\\
\label{eq:gaee}
 &\gaee=\ga\cdot{e^{-bA}}. 
\end{align}
The numerical constants $a=0.169$ 
and $b=0.008$ are obtained by fitting to 
those $\gae$ values that would be required 
to reproduce the $\mbbe(\trgg)$ values. 
At $A\approx1$, both $\gaep$ and $\gaee$ reduce to 
the free nucleon value $\ga=1.269$. 
A similar parametrization to the first choice 
(\ref{eq:gaep}) was considered in the previous 
IBM-2 study on the $\db$ decay \cite{barea2013}. 
The constant $a=0.169$ is also 
close to the one ($0.18$) used in the 
above reference. 

The resultant $\gaep$ and $\gaee$ values are shown  
in Fig.~\ref{fig:ga} as functions of $A$. 
The $\gaep$ value only gradually changes with $A$ 
within the range $0.5\lesssim\gaep\lesssim0.7$, 
corresponding to the quenching factor of 
$q\approx0.4-0.5$. 
The second choice, $\gaee$, takes more or less similar 
values to $\gaep$ in the mass range $A\approx100-130$, 
but is smaller for heavier nuclei, giving rise 
to a more drastic quenching than $\gaep$. 
For the $^{76}$Ge and $^{82}$Se decays, 
the present $\gae$ factors of both choices (\ref{eq:gaep}) 
and (\ref{eq:gaee}) are still much smaller than those 
that would be required to reproduce the $\mbbe$ values. 
In fact, one sees in Table~\ref{tab:nme} that 
even before the quenching the corresponding 
$\mbb$ is smaller than the data 
for $^{76}$Ge, and that is already close to the 
experimental value for $^{82}$Se. 
From these observations, 
the present $\gae$ values 
for the above two particular cases 
might not be adequate.

\begin{figure}[ht]
\begin{center}
\includegraphics[width=\linewidth]{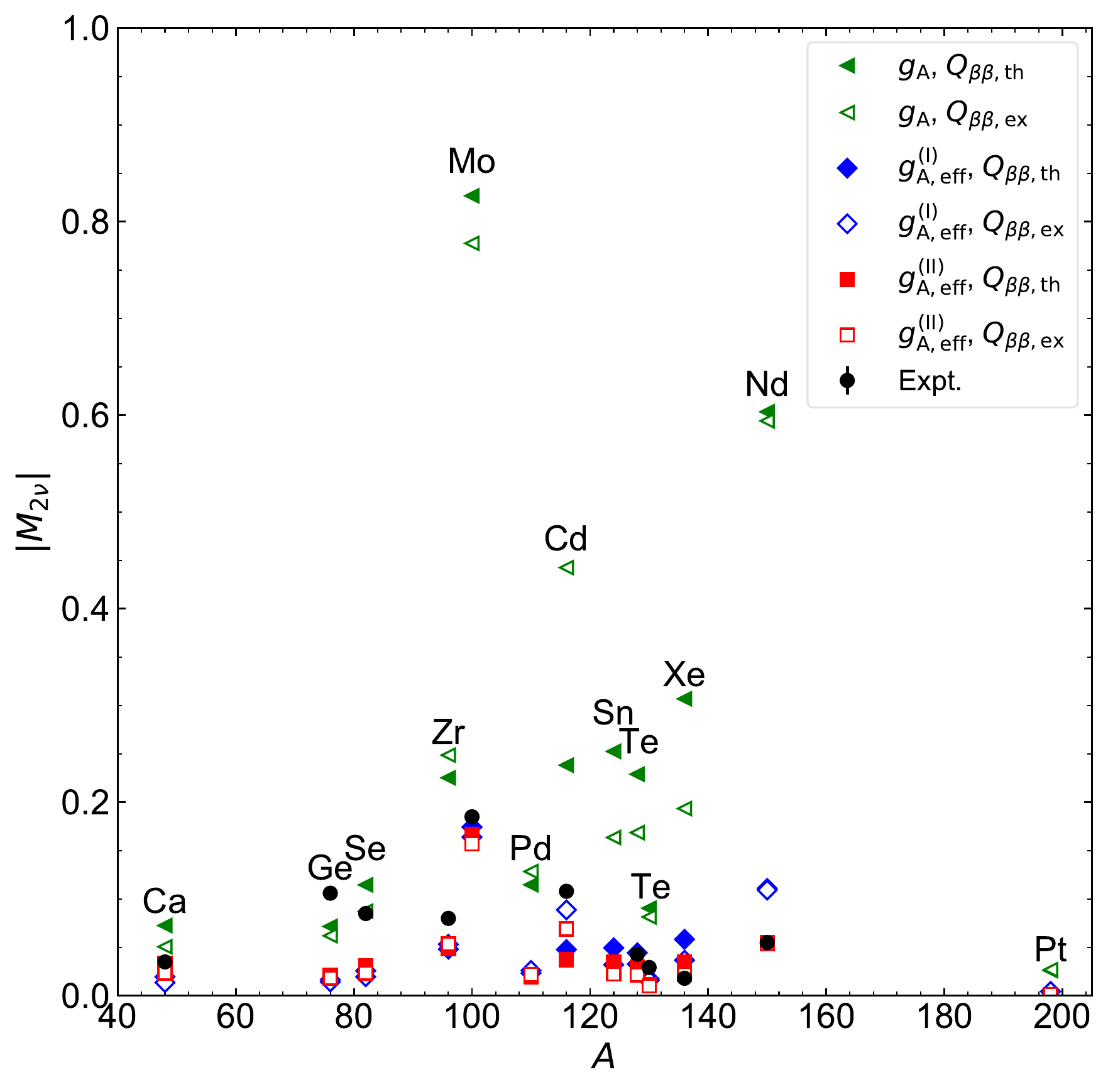}
\caption{$\trgg$ $\tnbb$-decay 
NMEs $\mbb$ (\ref{eq:mbb}) (triangles), and  
quenched NMEs $\mbbqp$ (diamonds) and $\mbbqe$ (squares), 
using the effective $\ga$ factors $\gaep$ 
and $\gaee$, respectively. 
The results obtained with the theoretical $\qbbt$ 
(filled symbols) and experimental $\qbbe$ (open symbols) 
values are compared. 
The solid circles represent the experimental NMEs or $\mbbe$.}
\label{fig:nme}
\end{center}
\end{figure}

In the third, fourth, sixth, and seventh columns of 
Table~\ref{tab:nme} listed are the values 
of the quenched NMEs, denoted as $\mbbqp$ and $\mbbqe$, 
obtained by using the two different choices of the 
effective $\ga$ factor $\gaep$ (\ref{eq:gaep}) 
and $\gaee$ (\ref{eq:gaee}), respectively. 
The quenched NMEs, $\mbbqp$ and $\mbbqe$, 
are both close to the experimental data. 
In general, the second choice $\gaee$ (\ref{eq:gaee}) 
appears to give a better agreement with experiment than 
the first one, $\gaep$ (\ref{eq:gaep}). 
A specific comment concerns that, 
for the $^{76}$Ge and $^{82}$Se decays, 
the quenching reduces too much the NMEs and 
only worsens the agreement with the experimental values, 
as compared to the unquenched NMEs. 
In addition, the NME for the 
$^{100}$Mo$\to^{100}$Ru$(0^{+}_{2})$ decay is 
considerably underestimated even without the quenching, 
irrespective of whether 
the $\qbbt$ or $\qbbe$ values is used. 
This may suggest that there are 
some missing elements in the 
calculation for the final nucleus $^{100}$Ru 
such as the intruder excitations, which may be 
needed to give a correct wave function for the 
$0^+_2$ state. 
For completeness, in Fig.~\ref{fig:nme} 
the values of the $\tnbb$-decay NMEs for the $\trgg$ decay, 
that are already listed in Table~\ref{tab:nme}, are shown 
in comparison with the experimental values $\mbbe$.

Alternatively, one can extract the $\gae$ values for 
the $\tnbb$ decay from the single-$\beta$ decay data. 
This can be done by fitting the calculated 
$\beta$-decay $\ft$ values to the experimental counterparts. 
Consider the $\tnbb$ decay of $^{100}$Mo 
an an illustrative example. 
This is an ideal case, in which the GT transitions for 
the single-$\beta$ 
and $\db$ decays of the intermediate nucleus is 
dominated by the transitions from 
the lowest $1^+$ state 
[see, Figs.~\ref{fig:dist_eo}--\ref{fig:dist_ee}]. 
For this decay chain, the $\gae$ values are calculated to be 
$\gaeb=0.863$ and $\gaeb=0.479$ 
for the EC and $\btm$ decays of the state 
$^{100}$Tc$\,(1^+_1)$, respectively. 
If their average $0.671$ is used as the effective $\gae$ 
for the $\tnbb$ decay, 
then the corresponding NME $\mbb$ is quenched to $0.231$, 
in a fair agreement with the  
$\mbbqp$ and $\mbbqe$ values shown in 
Table~\ref{tab:nme}. 
A similar result is obtained for the $^{128}$Te decay. 
In most of the other cases, however, as one sees 
in Figs.~\ref{fig:dist_eo}--\ref{fig:dist_ee} 
the GT matrix elements 
for both the single-$\beta$ and $\db$ decays 
are not as sharply populated at the lowest $1^+$ state 
as in the case of the $^{100}$Mo$\to^{100}$Ru 
or $^{128}$Te$\to^{128}$Xe ones. 
Furthermore, 
the present calculation for single-$\beta$ decays, 
which uses the free value 
for the $\ga$ factor, generally gives larger 
$\ft$ values for the EC processes of 
the intermediate nuclei. 
Experimental $\ft$ values for 
the $\btm$ decay are also not available, except 
for $^{100}$Tc, $^{110}$Ag, $^{116}$In, and $^{128}$I. 
For these reasons, within the present theoretical scheme, 
the prescription to extract $\db$-decay $\gae$ 
factors from the single-$\beta$ decays 
is not expected to give a reasonable agreement of the $\tnbb$ NME 
with experiment in a systematic way. 

\begin{figure}[ht]
\begin{center}
\includegraphics[width=\linewidth]{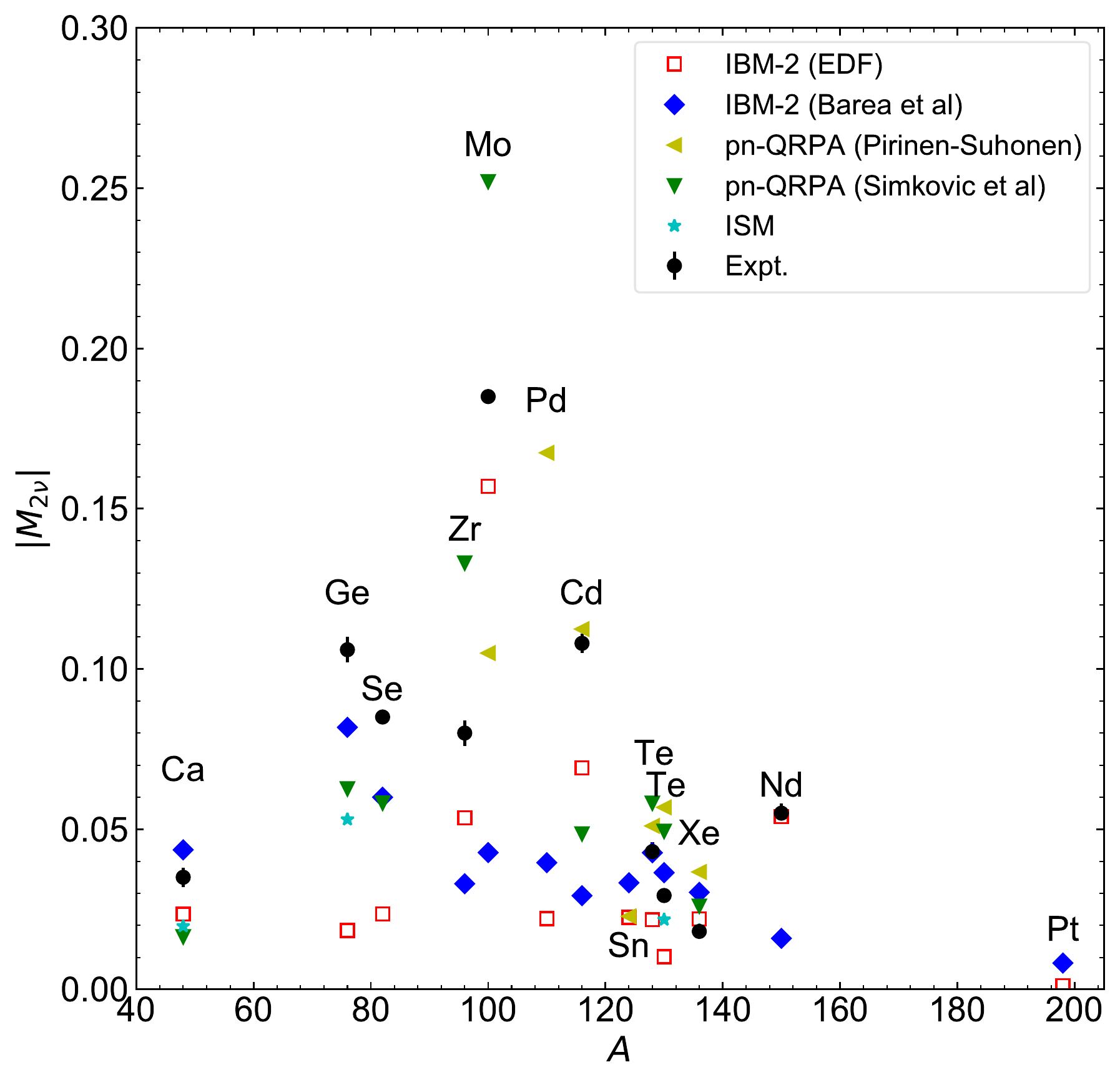}
\caption{Calculated $\tnbb$-decay NMEs (\ref{eq:mbb}) 
in the present study [denoted by ``IBM-2 (EDF)''], 
obtained with the effective $\ga$ factor $\gaee$ and 
with the $\qbbe$ values. 
The experimental NMEs ($\mbbe$) 
and those obtained from the proton-neutron QRPA 
(pn-QRPA) by \v{S}imkovi\'c et al. \cite{simkovic2018} and 
Pirinen and Suhonen \cite{pirinen2015}, from the IBM-2 by Barea et al. 
\cite{barea2015}, 
and from the ISM by Caurier et al. \cite{caurier2007} are also shown.}
\label{fig:nme-lit}
\end{center}
\end{figure}

\begin{table*}
\caption{\label{tab:tau}
Comparison of calculated 
$\tnbb$-decay half-lives $\tau_{1/2}^{(2\nu)}$ (in year) 
corresponding to the free value of the $\ga$ factor, and the 
effective values $\gaep$ (\ref{eq:gaep}) and $\gaee$ (\ref{eq:gaee}) factors, 
in two difference cases of the $\qbb$ values, $\qbbt$ 
(second to fourth columns)
and $\qbbe$ (fifth to seventh columns). 
The experimental values taken from Ref.~\cite{barabash2020} 
are shown in the eighth column.
}
 \begin{center}
 \begin{ruledtabular}
  \begin{tabular}{lccccccc}
\multirow{2}{*}{Decay}
&\multicolumn{3}{c}{$\tau_{1/2}^{(2\nu)}$ (yr), with $\qbbt$}
&\multicolumn{3}{c}{$\tau_{1/2}^{(2\nu)}$ (yr), with $\qbbe$}
&\multirow{2}{*}{Expt. \cite{barabash2020}}\\
\cline{2-4}
\cline{5-7}
& $\ga$ & $\gaep$ & $\gaee$ & $\ga$ & $\gaep$ & $\gaee$ & \\
\hline
$^{48}$Ca$\to^{48}$Ti & $1.22\times10^{19}$ & $1.67\times10^{20}$ & $5.66\times10^{19}$ & $2.50\times10^{19}$ & $3.43\times10^{20}$ & $1.16\times10^{20}$ & $5.3^{+1.2}_{-0.8}\times10^{19}$ \\ 
$^{76}$Ge$\to^{76}$Se & $4.04\times10^{21}$ & $7.54\times10^{22}$ & $4.59\times10^{22}$ & $5.39\times10^{21}$ & $1.01\times10^{23}$ & $6.14\times10^{22}$ & $(1.88\pm0.08)\times10^{21}$ \\ 
$^{82}$Se$\to^{82}$Kr & $4.77\times10^{19}$ & $9.38\times10^{20}$ & $6.58\times10^{20}$ & $8.20\times10^{19}$ & $1.61\times10^{21}$ & $1.13\times10^{21}$ & $(0.87^{+0.02}_{-0.01})\times10^{20}$ \\ 
$^{96}$Zr$\to^{96}$Mo & $2.89\times10^{18}$ & $6.32\times10^{19}$ & $6.24\times10^{19}$ & $2.37\times10^{18}$ & $5.19\times10^{19}$ & $5.12\times10^{19}$ & $(2.3\pm0.2)\times10^{19}$ \\ 
$^{100}$Mo$\to^{100}$Ru & $4.42\times10^{17}$ & $9.95\times10^{18}$ & $1.09\times10^{19}$ & $5.00\times10^{17}$ & $1.12\times10^{19}$ & $1.23\times10^{19}$ & $(7.06^{+0.15}_{-0.13})\times10^{18}$ \\ 
$^{100}$Mo$\to^{100}$Ru$(0^+_2)$ & $1.29\times10^{23}$ & $2.91\times10^{24}$ & $3.17\times10^{24}$ & $1.59\times10^{22}$ & $3.57\times10^{23}$ & $3.90\times10^{23}$ & $6.7^{+0.5}_{-0.4}\times10^{20}$ \\ 
$^{110}$Pd$\to^{110}$Cd & $5.51\times10^{20}$ & $1.32\times10^{22}$ & $1.86\times10^{22}$ & $4.40\times10^{20}$ & $1.06\times10^{22}$ & $1.49\times10^{22}$ & ${}$ \\ 
$^{116}$Cd$\to^{116}$Sn & $6.37\times10^{18}$ & $1.58\times10^{20}$ & $2.61\times10^{20}$ & $1.85\times10^{18}$ & $4.59\times10^{19}$ & $7.56\times10^{19}$ & $(2.69\pm0.09)\times10^{19}$ \\ 
$^{124}$Sn$\to^{124}$Te & $2.83\times10^{19}$ & $7.37\times10^{20}$ & $1.50\times10^{21}$ & $6.76\times10^{19}$ & $1.76\times10^{21}$ & $3.57\times10^{21}$ & ${}$ \\ 
$^{128}$Te$\to^{128}$Xe & $7.09\times10^{22}$ & $1.89\times10^{24}$ & $4.26\times10^{24}$ & $1.31\times10^{23}$ & $3.48\times10^{24}$ & $7.86\times10^{24}$ & $(2.25\pm0.09)\times10^{24}$ \\ 
$^{130}$Te$\to^{130}$Xe & $7.98\times10^{19}$ & $2.14\times10^{21}$ & $5.12\times10^{21}$ & $9.85\times10^{19}$ & $2.65\times10^{21}$ & $6.31\times10^{21}$ & $(7.91\pm0.21)\times10^{20}$ \\ 
$^{136}$Xe$\to^{136}$Ba & $7.41\times10^{18}$ & $2.05\times10^{20}$ & $5.75\times10^{20}$ & $1.86\times10^{19}$ & $5.16\times10^{20}$ & $1.45\times10^{21}$ & $(2.18\pm0.05)\times10^{21}$ \\ 
$^{150}$Nd$\to^{150}$Sm & $7.54\times10^{16}$ & $2.23\times10^{18}$ & $9.16\times10^{18}$ & $7.78\times10^{16}$ & $2.30\times10^{18}$ & $9.45\times10^{18}$ & $(9.34\pm0.65)\times10^{18}$ \\ 
$^{150}$Nd$\to^{150}$Sm$(0^+_2)$ & $5.21\times10^{17}$ & $1.54\times10^{19}$ & $6.33\times10^{19}$ & $5.84\times10^{17}$ & $1.73\times10^{19}$ & $7.10\times10^{19}$ & $1.2^{+0.3}_{-0.2}\times10^{20}$ \\ 
$^{198}$Pt$\to^{198}$Hg & $9.59\times10^{22}$ & $3.42\times10^{24}$ & $5.43\times10^{25}$ & $8.95\times10^{22}$ & $3.20\times10^{24}$ & $5.09\times10^{25}$ & ${}$ \\
  \end{tabular}
 \end{ruledtabular}
 \end{center}
\end{table*}

Figure~\ref{fig:nme-lit} presents the calculated values for 
the $\trgg$ $\tnbb$ decay NMEs. Here the experimental 
$\qbbe$ values are used, and the effective $\gae$ factors of 
the second choice $\gaee$ (\ref{eq:gaee}) are considered. 
The same figure compares the results using the experimental NMEs 
$\mbbe$ with those of the earlier theoretical predictions 
within the proton-neutron QRPA (pn-QRPA) calculations by 
\v{S}imkovi\'c et al \cite{simkovic2018} 
and by Pirinen and Suhonen \cite{pirinen2015}, 
the IBM-2 calculations with closure approximation 
by Barea et al. \cite{barea2015}, 
and the interacting shell model (ISM) by Caurier et al 
\cite{caurier2007}. 
All the theoretical NMEs shown in Fig.~\ref{fig:nme-lit} 
are obtained in the same way as those calculated 
here, i.e., both the GT and, if available, Fermi 
matrix elements are taken in the calculation of $\mbb$, 
which are then quenched with the effective $\ga$ 
factors considered in these references. 
As one sees in Fig.~\ref{fig:nme-lit}, 
the prediction of the NMEs appears to be quite different from 
one theoretical approach to another most notably, for the 
$^{100}$Mo$\,\to\,^{100}$Ru decay. 
For most of the considered $\tnbb$ decays, 
the present mapped IBM-2 values [denoted by 
``IBM-2 (EDF)'' in the figure] are close to some of 
the previous theoretical predictions. 
The exception is perhaps the NMEs for 
the $^{76}$Ge and $^{82}$Se decays, since the mapped IBM-2 
results substantially deviate from the other theoretical values.

In the IBM-2 calculations of Ref.~\cite{barea2015}, 
the Hamiltonian parameters for the even-even nuclei 
were taken from the earlier IBM-2 
fitting calculations, whereas the $\db$ operators 
were derived microscopically 
within the generalized seniority scheme 
of the Otsuka-Arima-Iachello mapping 
procedure \cite{OAI}. 
In addition, the spurious isospin-symmetry breaking 
in the Fermi transition was effectively taken into account 
by the procedure mentioned in Sec.~\ref{sec:gtf}. 
The values shown in 
Fig.~\ref{fig:nme-lit} are obtained by using the dimensionless 
GT and Fermi matrix elements shown in Table~XII of \cite{barea2015}, 
dividing them by the energy denominator $\tilde{A}^{CA}_{GT}$, 
found in Table XV of Ref.~\cite{barea2013}, 
for both the GT and Fermi matrix elements, 
and applying the effective $\ga$ factors $\gae=1.269A^{-0.18}$. 
There are some qualitative differences between the present 
IBM-2 calculation and the one in Ref.~\cite{barea2015}. 
For instance, the mapped IBM-2 produces large NMEs for the 
$^{100}$Mo and $^{96}$Zr, while they are considerably small 
in \cite{barea2015}. On the other hand, the values of 
the NMEs obtained from the results in the above reference 
for $^{76}$Ge and $^{82}$Se are closer to the data, i.e., 
$|\mbbe|=0.106\pm0.004$ and $0.085\pm0.001$ 
for $^{76}$Ge and $^{82}$Se, respectively, than 
the mapped IBM-2 results.

In Ref.~\cite{barea2013}, 
a similar mass-dependent $\gae$ factor to the one in the 
present calculation, $\gae=\ga{A^{-0.12}}$, was 
considered for the GT matrix elements calculated within 
the ISM \cite{caurier2007}. 
The ISM values included in Fig.~\ref{fig:nme-lit} are here 
obtained by using this parametrization for $\gae$ 
and the same energy denominator 
$\tilde{A}^{CA}_{GT}$ for the closure approximation. 
More recently, the pair-truncated shell-model calculation on 
the $\tnbb$ decays of $^{76}$Ge and $^{82}$Se \cite{yoshinaga2018} 
without the closure approximation 
obtained the $\gae$ factors of 1.41 and 1.66, respectively, 
greater than the free value.

For the NMEs within the pn-QRPA framework 
of Ref.~\cite{pirinen2015}, the GT and Fermi 
matrix elements listed in Table~V in that reference 
are used, 
employing the effective $\ga$ values that are determined by 
the ``linear $\ga$ model'' for single-$\beta$ decays 
\cite{pirinen2015}. 
These QRPA values exhibit a more or less similar 
behaviour with the mass number $A$ to the present ones, 
except for the $^{110}$Pd decay.

A more recent pn-QRPA calculation of Ref.~\cite{simkovic2018} 
with closure approximation used the mass-independent  
quenching factor $q=0.712$. 
In the present case, a more drastic quenching is made 
for the same mass region, varying from $q=0.54$ 
($^{76}$Ge) to $q=0.34$ ($^{136}$Xe). 
The corresponding QRPA NMEs shown in Fig.~\ref{fig:nme-lit}, which 
are here obtained by using the GT and Fermi matrix elements 
reported in Table~I of Ref.~\cite{simkovic2018}, 
are by roughly several factors larger than 
the present mapped IBM-2 results, but exhibit a similar 
trend with $A$.

\subsection{Half-lives}

The half-lives $\taubb$ (\ref{eq:taubb}) of the considered 
$\tnbb$ decays are computed by using the NMEs shown 
in Table~\ref{tab:nme} and 
the phase-space factors $G_{2\nu}$ calculated by Kotila 
et al. \cite{kotila2012}. 
Table~\ref{tab:tau} summarizes the calculated $\taubb$ values 
with different $\gae$ factors and $\qbb$ values. 
As one can see in the fourth and seventh columns of Table~\ref{tab:tau}, 
if the $\qbbe$ values combined  
with the effective $\gae$ factor $\gaee$ are adopted, 
a reasonable overall agreement between the predicted and experimental 
$\taubb(\trgg)$ \cite{barabash2020} values is reached typically  
within one order of magnitude.

For the $\trgg$ decays of $^{76}$Ge and $^{82}$Se, and the $\trge$ decay 
of $^{100}$Mo, however, 
either quenching here worsens the agreement with the experimental data. 
Particularly for the $^{100}$Mo$\to^{100}$Ru$(0^{+}_{2})$ decay, 
the corresponding half-life $\taubb$ is predicted to be by 3 to 4 
orders of magnitude longer than the experimental value 
($6.7^{+0.5}_{-0.4}\times10^{20}$ yr), 
regardless of which of the effective $\gae$ factors, 
$\gaep$ and $\gaee$, is employed. 

In addition, the calculated half-life 
for the  $^{100}$Mo$\to^{100}$Ru$(0^+_2)$ decay 
is much longer than the one for the 
$\trgg$ decay. 
Their ratio 
\begin{align}
\label{eq:tauratio1}
 \frac{\taubb\left[^{100}\mathrm{Mo}(0^{+}_{1})\to^{100}\mathrm{Ru}(0^{+}_{2})\right]}
{\taubb\left[^{100}\mathrm{Mo}(0^{+}_{1})\to^{100}\mathrm{Ru}(0^{+}_{1})\right]}
=31707,
\end{align}
with the experimental $\qbb$ value $\qbbe=3.03$ MeV, is indeed 
quite large, and overestimates the observed one 
\cite{barabash2020} 94.9$^{+7.4}_{-5.9}$ 
by three orders of magnitude.  
Note also that, in Eq.~(\ref{eq:tauratio1}), 
since the $\mbb(\trgg)$ and $\mbb(\trge)$ matrix elements are 
quenched by the same effective $\gae$ values, the 
ratio is independent of what kind of quenching is made 
for the NMEs.

On the other hand, 
the calculated half-life of the $\trge$ decay of $^{150}$Nd, 
with the NME quenched by $\gaee$ and 
in both cases of the $\qbbt$ and $\qbbe$ values, 
agrees rather well with the experimental data, 
in comparison to the $\trge$ decay of $^{100}$Mo. 
%
For the $^{150}$Nd$\,\to\,^{150}$Sm decay, the $\trge$ transition is 
also suggested to be not as significantly slower than the $\trgg$ one 
as in the case of the $^{100}$Mo$\,\to\,^{100}$Ru decay. 
In fact, the ratio of the corresponding half-lives $\taubb$ 
is calculated be 
\begin{align}
\frac{\taubb\left[^{150}\mathrm{Nd}(0^{+}_{1})\to^{150}\mathrm{Sm}(0^{+}_{2})\right]}{\taubb\left[^{150}\mathrm{Nd}(0^{+}_{1})\to^{150}\mathrm{Sm}(0^{+}_{1})\right]}=7.5, 
\end{align} 
with the experimental $\qbb$ value $\qbbe=3.37$ MeV, 
which is indeed much smaller than the ratio in Eq.~(\ref{eq:tauratio1}), 
and is also in a fair agreement with the 
corresponding experimental value, 12.8$^{+3.3}_{-2.3}$.

\subsection{Sensitivity to model assumptions}

\subsubsection{Choice of the single-particle energies\label{sec:comp-spe}}

In this section, the sensitivity 
of the calculated results to the choice of the spherical 
single-particle energies (SPEs) for odd particles is 
investigated. 
Here the $^{76}$Ge$\to^{76}$Se and 
$^{82}$Se$\to^{82}$Kr decays are taken as an example. 
For these two cases considerably 
small $\mbb$ have been obtained, and it would be meaningful 
to confirm if the modification of the SPEs improves the result. 

Another set of calculations is then performed 
by employing the SPEs of the effective 
shell-model interaction JUN45 \cite{honma2009} for 
the neutron and proton $N,Z=28-50$ major shells, 
which are fine tuned to the experimental data. 
Their values are listed in Table~\ref{tab:comp-spe-spe} 
of Appendix~\ref{sec:spe}. 
There are notable differences between 
the SPEs of \cite{honma2009} and those 
obtained from the RHB calculation. 
First, the unique-parity orbital $1g_{9/2}$ 
is much lower in energy in \cite{honma2009} 
than the RHB one. 
Second, energy splitting between 
the $1f_{5/2}$ and $2p_{3/2}$ orbitals, 
$\Delta\epsilon\equiv\epsilon_{1f_{5/2}}-\epsilon_{2p_{3/2}}$, 
is approximately 0.9 MeV for both 
neutrons and protons in Ref.~\cite{honma2009}, 
which is quite different from the one in 
the spherical RHB calculation: 
$\Delta\epsilon\approx0.05$ MeV and $-0.4$ MeV 
($-0.5$ MeV and $-1.3$ MeV) for the neutron 
and proton orbitals of $^{76}$As ($^{82}$Se), 
respectively. 

By using the phenomenological SPEs of \cite{honma2009}, 
the quasiparticle energies $\tilde\epsilon_{\jr}$ 
and the occupation probabilities 
$v_{\jr}^{2}$, 
required for $\hf^\rho$ and $\hbf^\rho$, respectively, 
are calculated within the BCS approximation \cite{IBFM},  
with a empirical pairing gap $\Delta=12A^{-1/2}$. 
The strength parameters for $\hbf^{\rho}$ in 
Table~\ref{tab:paraff} are slightly changed as 
$A_{\nu}^{(+)}=0\to-0.2$ MeV, $\Lambda_{\pi}^{(-)}=1.6\to0.4$ MeV, 
and $A_{\pi}^{(-)}=-0.8\to0$ MeV for $^{76}$As, and 
$\Lambda_{\pi}^{(-)}=0.8\to2.1$ MeV for $^{82}$Br, 
where the superscripts $\pm$ represent those values 
used for positive- or negative-parity orbitals. 

%
\begin{figure}[ht]
\begin{center}
\includegraphics[width=\linewidth]{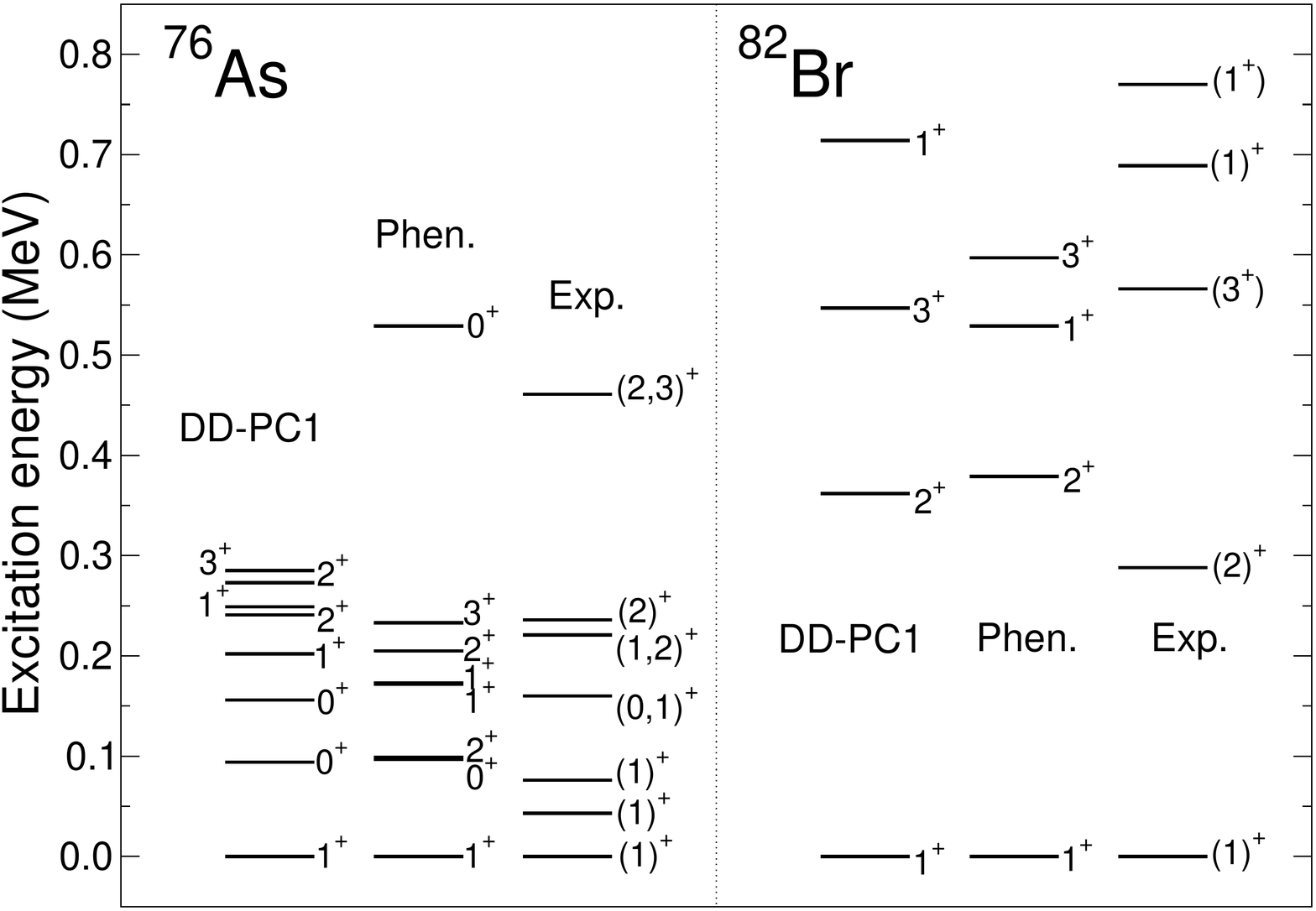}
\caption{Energy spectra of the low-lying 
positive-parity states of $^{76}$As and $^{82}$Br resulting 
from the IBFFM-2 calculations using the SPEs computed by 
the spherical RHB (denoted by ``DD-PC1'') and the phenomenological 
SPEs taken from Ref.~\cite{honma2009}. The corresponding experimental 
spectra are also shown for comparison.}
\label{fig:asbr}
\end{center}
\end{figure}

%
\begin{figure}[ht]
\begin{center}
\includegraphics[width=\linewidth]{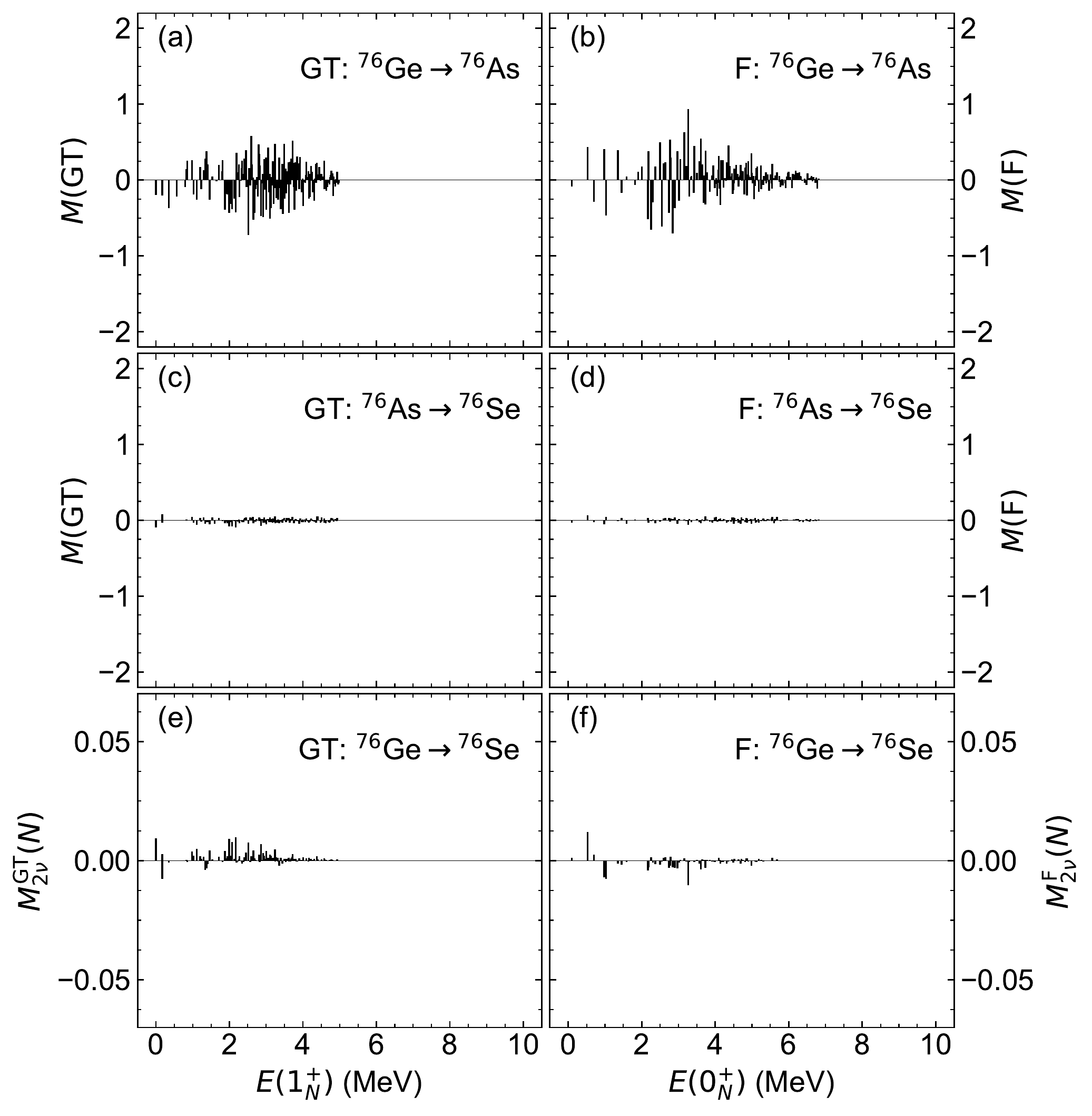}
\caption{Gamow-Teller and Fermi strength distributions 
for the single-$\beta$ decays $^{76}$Ge$\to^{76}$As (a,b) 
and $^{76}$As$\to^{76}$Se (c,d), and for the $\tnbb$ decay 
$^{76}$Ge$\to^{76}$Se (e,f) as functions of the excitation 
energies of the $N$th intermediate states $1^+_N$ and $0^+_N$.}
\label{fig:dist-spe-ge}
\end{center}
\end{figure}

%
\begin{figure}[ht]
\begin{center}
\includegraphics[width=\linewidth]{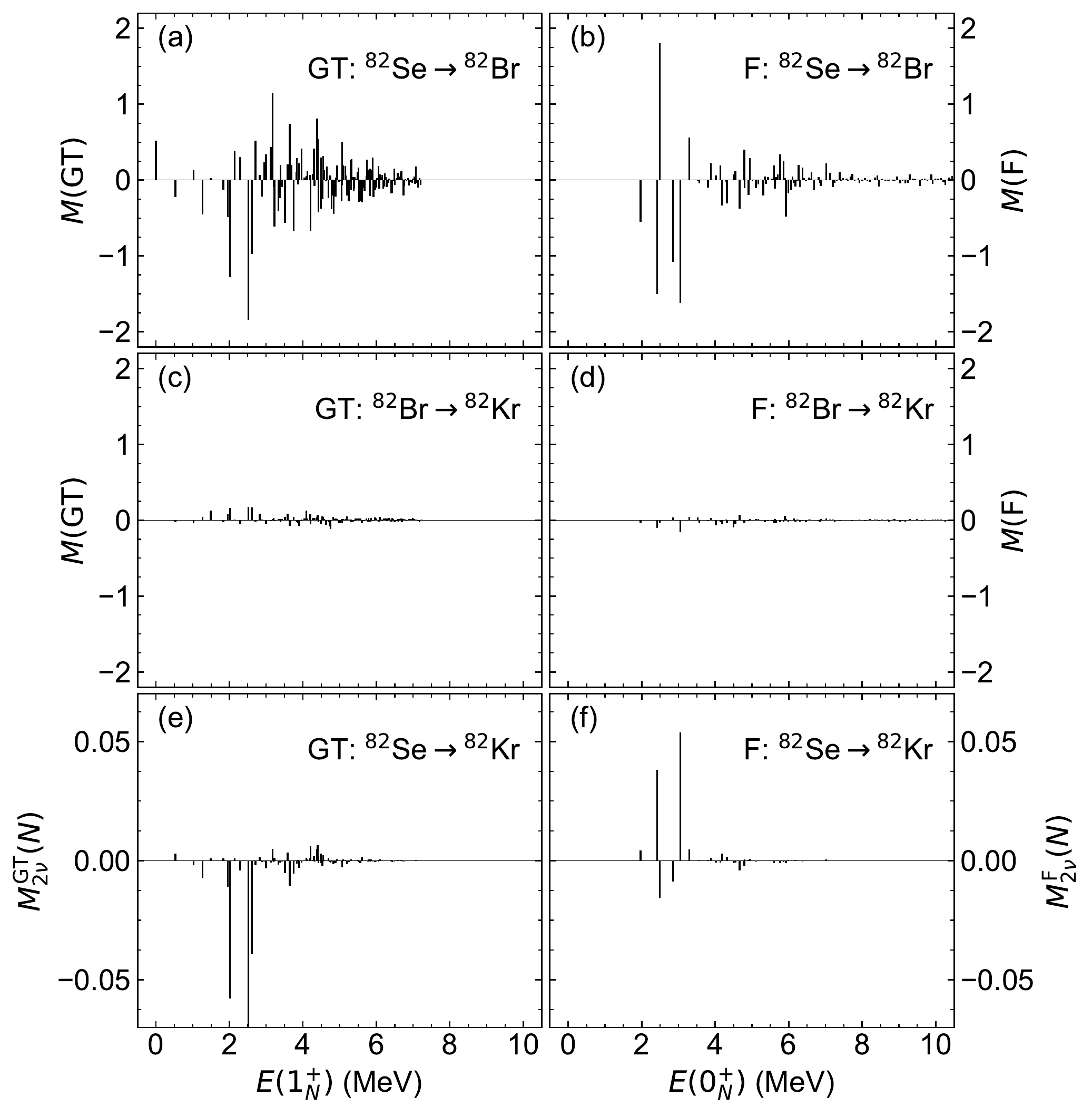}
\caption{Same as Fig.~\ref{fig:dist-spe-ge}, but for the 
single-$\beta$ decays 
$^{82}$Se$\to^{82}$Br (a,b) 
and $^{82}$Br$\to^{82}$Kr (c,d), and for the $\tnbb$ decay 
$^{82}$Se$\to^{82}$Kr (e,f).
}
\label{fig:dist-spe-se}
\end{center}
\end{figure}

In Fig.~\ref{fig:asbr} the excitation spectra 
for the $^{76}$As and $^{82}$Br nuclei calculated with the 
phenomenological SPEs are compared with those calculated 
self-consistently by the spherical RHB method. 
It is found that the quality of the IBFFM-2 description 
does not drastically differ between the two sets of 
calculations with different SPEs, except perhaps for the 
energy levels of the $2^+_1$ and $0^+_2$ states. 
The calculated magnetic dipole moment for the ground state 
$\mu(1^+_1)=0.388\;\mu_N$ for the calculation with the 
self-consistent SPEs (cf. Table~\ref{tab:comp-spe-spe}) 
compares well with the data ($+0.559\pm0.005\;\mu_N$). 
In the calculation with the phenomenological SPEs, 
the value $\mu(5^+_1)=1.157\;\mu_N$ is obtained.

Figures~\ref{fig:dist-spe-ge} and \ref{fig:dist-spe-se} 
show the GT $\mgt$ and Fermi $\mf$ strength distributions 
for the single-$\beta$ decays between the even-even and 
odd-odd nuclei, and the contributions of the terms 
${\mgt}(N)$ and $\mf(N)$ in the $\tnbb$ decay NMEs 
of $^{76}$Ge and $^{82}$Se. 
Figures~\ref{fig:dist-spe-ge}(a), \ref{fig:dist-spe-ge}(b), 
and \ref{fig:dist_eo} show 
that systematics of both the GT and Fermi strengths for the 
$^{76}$Ge$\to^{76}$As decay are similar between the 
two calculations in that they are more or less 
evenly distributed. 
For the $^{82}$Se$\to^{82}$Br decay, the phenomenological 
and RHB SPEs also give rise to similar trends, but the contributions 
from low-lying states to the GT and Fermi strengths 
are slightly larger in the former than in the latter 
[Figs.~\ref{fig:dist-spe-se}(a), \ref{fig:dist-spe-se}(b), 
and \ref{fig:dist_eo}]. 
For both the $^{76}$As and $^{82}$Br $\btm$ decays, 
the two calculations consistently give negligible 
GT and Fermi contributions 
[panels (c) and (d) of Figs.~\ref{fig:dist-spe-ge} 
and \ref{fig:dist-spe-se}, and Fig.~\ref{fig:dist_oe}]. 
From panels (e) and (f) of Figs.~\ref{fig:dist-spe-ge} 
and \ref{fig:dist-spe-se}, 
one finds that, especially for the $^{82}$Se decay, 
contributions of the low-energy intermediate states 
in the calculation that employs the phenomenological SPEs are 
even more significant than in the corresponding results 
obtained by using the RHB SPEs (cf. Fig.~\ref{fig:dist_ee}).

\begin{table}
\caption{\label{tab:spe-comp}
Gamow-Teller $\mgt$, Fermi $\mf$, and 
unquenched total $\mbb$ (\ref{eq:mbb}) NMEs for the 
$\trgg$ and $\trge$ $\tnbb$ decays 
of $^{76}$Ge and $^{82}$Se, 
obtained with the SPEs computed self-consistently (``DD-PC1'')
and those taken from Ref.~\cite{honma2009} (``Phen.''). 
Note that $\qbbe$ values are used for both calculations.}
 \begin{center}
 \begin{ruledtabular}
  \begin{tabular}{llcccc}
&&\multicolumn{2}{c}{$^{76}$Ge}&\multicolumn{2}{c}{$^{82}$Se}\\
\cline{3-4}
\cline{5-6}
&& DD-PC1 & Phen. & DD-PC1 & Phen. \\
\hline
\multirow{3}{*}{$0^+_1$}
&{$\mgt$} & $0.034$ & $0.069$ & $-0.045$ & $-0.103$ \\
&{$\mf$} & $-0.007$ & $-0.022$ & 0.015 & 0.037 \\
&{$|\mbb|$} & 0.062 & 0.134 & 0.087 & 0.203 \\
\cline{1-6}
\multirow{3}{*}{$0^+_2$}
&{$\mgt$} & $0.078$ & $0.118$ & $0.070$ & 0.073 \\
&{$\mf$} & $-0.069$ & $-0.101$ & $-0.064$ & $-0.108$ \\
&{$|\mbb|$} & 0.194 & 0.292 & 0.177 & 0.225 \\
  \end{tabular}
 \end{ruledtabular}
 \end{center}
\end{table}

Table~\ref{tab:spe-comp} shows the Gamow-Teller $\mgt$, 
Fermi $\mf$, and (unquenched) $\tnbb$-decay NMEs $\mbb$ 
of the $\trgg$ and $\trge$ 
transitions of $^{76}$Ge and $^{82}$Se, 
calculated with the phenomenological SPEs of Ref.~\cite{honma2009}, 
and with those computed self-consistently 
within the spherical RHB method. 
The calculation with the phenomenological SPEs 
generally gives larger NMEs than with the RHB ones 
for both $^{76}$Ge and $^{82}$Se. 
If a typical effective $\ga$ value, $\gae=1$, is used, 
then the $\mbb(\trgg)$ matrix elements 
for $^{76}$Ge ($^{82}$Se) computed with 
the phenomenological SPEs are reduced to $\approx0.083$ (0.126), 
which better agrees with experiment than with 
the SPEs provided by the RHB method.

\subsubsection{Choice of the EDF\label{sec:comp-edf}}

Second, NMEs are computed by using the results 
of the constrained Hartree-Fock-Bogoliubov (HFB) 
calculations based on the Gogny-D1M \cite{D1M} EDF. 
The $^{198}$Pt$\to^{198}$Hg decay is taken as an example, since 
the derived IBM-2 parameters for the 
initial and final even-even nuclei, 
and the SPEs for the odd-odd nucleus $^{198}$Au 
are available from the previous mapped IBM-2 results 
\cite{nomura2011sys,nomura2019dodd}. 
The $\tilde\epsilon_{\jr}$ 
and $v_{\jr}^{2}$ values for $^{198}$Au are obtained 
in the same manner as described 
in the previous section. 
The SPEs obtained from the Gogny-D1M HFB calculation, 
as well as the $\tilde\epsilon_{\jr}$ and $v_{\jr}^{2}$ 
values computed within the BCS approximation, 
are found in Table~\ref{tab:comp-edf}. 
The strength parameters for $\hbf^{\rho}$, 
available from Ref.~\cite{nomura2019dodd}, 
are used here for the IBFFM-2 
Hamiltonian without modification, 
but only the parameters for the residual 
neutron-proton interaction $\hff$ 
are changed as $\vd=-0.30\to-0.08$ MeV, $\vssd=-0.033\to0$ MeV, 
and $\vt=0\to0.1$ MeV.

\begin{figure}[ht]
\begin{center}
\includegraphics[width=.8\linewidth]{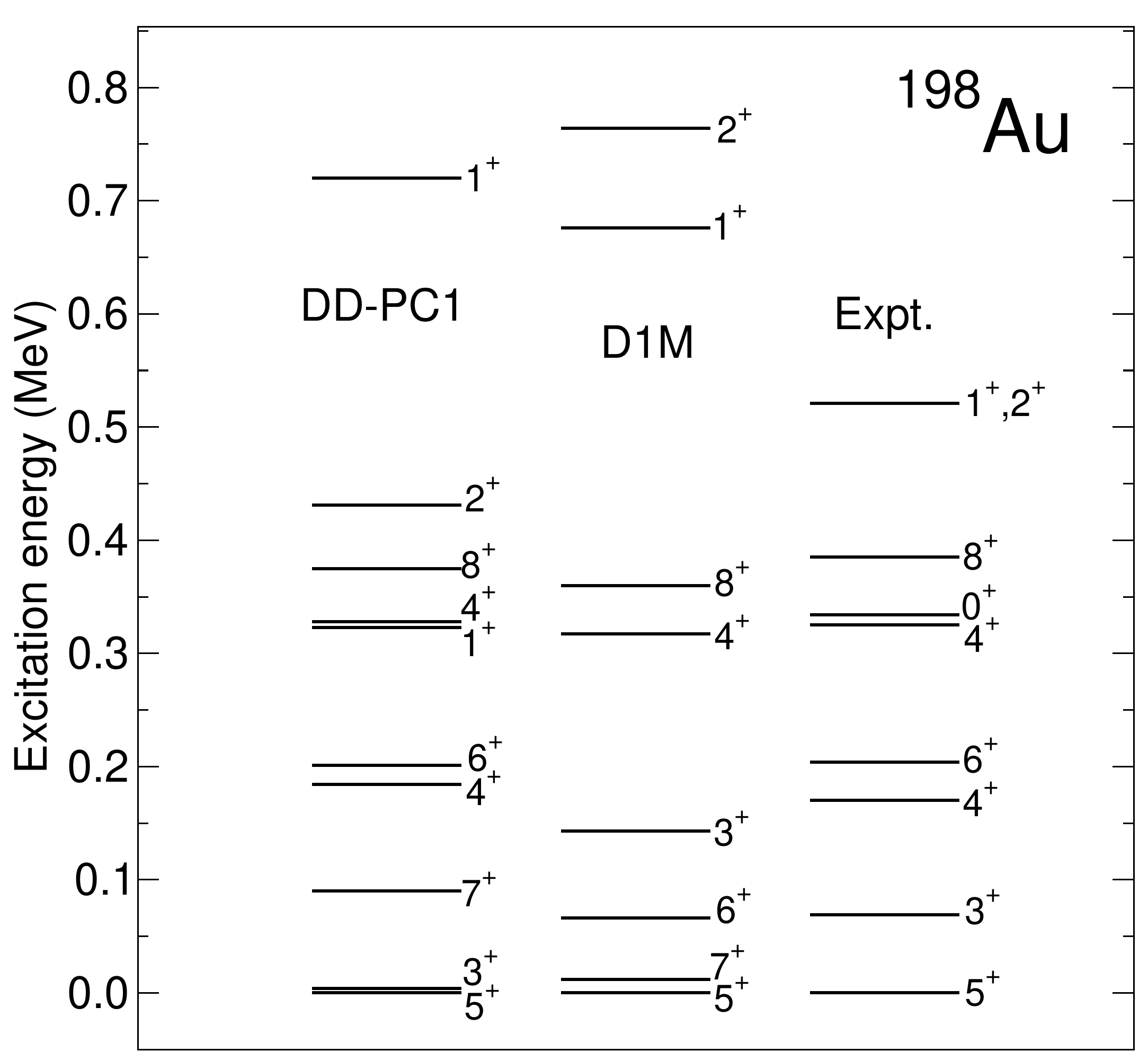}
\caption{Energy spectra for the low-lying 
positive-parity states of $^{198}$Au computed by the 
IBFFM-2 based on the DD-PC1 and Gogny-D1M EDFs. 
The corresponding experimental 
spectrum is also shown for comparison.}
\label{fig:comp-edf}
\end{center}
\end{figure}

The IBM-2 energy spectra for the $^{198}$Pt and $^{198}$Hg computed 
with the microscopic input from the Gogny-D1M HFB calculation are 
found in Figs.~4 and 5 of Ref.~\cite{nomura2011sys}, and in 
Fig.~2 of Ref.~\cite{nomura2019dodd}, respectively, 
which can be compared with the present results shown in 
Fig.~\ref{fig:ee}. 
The description of the low-energy spectra of the even-even 
nuclei appears to be qualitatively similar between the calculations 
with the Gogny-D1M and DD-PC1 EDFs. 
In addition, the calculated energy spectra of the 
low-lying positive-parity states of the intermediate nucleus 
$^{198}$Au from the DD-PC1 and D1M EDFs are compared 
with each other in Fig.~\ref{fig:comp-edf}. 
The description of the observed spectrum appears to be 
better when the DD-PC1 functional is used than for the 
D1M one. As for the magnetic dipole moment for the 
ground state $\mu(5^+_1)$, however, the IBFFM-2 with 
the D1M EDF reproduces correctly the experimental 
value \cite{stone2005}, while the calculation with the 
DD-PC1 EDF does not (cf. Appendix~\ref{sec:doo-em}).

%

Table~\ref{tab:comp-edf} compares the 
predicted $\mgt$, $\mf$, and total (unquenched) $\mbb$ matrix elements 
for the $\trgg$ and $\trge$ decays of $^{198}$Pt, that are 
obtained with the relativistic DD-PC1 
and nonrelativistic Gogny-D1M EDFs. 
In both cases, the theoretical $\qbb$ values, $\qbbt$, 
are used. 
For the Gogny-D1M EDF, the $\qbbt$ value of $\qbbt=0.667$ MeV is obtained, 
which is about half the value for the DD-PC1 EDF, $\qbbt=1.290$ MeV. 
As shown in Table~\ref{tab:comp-edf}, 
for both the $0^{+}_{1}\to0^{+}_{1}$ and 
$0^{+}_{1}\to0^{+}_{2}$ transitions, the calculation with the 
Gogny-D1M EDF produces $\mbb$ that is by more than a factor 
of 4 to 5 larger than the one with the DD-PC1 EDF. 
The discrepancy in the prediction of NME appears to be, 
to some extent, due to the difference between the SPEs 
provided by the two functionals. 
For instance, the energy splitting between  
the lowest neutron orbital $1h_{9/2}$ and the 
second lowest orbital $2f_{7/2}$ is
$\Delta{\epsilon}=\epsilon_{2f_{7/2}}-\epsilon_{1h_{9/2}}\approx0.5$ MeV 
for the Gogny D1M, while the energy gap is much 
larger for the DD-PC1, $\Delta\epsilon\approx2.5$ MeV 
[see Table~\ref{tab:comp-edf-spe} in Appendix~\ref{sec:spe}]. 
It should be nevertheless kept in mind that 
several other model assumptions are made 
in each step of the theoretical procedure. 
To identify the major cause of the 
discrepancy between the results from 
different functionals, 
a more systematic investigation may be in order.

\begin{table}
\caption{\label{tab:comp-edf}
Comparison of Gamow-Teller $\mgt$, Fermi $\mf$, 
and unquenched total $\mbb$ NMEs for the 
$0^{+}_{1}\to0^{+}_{1}$ and $0^{+}_{1}\to0^{+}_{2}$ 
$\tnbb$ decays of $^{198}$Pt, calculated with 
the DD-PC1 and Gogny-D1M EDFs. 
$\qbbt=1.2895$ MeV and $\qbbt=0.6672$ MeV are 
used for the DD-PC1 and Gogny-D1M EDFs, respectively.}
 \begin{center}
 \begin{ruledtabular}
  \begin{tabular}{lcccccc}
\multirow{2}{*}{EDF}
&\multicolumn{3}{c}{$0^{+}_{1}$}
&\multicolumn{3}{c}{$0^{+}_{2}$}\\
\cline{2-4}
\cline{5-7}
&{$\mgt$}
&{$\mf$}
&{$|\mbb|$}
&{$\mgt$}
&{$\mf$}
&{$|\mbb|$}\\
\hline
DD-PC1 & $-0.016$ & $0.001$ & 0.026 & $-0.008$ & $-0.000$ & 0.012 \\
D1M & $-0.074$ & $0.000$ & 0.120 & $-0.034$ & $-0.000$ & 0.054\\
  \end{tabular}
 \end{ruledtabular}
 \end{center}
\end{table}

\subsubsection{Truncation of the intermediate state energy\label{sec:comp-ex}}

Throughout this study, the fixed energy 
cutoff $E_{x}=10$ MeV is considered 
for the intermediate $1^{+}$ and $0^{+}$ states 
of the odd-odd nuclei included in the $\tnbb$-decay NMEs. 
In most cases, both the $\mgt$ and $\mf$ matrix 
elements saturate, if the energy cutoff is made 
at $E_{x}\gtrsim6$ MeV. 

An exception is perhaps the $^{48}$Ca decay, 
since as shown in Figs.~\ref{fig:dist_eo}, 
\ref{fig:dist_oe}, and \ref{fig:dist_ee}, there 
are sizable contributions to the GT and Fermi strengths 
from those states with $E_{x}\gtrsim5$ MeV. 
The present IBFFM-2 calculation 
yields 75 $1^{+}$ and 30 $0^{+}$ states below $E_x=10$ MeV 
in the intermediate nucleus $^{48}$Sc, but 
there are also more states at $E_{x}>10$ MeV. 
If the truncation is set to $E_{x}=20$ MeV, for instance, 
the resultant $\mgt$ and $\mf$ values turn out to be 
by $\approx20\,\%$ smaller than those with $E_{x}=10$ MeV.  

In the previous IBFFM-2 calculation by Yoshida and Iachello 
\cite{yoshida2013} for the $\tnbb$ decays of 
$^{128}$Te and $^{130}$Te,  
the truncation was set to be a much lower energy, $E_{x}=3$ MeV. 
If the same energy cutoff $E_x=3$ MeV is applied to the 
present calculation for the $^{130}$Te decay, 
then the GT matrix element $|\mgt|$, for instance, 
is reduced by a factor of 3. 
While the calculation in Ref.~\cite{yoshida2013} obtained 
100 $1^{+}$ and 50 $0^{+}$ states below $E_{x}=3$ MeV, 
here only 20 $1^{+}$ and 6 $0^{+}$ states 
are found in the same energy range. 
Furthermore, the GT part of the $\tnbb$ decay NME 
was shown \cite{yoshida2013} to be accounted for almost 
solely by the lowest $1^{+}$ intermediate state. 
On the other hand, the corresponding GT strength 
is here more evenly distributed up to $E_x=8$ MeV 
(see Fig.~\ref{fig:dist_ee}). 
These differences between the two calculations 
seem to have originated from 
the different starting points to determine 
the IBM-2 and IBFFM-2 Hamiltonians. 
In Ref.~\cite{yoshida2013}, these ingredients are more or 
less taken from the empirical data, whereas the 
present calculation is largely based on the nuclear EDF.

\section{Concluding remarks\label{sec:summary}}

The $\tnbb$ decay has been investigated within 
the IBM-2 and IBFFM-2 that are based on the nuclear 
density functional theory. 
By mapping the quadrupole triaxial deformation 
energy surface, computed by the constrained RHB method 
that employs the universal functional DD-PC1 and the separable 
pairing interaction, onto the 
corresponding bosonic energy surface, the IBM-2 Hamiltonian 
for the even-even initial- and final-state nuclei 
of the considered $\db$ decays have been determined. 
The same EDF calculation has provided the essential 
ingredients of the IBFFM-2 for describing the 
intermediate odd-odd nuclei and the GT and Fermi 
transition operators.

The EDF-based IBM-2 and IBFFM-2 give 
reasonable descriptions of the excitation spectra and 
electromagnetic transition rates for the low-lying states 
of the relevant even-even and odd-odd nuclei. 
The GT and Fermi parts of the 
$\tnbb$ decays are, in most cases, predicted 
to be dominated by 
the contributions from the low-lying intermediate 
states, the illustrative examples being the 
$^{100}$Mo$\to^{100}$Ru, 
$^{116}$Cd$\to^{116}$Sn, $^{128}$Te$\to^{128}$Xe, and 
$^{150}$Nd$\to^{150}$Sm decays. 
Two different parametrizations for the effective values of 
the axial vector coupling constants $\ga$ 
for $\tnbb$ decay have been introduced: 
one that is modest, decreasing gradually 
from $\gae\approx0.7$ to 0.5, 
and the other varying more drastically 
from $\gae\approx0.9$ to 0.3 as functions of 
the mass number $A$. 
The resultant $\tnbb$-decay NMEs, 
with these $\gae$ factors 
taken into account, exhibit a similar 
trend with $A$ to the observed ones. 
As shown in Fig.~\ref{fig:nme-lit}, 
the previous theoretical predictions on the 
NMEs are quite at variance with each other. 
For most of the $\tnbb$-decay candidates, 
the present values of the NMEs more or less fall into the range of 
various other theoretical values. 
However, as compared to the experiment and majority of 
the earlier theoretical calculations, 
the present calculations have given particularly small NMEs 
for the $^{76}$Ge and $^{82}$Se $\trgg$ decays even 
without the quenching. 
The prediction of the NMEs is shown to be rather 
sensitive to the spherical SPEs for the odd particles, 
and to the choice of the underlying EDF.

This paper presents a way of calculating 
simultaneously the low-lying states of 
the relevant even-even 
and odd-odd nuclei, and the $\beta$ 
and $\db$ decay NMEs, 
based largely on the nuclear EDF. 
Within this theoretical scheme, 
one can go further to explore the $\znbb$ decay. 
At the same time, it remains an open problem to 
identify major sources of uncertainties in 
the prediction of the NMEs within the 
employed theoretical scheme, 
whether it is the SPEs, the use of the specific  
form of the IBM-2 or IBFFM-2 Hamiltonian, the 
deficiency of the underlying EDF, 
or some combinations of these.
As an instance, correlations between the choice of 
the EDF as a microscopic input and the form of the Hamiltonian 
or building blocks of the IBM-2 would be among the 
most relevant factors that significantly affect the 
quality of the wave functions for the initial and final 
states of a given decay process. 
While the current study has resorted to a 
particular type of the nuclear EDF and pairing interaction, 
in some cases there appear notable differences 
between the topology of the 
triaxial quadrupole potential energy surface obtained 
from different classes of the EDF, such as the ones in 
the relativistic and nonrelativistic regimes. 
The present form of the IBM-2 Hamiltonian 
does not account for some important features of 
the SCMF energy surfaces, including the triaxial minimum 
or competing mean-field minima, and a more reliable 
prediction of the initial- and final-state wave functions 
would require the inclusions of some 
higher-order boson terms in the Hamiltonian 
or the configuration mixing between normal and intruder states. 
The configuration mixing would be of particular importance 
for an accurate description of the excited $0^+$ states.
By comparing the low-energy spectra 
and electromagnetic transition rates resulting 
from different versions of the IBM-2 calculations with different 
microscopic inputs, one could quantify which 
of these extensions or terms of the Hamiltonian are 
most relevant to change the spectroscopic predictions 
and the final results for the NMEs. 
These problems will be taken up and 
investigated in a more systematic manner elsewhere.

\acknowledgments
The author would like to thank N. Yoshida 
for helping him with the numerical calculations 
for single-$\beta$ and $\db$ decays.
This work is financed within the Tenure Track Pilot Programme of 
the Croatian Science Foundation and the \'Ecole Polytechnique 
F\'ed\'erale de Lausanne, and the Project TTP-2018-07-3554 Exotic 
Nuclear Structure and Dynamics, with funds of the Croatian-Swiss 
Research Programme.

\appendix

\section{Parameters for the IBM-2 Hamiltonian\label{sec:parab}}

\begin{table}
\caption{\label{tab:parab}
Derived strength parameters for the 
IBM-2 Hamiltonian $\hb$ for even-even nuclei. 
}
 \begin{center}
 \begin{ruledtabular}
  \begin{tabular}{lccccccc}
Nucleus 
& $\epsilon$ & $-\kappa$ 
& $-\kappa_{\nu}$ & $-\kappa_{\pi}$ 
& $\chi_{\nu}$ & $\chi_{\pi}$ & $\kappa'$ \\
& (MeV) & (MeV) & (keV) & (keV) & & & (keV)\\
\hline
$^{46}$Ca & 1.50 & & $57$ & & $-1.30$ & & \\
$^{48}$Ca & 1.50 & & $57$ & & & & \\
$^{48}$Ti & 1.25 & $0.59$ & & & $-1.10$ & $-1.10$ & \\
$^{76}$Ge & 0.60 & 0.38 & & & $-$0.90 & $-$0.50 & \\
$^{76}$Se & 0.84 & 0.22 & & & 0.90 & 0.50 & 21 \\
$^{82}$Se & 0.80 & 0.70 & & & $-$1.00 & $-$1.00 &  \\
$^{82}$Kr & 1.16 & 0.35 & & & $-$0.40 & $-$0.40 &  \\
$^{96}$Zr & 0.80 & 0.35 & & & $-$0.45 & 0.47 & 21 \\
$^{96}$Mo & 0.69 & 0.44 & & & $-$0.65 & 0.45 & 9.5 \\
$^{100}$Mo & 0.58 & 0.35 & & & $-$0.50 & 0.45 &  \\
$^{100}$Ru & 0.50 & 0.44 & & & $-$0.70 & $-0.40$ &  \\
$^{110}$Pd & 0.56 & 0.29 & & & $-$0.30 & 0.47 &  \\
$^{110}$Cd & 0.40 & 0.51 & & & $-$1.10 & $-$0.45 & 55 \\
$^{116}$Sn & 1.50 & & $57$ & & 0.80 & & \\
$^{116}$Cd & 0.85 & 0.23 & & & $-0.30$ & 0.40 & \\
$^{118}$Sn & 1.50 & & $57$ & & 0.80 & & \\
$^{124}$Sn & 1.50 & & $57$ & & 0.80 & & \\
$^{124}$Te & 0.51 & 0.48 & & & 0.44 & $-$0.13 &  \\
$^{128}$Te & 0.78 & 0.48 & & & 0.40 & $-$0.90 &  \\
$^{128}$Xe & 0.42 & 0.42 & & & 0.40 & $-$0.80 &  \\
$^{130}$Te & 0.95 & 0.48 & & & 0.30 & $-$0.78 &  \\
$^{130}$Xe & 0.56 & 0.44 & & & 0.25 & $-$0.86 &  \\
$^{136}$Xe & 1.50 & & & $57$ & & $-0.80$ & \\
$^{136}$Ba & 1.06 & 0.34 & & & $-0.28$ & $-0.95$ &  \\
$^{148}$Nd & 0.43 & 0.27 & & & $-$0.78 & $-$0.54 &  \\
$^{150}$Nd & 0.21 & 0.24 & & & $-$0.80 & $-$0.50 & $-$8 \\
$^{150}$Sm & 0.10 & 0.21 & & & $-$0.70 & $-$0.55 & 9.5 \\
$^{198}$Pt & 0.26 & 0.32 & & & 0.25 & 0.45 &  \\
$^{198}$Hg & 0.60 & 0.47 & & & 1.00 & 0.90 &  \\
$^{200}$Hg & 0.55 & 0.41 & & & 1.20 & 1.00 &  \\
  \end{tabular}
 \end{ruledtabular}
 \end{center}
\end{table}

Table~\ref{tab:parab} lists the adopted 
IBM-2 parameters for the relevant even-even nuclei, 
determined based on the SCMF 
calculations. 
It is noted that the nuclei $^{46}$Ca, $^{118}$Sn, $^{148}$Nd, 
and $^{200}$Hg are not involved in the 
studied $\db$ decays, 
but are here considered only as the boson cores 
for the IBFFM-2 calculations 
(see compositions of the odd-odd nuclei in Table~\ref{tab:odda}). 
The calculated low-energy spectra of these nuclei are 
in a good agreement with the experimental data to 
the same extent as those shown in Fig.~\ref{fig:ee}. 
Note also that, for majority of the studied even-even nuclei, 
the $\hat L\cdot\hat L$ term does not play a major role, 
and is considered only in a few nuclei that exhibit a 
relatively large deformation ($\beta\gtrsim0.15$) in 
the energy surface, such as $^{76}$Se and $^{150}$Nd 
(see Fig.~\ref{fig:pesdft}). 
For the semi-magic nuclei $^{46,48}$Ca, $^{116,118,124}$Sn, 
and $^{136}$Xe, the SCMF energy surface shows only 
spherical minimum, which makes 
it difficult to uniquely determine 
the Hamiltonian parameters. 
In these cases, therefore, common 
values $\epsilon_{d}=1.5$ MeV and 
$\kappa_{\rho}=-0.057$ MeV are used.

\section{Single-particle energies and occupation probabilities for the IBFFM-2\label{sec:spe}}

\begin{table*}
\caption{\label{tab:spe1}
Neutron and proton spherical single-particle 
energies $\epsilon_{\jr}$ (in MeV) 
and occupation probabilities $v^{2}_{\jr}$ obtained from 
the spherical RHB calculations 
for the odd-odd nuclei $^{48}$Sc, $^{76}$As, and $^{82}$Br. 
}
 \begin{center}
 \begin{ruledtabular}
  \begin{tabular}{lccccccccccc}
\multirow{2}{*}{Nucleus}
& & \multicolumn{5}{c}{Neutron orbital} & \multicolumn{5}{c}{Proton orbital} \\
\cline{3-7}
\cline{8-12}
& & $2p_{1/2}$ & $2p_{3/2}$ & $1f_{5/2}$ & $1f_{7/2}$ & $1g_{9/2}$ 
&   $2p_{1/2}$ & $2p_{3/2}$ & $1f_{5/2}$ & $1f_{7/2}$ & $1g_{9/2}$ \\
\hline
\multirow{2}{*}{$^{48}$Sc}
& $\epsilon_{\jr}$
& $-3.239$ & $-5.163$ & $-2.233$ & $-10.381$ & 
& $1.253$ & $-0.759$ & $0.726$ & $-7.785$ & \\ 
& $v^{2}_{\jr}$
& $0.005$ & $0.011$ & $0.007$ & $0.865$ & 
& $0.001$ & $0.002$ & $0.004$ & $0.129$ & \\
\hline
\multirow{2}{*}{$^{76}$As}
& $\epsilon_{\jr}$
& $-12.071$ & $-14.145$ & $-14.097$ & & $-7.950$ 
& $-6.422$ & $-8.203$ & $-8.570$ & & $-2.729$ \\ 
& $v^{2}_{\jr}$
& $0.966$ & $0.985$ & $0.982$ & & $0.316$ 
& $0.116$ & $0.411$ & $0.515$ & & $0.017$ \\
\hline
\multirow{2}{*}{$^{82}$Br}
& $\epsilon_{\jr}$
& $-13.480$ & $-15.487$ & $-16.010$ & & $-9.534$ 
& $-7.839$ & $-9.462$ & $-10.739$ & & $-4.747$ \\ 
& $v^{2}_{\jr}$
& $0.980$ & $0.990$ & $0.990$ & & $0.704$ 
& $0.137$ & $0.462$ & $0.793$ & & $0.019$ \\
  \end{tabular}
 \end{ruledtabular}
 \end{center}
\end{table*}

\begin{table*}
\caption{\label{tab:spe2}
Same as Table~\ref{tab:spe1}, but for the $^{96}$Nb, $^{100}$Tc, 
$^{110}$Ag, and $^{116}$In nuclei.
}
 \begin{center}
 \begin{ruledtabular}
  \begin{tabular}{lccccccccccc}
\multirow{2}{*}{Nucleus}
& & \multicolumn{5}{c}{Neutron orbital} & \multicolumn{5}{c}{Proton orbital} \\
\cline{3-7}
\cline{8-12}
& & $3s_{1/2}$ & $2d_{3/2}$ & $2d_{5/2}$ & $1g_{7/2}$ & $1h_{11/2}$ 
& $2p_{1/2}$ & $2p_{3/2}$ & $1f_{5/2}$ & $1g_{7/2}$ & $1g_{9/2}$ \\
\hline
\multirow{2}{*}{$^{96}$Nb}
& $\epsilon_{\jr}$
& $-4.655$ & $-4.249$ & $-6.596$ & $-5.645$ & $-1.380$ 
& $-9.685$ & $-11.406$ & $-13.164$ & $0.582$ & $-7.167$ \\ 
& $v^{2}_{\jr}$
& $0.073$ & $0.068$ & $0.474$ & $0.211$ & $0.014$ 
& $0.864$ & $0.962$ & $0.981$ & $0.004$ & $0.150$ \\
\hline
\multirow{2}{*}{$^{100}$Tc}
& $\epsilon_{\jr}$
& $-5.479$ & $-5.182$ & $-7.489$ & $-6.955$ & $-2.605$ 
& $-10.117$ & $-11.839$ & $-13.716$ & $-0.064$ & $-7.600$ \\ 
& $v^{2}_{\jr}$
& $0.095$ & $0.092$ & $0.544$ & $0.381$ & $0.022$ 
& $0.909$ & $0.967$ & $0.980$ & $0.008$ & $0.336$ \\
\hline
\multirow{2}{*}{$^{110}$Ag}
& $\epsilon_{\jr}$
& $-7.205$ & $-7.081$ & $-9.274$ & $-9.654$ & $-5.147$ 
& $-11.703$ & $-13.384$ & $-15.563$ & $-2.169$ & $-9.236$ \\ 
& $v^{2}_{\jr}$
& $0.192$ & $0.184$ & $0.784$ & $0.841$ & $0.040$ 
& $0.963$ & $0.983$ & $0.989$ & $0.010$ & $0.710$ \\ 
\hline
\multirow{2}{*}{$^{116}$In}
& $\epsilon_{\jr}$
& $-7.912$ & $-7.964$ & $-10.143$ & $-10.959$ & $-6.368$ 
& $-12.798$ & $-14.497$ & $-16.617$ & $-3.362$ & $-10.239$ \\ 
& $v^{2}_{\jr}$
& $0.437$ & $0.456$ & $0.911$ & $0.948$ & $0.102$ 
& $0.988$ & $0.994$ & $0.996$ & $0.005$ & $0.901$ \\
  \end{tabular}
 \end{ruledtabular}
 \end{center}
\end{table*}

\begin{table*}
\caption{\label{tab:spe3}
Same as Table~\ref{tab:spe1}, but for the $^{124}$Sb, $^{128}$I, 
$^{130}$I, and $^{136}$Cs nuclei.
}
 \begin{center}
 \begin{ruledtabular}
  \begin{tabular}{lccccccccccc}
\multirow{2}{*}{Nucleus}
& & \multicolumn{5}{c}{Neutron orbital} & \multicolumn{5}{c}{Proton orbital} \\
\cline{3-7}
\cline{8-12}
& & $3s_{1/2}$ & $2d_{3/2}$ & $2d_{5/2}$ & $1g_{7/2}$ & $1h_{11/2}$ 
& $3s_{1/2}$ & $2d_{3/2}$ & $2d_{5/2}$ & $1g_{7/2}$ & $1h_{11/2}$ \\
\hline
\multirow{2}{*}{$^{124}$Sb}
& $\epsilon_{\jr}$
& $-8.718$ & $-8.987$ & $-11.180$ & $-12.478$ & $-7.658$ 
& $-1.197$ & $-1.957$ & $-4.110$ & $-5.526$ & $-1.108$ \\ 
& $v^{2}_{\jr}$
& $0.735$ & $0.785$ & $0.954$ & $0.972$ & $0.398$ 
& $0.002$ & $0.004$ & $0.015$ & $0.114$ & $0.004$ \\
\hline
\multirow{2}{*}{$^{128}$I}
& $\epsilon_{\jr}$
& $-9.362$ & $-9.734$ & $-11.949$ & $-13.646$ & $-8.606$ 
& $-1.434$ & $-2.193$ & $-4.332$ & $-6.087$ & $-1.496$ \\ 
& $v^{2}_{\jr}$
& $0.763$ & $0.826$ & $0.961$ & $0.979$ & $0.537$ 
& $0.006$ & $0.012$ & $0.039$ & $0.339$ & $0.011$ \\
\hline
\multirow{2}{*}{$^{130}$I}
& $\epsilon_{\jr}$
& $-9.449$ & $-9.860$ & $-12.088$ & $-13.845$ & $-8.737$ 
& $-2.038$ & $-2.834$ & $-4.954$ & $-6.889$ & $-2.246$ \\ 
& $v^{2}_{\jr}$
& $0.841$ & $0.885$ & $0.973$ & $0.984$ & $0.662$ 
& $0.005$ & $0.010$ & $0.033$ & $0.345$ & $0.010$ \\
\hline
\multirow{2}{*}{$^{136}$Cs}
& $\epsilon_{\jr}$
& $-10.141$ & $-10.690$ & $-12.865$ & $-15.120$ & $-9.820$ 
& $-2.819$ & $-3.676$ & $-5.685$ & $-8.113$ & $-3.377$ \\ 
& $v^{2}_{\jr}$
& $0.959$ & $0.975$ & $0.994$ & $0.997$ & $0.933$ 
& $0.006$ & $0.013$ & $0.039$ & $0.583$ & $0.014$ \\
  \end{tabular}
 \end{ruledtabular}
 \end{center}
\end{table*}

\begin{table*}
\caption{\label{tab:spe4}
Same as Table~\ref{tab:spe1}, but for the $^{150}$Pm, 
and $^{198}$Au nuclei.
}
 \begin{center}
 \begin{ruledtabular}
  \begin{tabular}{lccccccccccccc}
\multirow{2}{*}{Nucleus}
& & \multicolumn{6}{c}{Neutron orbital} & \multicolumn{6}{c}{Proton orbital} \\
\cline{3-8}
\cline{9-14}
& & $3p_{1/2}$ & $3p_{3/2}$ & $2f_{5/2}$ & $2f_{7/2}$ & $1h_{9/2}$ & $1i_{13/2}$ 
& $3s_{1/2}$ & $2d_{3/2}$ & $2d_{5/2}$ & $1g_{7/2}$ & $1h_{9/2}$ & $1h_{11/2}$ \\
\hline
\multirow{2}{*}{$^{150}$Pm}
& $\epsilon_{\jr}$
& $-1.413$ & $-2.045$ & $-2.452$ & $-4.585$ & $-5.491$ & $-2.043$ 
& $-3.965$ & $-4.782$ & $-6.625$ & $-9.778$ & $2.366$ & $-4.946$ \\ 
& $v^{2}_{\jr}$
& $0.012$ & $0.016$ & $0.033$ & $0.186$ & $0.499$ & $0.027$ 
& $0.030$ & $0.072$ & $0.376$ & $0.958$ & $0.004$ & $0.063$ \\
\hline
\multirow{2}{*}{$^{198}$Au}
& $\epsilon_{\jr}$
& $-7.037$ & $-7.405$ & $-8.424$ & $-10.687$ & $-13.191$ & $-8.954$ 
& $-6.853$ & $-8.421$ & $-9.902$ & $-14.382$ & $-3.198$ & $-9.213$ \\ 
& $v^{2}_{\jr}$
& $0.264$ & $0.391$ & $0.779$ & $0.968$ & $0.990$ & $0.887$ 
& $0.250$ & $0.832$ & $0.961$ & $0.995$ & $0.010$ & $0.942$ \\
  \end{tabular}
 \end{ruledtabular}
 \end{center}
\end{table*}

Tables~\ref{tab:spe1}, \ref{tab:spe2}, \ref{tab:spe3}, and 
\ref{tab:spe4} 
show the employed neutron and proton spherical SPEs 
$\epsilon_{\jr}$ and occupation probabilities $v^{2}_{\jr}$ 
within the considered fermion configuration spaces for the 
intermediate odd-odd nuclei. 
They are obtained by the spherical 
RHB calculation with the 
DD-PC1 EDF constrained to zero deformation 
(see Ref.~\cite{nomura2016odd} for details). 
As seen in Table~\ref{tab:spe1}, 
only the configuration space for $^{48}$Sc includes the 
$1f_{7/2}$ orbital for both proton and neutron, 
since in this case the $^{40}$Ca nucleus is taken as the boson core, 
and the low-lying states of $^{48}$Sc are assumed to be 
mainly composed of the proton and neutron $1f_{7/2}$ 
single-particle configurations. 
As seen in Table~\ref{tab:spe2}, the proton $1g_{7/2}$ 
orbital, which belongs to the next major shell 
$Z=50-82$, is included in the calculation. 
This gives rise to a non-vanishing contribution 
to the Fermi matrix element by the coupling between the 
neutron and proton $1g_{7/2}$ orbitals. 
The $\pi 1g_{7/2}$ orbital is included 
in order to see to what extent the Fermi transition, 
arising in that way, contributes to the final result on the 
$\mbb$ matrix element. 
For the same reason, the proton single-particle spaces for 
$^{150}$Pm and $^{198}$Au include the $1h_{9/2}$ orbital, 
which basically comes from the next major oscillator shell $82<Z<126$.

\begin{table*}
\caption{\label{tab:comp-spe-spe}
Phenomenological single-particle energies $\epsilon_{\jr}$ (in MeV) 
that are taken from Ref.~\cite{honma2009} (denoted by ``Phen.''), 
and the quasiparticle 
energies $\tilde\epsilon_{\jr}$ (in MeV) and occupation probabilities 
$v^2_{\jr}$ obtained from the BCS calculation for $^{76}$As 
and $^{82}$Br. The corresponding quantities computed 
self-consistently by the spherical RHB method are also 
shown for comparison. See the main text in Sec.~\ref{sec:comp-spe}.}
 \begin{center}
 \begin{ruledtabular}
  \begin{tabular}{lllcccccccc}
\multirow{2}{*}{Nucleus} & & 
& \multicolumn{4}{c}{Neutron orbital} 
& \multicolumn{4}{c}{Proton orbital} \\
\cline{4-7}
\cline{8-11}
& & & 
$2p_{1/2}$ & $2p_{3/2}$ & $1f_{5/2}$ & $1g_{9/2}$ &
$2p_{1/2}$ & $2p_{3/2}$ & $1f_{5/2}$ & $1g_{9/2}$ \\
\hline
\multirow{6}{*}{$^{76}$As} 
& \multirow{2}{*}{$\epsilon_{\jr}$} 
& Phen.
& $-7.839$ & $-9.828$ & $-8.709$ & $-6.262$ 
& $-7.839$ & $-9.828$ & $-8.709$ & $-6.262$ \\ &
& RHB
& $-12.071$ & $-14.145$ & $-14.097$ & $-7.950$ 
& $-6.422$ & $-8.203$ & $-8.570$ & $-2.729$ \\
\cline{3-11}
& \multirow{2}{*}{$\tilde\epsilon_{\jr}$}
& Phen.
& $1.880$ & $3.548$ & $2.554$ & $1.408$ 
& $2.063$ & $1.449$ & $1.529$ & $3.404$ \\ &
& RHB
& $3.570$ & $5.251$ & $5.813$ & $1.456$ 
& $2.671$ & $1.739$ & $1.792$ & $5.896$ \\
\cline{3-11}
& \multirow{2}{*}{$v^2_{\jr}$}
& Phen.
& $0.841$ & $0.961$ & $0.921$ & $0.395$ 
& $0.128$ & $0.656$ & $0.282$ & $0.043$ \\ &
& RHB
& $0.966$ & $0.985$ & $0.982$ & $0.316$ 
& $0.116$ & $0.411$ & $0.515$ & $0.017$ \\
\hline
\multirow{6}{*}{$^{82}$Br}
& \multirow{2}{*}{$\epsilon_{\jr}$}
& Phen.
& $-7.839$ & $-9.828$ & $-8.709$ & $-6.262$ 
& $-7.839$ & $-9.828$ & $-8.709$ & $-6.262$ \\ &
& RHB
& $-13.480$ & $-15.487$ & $-16.010$ & $-9.534$ 
& $-7.839$ & $-9.462$ & $-10.739$ & $-4.747$ \\
\cline{3-11}
& \multirow{2}{*}{$\tilde\epsilon_{\jr}$}
& Phen.
& $2.673$ & $4.510$ & $3.456$ & $1.520$ 
& $1.637$ & $1.677$ & $1.328$ & $2.863$ \\ &
& RHB
& $4.310$ & $5.900$ & $7.184$ & $1.394$ 
& $2.368$ & $1.633$ & $1.963$ & $4.974$ \\ 
\cline{3-11}
& \multirow{2}{*}{$v^2_{\jr}$}
& Phen.
& $0.934$ & $0.978$ & $0.962$ & $0.745$ 
& $0.207$ & $0.807$ & $0.466$ & $0.057$ \\ & 
& RHB
& $0.980$ & $0.990$ & $0.990$ & $0.704$ 
& $0.137$ & $0.462$ & $0.793$ & $0.019$ \\
  \end{tabular}
 \end{ruledtabular}
 \end{center}
\end{table*}

\begin{table*}
\caption{\label{tab:comp-edf-spe}
Single-particle $\epsilon_{\jr}$ and quasiparticle 
$\tilde\epsilon_{\jr}$ energies (both in MeV), 
and occupation probabilities 
$v^2_{\jr}$ employed for the IBFFM-2 calculations based on the 
Gogny-D1M and DD-PC1 EDFs. The $\tilde\epsilon_{\jr}$ 
and $v^2_{\jr}$ values for the Gogny-D1M EDF are obtained 
from the BCS calculation. See the main text in Sec.~\ref{sec:comp-edf}. 
}
 \begin{center}
 \begin{ruledtabular}
  \begin{tabular}{llcccccccccccc}
& & 
\multicolumn{6}{c}{Neutron orbital} & 
\multicolumn{6}{c}{Proton orbital} \\
\cline{3-8}
\cline{9-14}
& &
$3p_{1/2}$ & $3p_{3/2}$ & $2f_{5/2}$ & $2f_{7/2}$ & $1h_{9/2}$ & $1i_{13/2}$ &
$3s_{1/2}$ & $2d_{3/2}$ & $2d_{5/2}$ & $1g_{7/2}$ & $1h_{9/2}$ & $1h_{11/2}$ \\
\hline
\multirow{2}{*}{$\epsilon_{\jr}$} 
& D1M
& $-7.285$ & $-8.234$ & $-8.344$ & $-11.147$ & $-11.690$ & $-8.855$ 
& $-7.738$ & $-8.615$ & $-10.313$ & $-12.882$ & $-2.505$ & $-8.565$ \\
& DD-PC1
& $-7.037$ & $-7.405$ & $-8.424$ & $-10.687$ & $-13.191$ & $-8.954$ 
& $-6.853$ & $-8.421$ & $-9.902$ & $-14.382$ & $-3.198$ & $-9.213$ \\
\cline{2-14}
\multirow{2}{*}{$\tilde\epsilon_{\jr}$}
& D1M
& $1.090$ & $0.895$ & $0.934$ & $3.295$ & $3.823$ & $1.234$ 
& $0.861$ & $1.313$ & $2.828$ & $5.335$ & $5.182$ & $1.275$ \\
& DD-PC1
& $1.301$ & $1.161$ & $1.352$ & $3.166$ & $5.632$ & $1.668$ 
& $1.329$ & $1.333$ & $2.559$ & $6.912$ & $4.375$ & $1.899$ \\
\cline{2-14}
\multirow{2}{*}{$v^2_{\jr}$}
& D1M
& $0.189$ & $0.651$ & $0.704$ & $0.983$ & $0.987$ & $0.861$ 
& $0.570$ & $0.880$ & $0.977$ & $0.994$ & $0.007$ & $0.872$ \\
& DD-PC1
& $0.264$ & $0.391$ & $0.779$ & $0.968$ & $0.990$ & $0.887$ 
& $0.250$ & $0.832$ & $0.961$ & $0.995$ & $0.010$ & $0.942$ \\
  \end{tabular}
 \end{ruledtabular}
 \end{center}
\end{table*}

Table~\ref{tab:comp-spe-spe} lists the phenomenological SPEs 
$\epsilon_{\jr}$ of Ref.~\cite{honma2009}, and the 
quasiparticle energies $\tilde\epsilon_{\jr}$ 
and occupation probabilities $v^2_{\jr}$ for $^{76}$As 
and $^{82}$Se calculated within the BCS approximation, 
in comparison to the RHB results. 
Table~\ref{tab:comp-edf-spe} summarizes the 
$\epsilon_{\jr}$, $\tilde\epsilon_{\jr}$, and $v^2_{\jr}$ 
for $^{198}$Au, obtained with the RHB method with 
the DD-PC1 EDF, and those based on the HFB calculation 
using the Gogny-D1M EDF.

\section{Parameters for the IBFFM-2 Hamiltonian\label{sec:paraff}}

\begin{table}
\caption{\label{tab:paraff}
Adopted IBFFM-2 parameters for the boson-fermion 
interactions $\hbf^{\rho}$ 
(\ref{eq:hbf})--(\ref{eq:exc-semimagic}), and for the 
residual neutron-proton interaction $\hff$ (\ref{eq:hff}), 
for the intermediate odd-odd nuclei. 
For each odd-odd nucleus, the values of the strength parameters 
$\Gamma_{\rho}$, $\Lambda_{\rho}$, and $A_{\rho}$ 
(with $\rho=\nu$ and $\pi$) that appear in the 
upper and lower rows refer to the ones employed for 
computing the positive- and negative-parity states of the neighboring 
odd-mass nuclei, respectively. 
All the parameters are in MeV units.}
 \begin{center}
 \begin{ruledtabular}
  \begin{tabular}{lccccccccc}
Nucleus 
& $\Gamma_{\nu}$ & $\Lambda_{\nu}$ & $A_{\nu}$ &
$\Gamma_{\pi}$ & $\Lambda_{\pi}$ & $A_{\pi}$ &
$\vd$ & $\vssd$ & $\vt$ \\
\hline
\multirow{2}{*}{$^{48}$Sc} &
& & & & & &
\multirow{2}{*}{0.6} &
\multirow{2}{*}{} &
\multirow{2}{*}{} \\ & 
0.3 & 1.0 & & 0.3 & & &  & & \\
\hline
\multirow{2}{*}{$^{76}$As} &
0.3 & 2.0 & & 1.0 & & &
\multirow{2}{*}{0.8} &
\multirow{2}{*}{} &
\multirow{2}{*}{0.02} \\ & 
0.3 & 7.5 & $-0.6$ & 0.3 & 1.6 & $-0.8$ &  & & \\
\hline
\multirow{2}{*}{$^{82}$Br} &
0.3 & 2.1 & & 2.5 & & & 
\multirow{2}{*}{} &
\multirow{2}{*}{$-0.4$} &
\multirow{2}{*}{} \\ & 
0.3 & 0.8 & & 0.3 & 0.8 & & & & \\
\hline
\multirow{2}{*}{$^{96}$Nb} &
0.3 & 0.4 & & 0.3 & 0.9 & $-0.5$ & 
\multirow{2}{*}{0.8} &
\multirow{2}{*}{} &
\multirow{2}{*}{} \\ & 
0.3 & & $-1.5$ & 0.3 & 1.6 & $-0.3$ & & & \\
\hline
\multirow{2}{*}{$^{100}$Tc} &
0.3 & 0.35 & & 0.3 & 0.9 & & 
\multirow{2}{*}{$-0.08$} &
\multirow{2}{*}{} &
\multirow{2}{*}{0.05} \\ & 
0.3 & & & 0.3 & 5.0 & $-1.0$ & & & \\
\hline
\multirow{2}{*}{$^{110}$Ag} &
0.3 & 2.8 & & 0.3 & 8.0 & & 
\multirow{2}{*}{$-0.08$} &
\multirow{2}{*}{} &
\multirow{2}{*}{0.8} \\ & 
0.3 & & & 0.3 & 1.4 & & & & \\
\hline
\multirow{2}{*}{$^{116}$In} &
0.3 & 0.2 & $-0.15$ & 0.3 & & & 
\multirow{2}{*}{$-0.8$} &
\multirow{2}{*}{} &
\multirow{2}{*}{0.4} \\ & 
0.3 & 0.2 & $-0.15$ & 1.0 & & & & & \\
\hline
\multirow{2}{*}{$^{124}$Sb} &
0.3 & 10.0 & $-1.0$ & 1.2 & & & 
\multirow{2}{*}{$-0.03$} &
\multirow{2}{*}{} &
\multirow{2}{*}{0.135} \\ & 
0.3 & 2.0 & $-0.60$ & 1.2 & & & & & \\
\hline
\multirow{2}{*}{$^{128}$I} &
0.3 & 6.5 & & 0.3 & 0.6 & $-1.0$ & 
\multirow{2}{*}{} &
\multirow{2}{*}{$-0.51$} &
\multirow{2}{*}{} \\ & 
0.3 & 0.9 & $-0.2$ & 0.3 & & $-1.05$ & & & \\
\hline
\multirow{2}{*}{$^{130}$I} &
0.3 & 7.6 & & 0.3 & 0.8 & $-0.75$ & 
\multirow{2}{*}{$-0.08$} &
\multirow{2}{*}{} &
\multirow{2}{*}{0.01} \\ & 
0.3 & 0.9 & $-0.5$ & 0.3 & & $-1.05$ & & & \\
\hline
\multirow{2}{*}{$^{136}$Cs} &
0.6 & & & 0.6 & 0.2 & $-0.9$ & 
\multirow{2}{*}{$-0.08$} &
\multirow{2}{*}{} &
\multirow{2}{*}{0.09} \\ & 
0.6 & & & 0.6 & 0.2 & $-0.9$ & & & \\
\hline
\multirow{2}{*}{$^{150}$Pm} &
0.3 & 10.0 & $-0.8$ & 0.3 & 0.4 & $-1.0$ & 
\multirow{2}{*}{$-0.08$} &
\multirow{2}{*}{} &
\multirow{2}{*}{0.14} \\ & 
0.3 & 0.6 & $-1.0$ & 3.0 & & & & \\
\hline
\multirow{2}{*}{$^{198}$Au} &
0.3 & 1.6 & & 0.3 & 0.6 & & 
\multirow{2}{*}{$-0.08$} &
\multirow{2}{*}{} &
\multirow{2}{*}{0.05} \\ & 
0.3 & 1.6 & $-0.4$ & 0.3 & 0.6 & & & & \\
  \end{tabular}
 \end{ruledtabular}
 \end{center}
\end{table}

Table~\ref{tab:paraff} lists the adopted 
strength parameters for the boson-fermion interaction 
$\hbf^\rho$ (\ref{eq:hbf})--(\ref{eq:exc-semimagic}) 
and the residual neutron-proton interaction 
$\hff$ (\ref{eq:hff}) in the IBFFM-2 Hamiltonian. 
For many of the odd-odd nuclei, the strength parameters for the 
dynamical terms are assumed to take a fixed value 
$\Gamma_{\rho}=0.3$ MeV. 
In most cases, only the delta and tensor terms are considered. 
The parameters for $\hff$ also do not differ too much 
from one nucleus to another 
within the same mass region. For instance, 
the parameter $\vd$ is stable in the 
mass $A\gtrsim100$ region, except for $^{116}$In.

\section{Coefficients for one-particle transfer operators\label{sec:opt}}

The present formulation of the one-particle transfer 
operators is based on the one developed in the 
earlier IBFM-2 studies for 
the single-$\beta$ decays of mass $A\approx130$ nuclei 
\cite{dellagiacoma1988phdthesis,DELLAGIACOMA1989}. 
Within the generalized seniority considerations, 
the coefficients $\zeta_{j}$, $\zeta_{jj'}$, 
$\theta_{j}$, and $\theta_{jj'}$ 
in Eqs.~(\ref{eq:creation1}) and (\ref{eq:creation2}) 
are assumed to take the forms 
\begin{subequations}
 \begin{align}
\label{eq:zeta1}
\zeta_{\jr}&= 
u_{\jr} \frac{1}{K_{\jr}'}, \\
\label{eq:zeta2}
\zeta_{\jr\jr'}
&= -v_{\jr} 
\beta_{\jr'\jr}
\sqrt{\frac{10}{N_{\rho}(2\jr+1)}}\frac{1}{K K_{\jr}'} , \\ 
\label{eq:theta1}
\theta_{\jr}
&= \frac{v_{\jr}}{\sqrt{N_{\rho}}} 
\frac{1}{K_{\jr}''},\\
\label{eq:theta2}
\theta_{\jr\jr'}
&= u_{\jr} 
\beta_{\jr'\jr}
\sqrt{\frac{10}{2\jr+1}} \frac{1}{K K_{\jr}''}.
\end{align}
\end{subequations}
The occupation $v_{\jr}$ and unoccupation 
$u_{\jr}$ amplitudes are here provided by the spherical 
RHB calculation. 
The factors $K$, $K'_{\jr}$, and $K''_{\jr}$ 
are given by
\begin{subequations} 
\begin{align}
&K = \left( \sum_{\jr\jr'} 
\beta_{\jr\jr'}^{2} \right)^{1/2},\\
&K_{\jr}' = \left[ 1 + 2 
\left(\frac{v_{\jr}}{u_{\jr}}\right)^{2} \frac{\braket{(\hat 
n_{s_\rho}+1)\hat n_{d_\rho}}_{0^+_1}} {N_\rho(2\jr+1)} \frac{\sum_{\jr'} 
\beta_{\jr'\jr}^{2}}{K^{2}} \right]^{1/2} ,\\
&K_{\jr}'' = \left[ 
\frac{\braket{\hat n_{s_\rho}}_{0^+_1}}{N_\rho} 
+2\left(\frac{u_{\jr}}{v_{\jr}}\right)^{2} \frac{\braket{\hat 
n_{d_\rho}}_{0^+_1}}{2\jr+1} \frac{\sum_{\jr'} \beta_{\jr'\jr}^{2}}{K^{2}} 
\right]^{1/2},
\end{align} 
\end{subequations}
where $\hat n_{s_{\rho}}$ is the number operator 
for the $s_\rho$ boson and $\braket{\cdots}_{0^+_1}$ 
represents the expectation 
value of a given operator in the $0^+_1$ ground state 
of the even-even nucleus. 
A more detailed account on the $\beta$-decay 
operators within the IBFM-2 framework is found in 
Refs.~\cite{dellagiacoma1988phdthesis,DELLAGIACOMA1989,IBFM}.

\section{Electromagnetic properties of the even-even and odd-odd nuclei\label{sec:doo-em}}

The $E2$ operator $\hat T^{(E2)}$ 
in the IBFFM-2 takes the form \cite{IBFM}:
\begin{align}
 \label{eq:e2}
\hat T^{(E2)}
= \hat T^{(E2)}_\mathrm{B}
+ \hat T^{(E2)}_\mathrm{F}, 
\end{align}
where the first and second terms are the 
boson and fermion parts, given respectively as
\begin{align}
 \label{eq:e2b}
\hat T^{(E2)}_\mathrm{B}
=e_\nu^\mathrm{B}\hat Q_\nu
+e_\pi^\mathrm{B}\hat Q_\pi
\end{align}
and 
\begin{align}
 \label{eq:e2f}
\hat T^{(E2)}_\mathrm{F}
=-\frac{1}{\sqrt{5}}
&\sum_{\rho=\nu,\pi}
\sum_{\jr\jr'}
(u_{\jr}u_{\jr'}-v_{\jr}v_{\jr'})
\nonumber\\
&\times
\left\langle
\ell_\rho\frac{1}{2}\jr 
\bigg\| 
e^\mathrm{F}_\rho r^2 Y^{(2)} 
\bigg\|
\ell_\rho'\frac{1}{2}\jr'
\right\rangle
(a_{\jr}^\dagger\times\tilde a_{\jr'})^{(2)}.
\end{align}
The boson effective charges 
$e^\mathrm{B}_\nu = e^\mathrm{B}_\pi$ are adjusted to 
reproduce the experimental 
$B(E2; 2^+_1\rightarrow 0^+_1)$ value 
for the even-even boson core nucleus summarized in 
Table~\ref{tab:odda}. 
The standard neutron and proton effective charges 
$e^\mathrm{F}_\nu =0.5$ $e$b,  
$e^\mathrm{F}_\pi =1.5$ $e$b 
are adopted from the earlier IBFFM-2 
calculation on the odd-odd Cs nuclei \cite{nomura2020cs}. 
The $M1$ transition operator $\hat T^{(M1)}$ reads 
\begin{align}
 \label{eq:m1}
\hat T^{(M1)}
=\sqrt{\frac{3}{4\pi}}
&\sum_{\rho=\nu,\pi}
\Biggl[
g_\rho^\mathrm{B}\hat L_\rho
-\frac{1}{\sqrt{3}}
\sum_{\jr\jr'}
(u_{\jr}u_{\jr'}+v_{\jr}v_{\jr'})
\nonumber \\
&\times
\left\langle \jr \| g_\ell^\rho{\mathbf{\ell}}+g_s^\rho{\bf s} 
\| \jr' \right\rangle
(a_{\jr}^\+\times\tilde a_{\jr'})^{(1)}
\Biggr].
\end{align}
The empirical $g$ factors for the neutron and
proton bosons, $g_\nu^\mathrm{B}=0\,\mu_N$ and 
$g_\pi^\mathrm{B}=1.0\,\mu_N$, respectively, are adopted. 
For the neutron (or proton) $g$ factors, the standard 
Schmidt values $g_\ell^\nu=0\,\mu_N$ and $g_s^\nu=-3.82\,\mu_N$
(or $g_\ell^\pi=1.0\,\mu_N$ and $g_s^\pi=5.58\,\mu_N$) 
are used, with $g_s^\rho$ quenched by 30\% 
with respect to the free value. 

%
\begin{figure}[ht]
\begin{center}
\includegraphics[width=\linewidth]{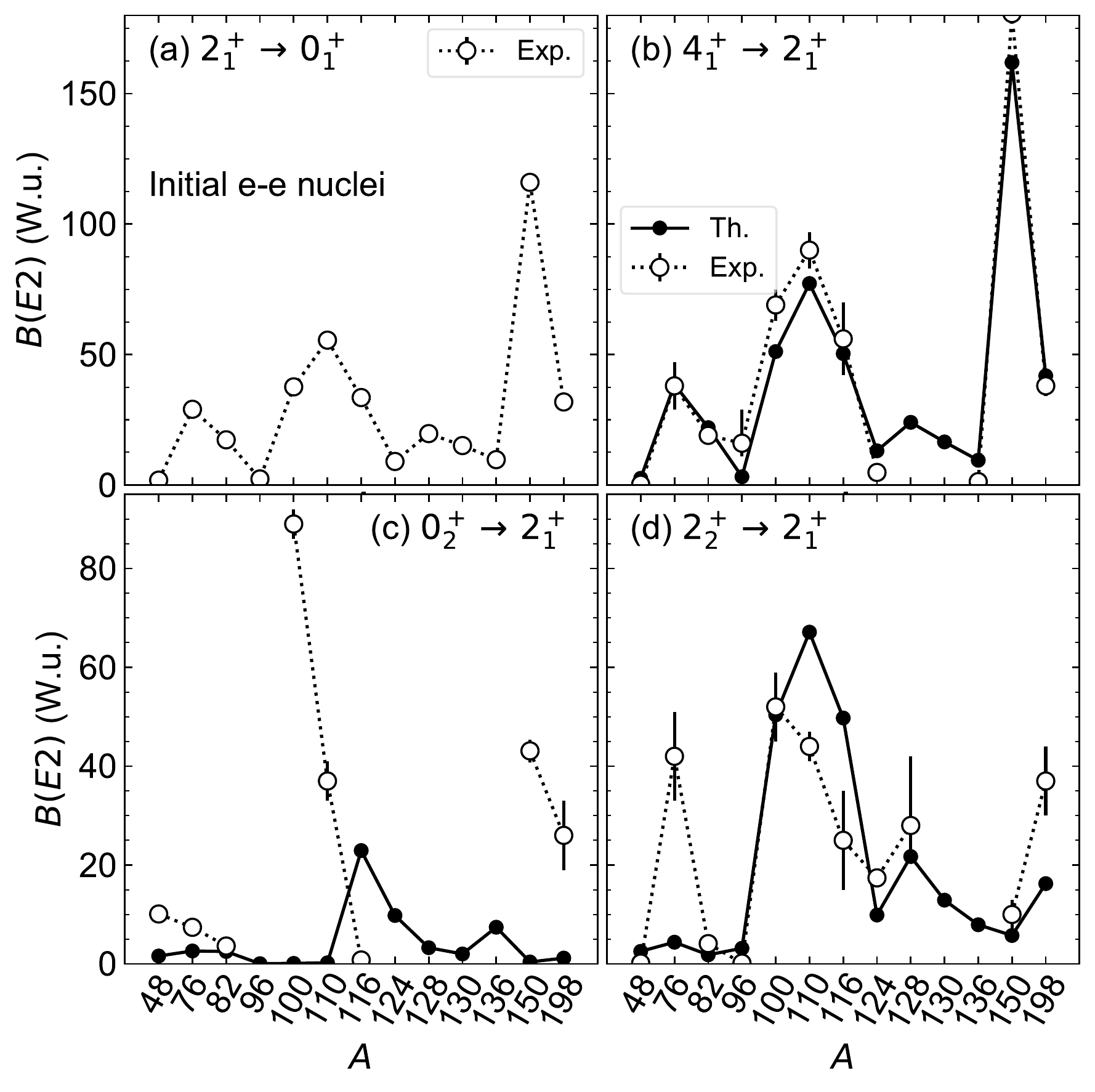}
\caption{Theoretical and experimental $B(E2)$ transition 
strengths in Weisskopf units (W.u.) 
for the transitions (a) $2^+_1\to0^+_1$, 
(b) $4^+_1\to2^+_1$, (c) $0^+_2\to2^+_1$ and (d) $2^+_2\to2^+_1$ 
for the even-even initial nuclei with mass $A$ of the 
considered $\tnbb$ decays. The experimental data are 
taken from Ref.~\cite{data}. The effective boson 
charges are adjusted to reproduce the experimental 
$B(E2;2^+_1\to0^+_1)$ value for each nucleus, and therefore 
the theoretical values are not shown in panel (a).}
\label{fig:ee1-em}
\end{center}
\end{figure}

%
\begin{figure}[ht]
\begin{center}
\includegraphics[width=\linewidth]{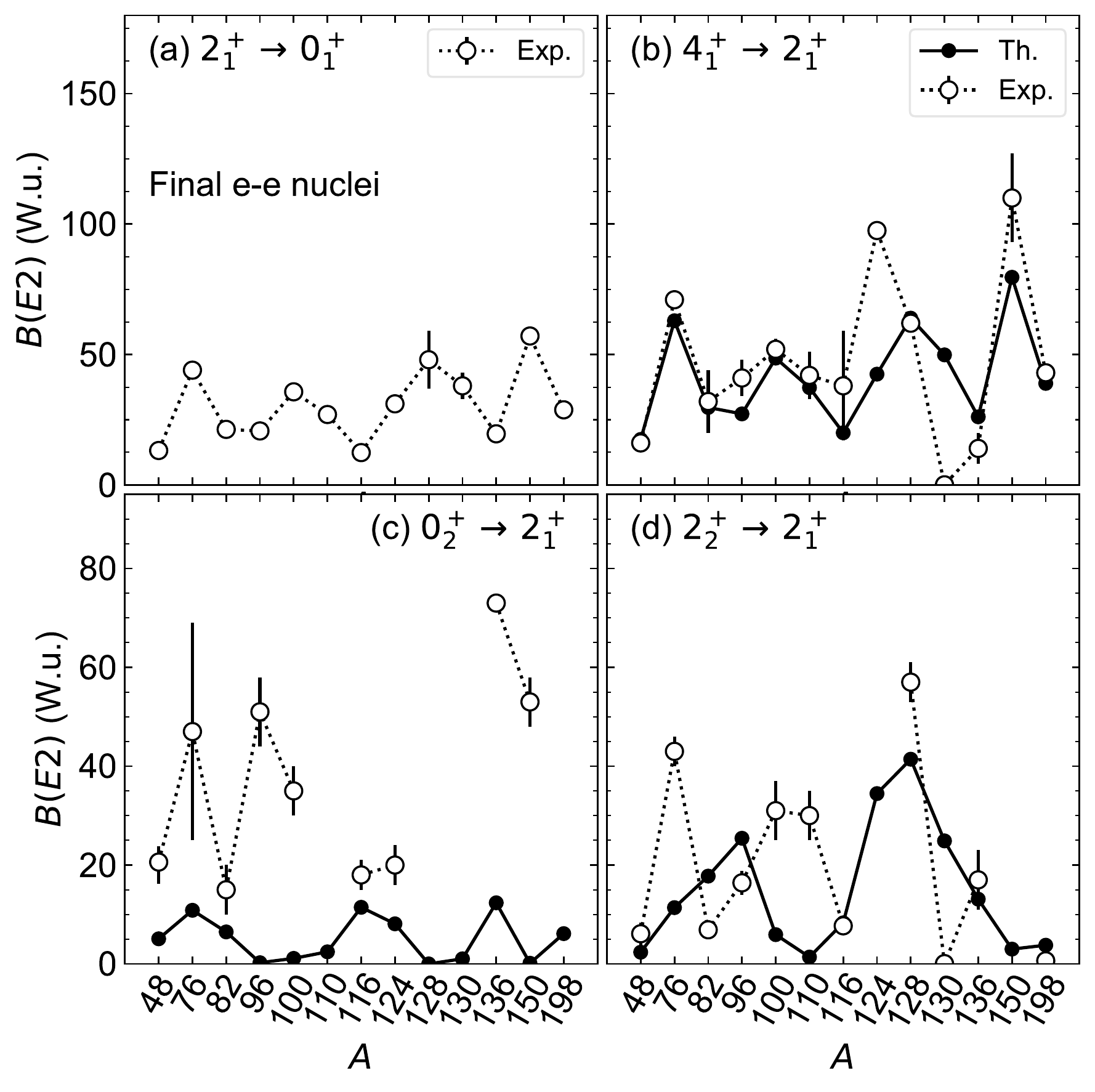}
\caption{Same as Fig.~\ref{fig:ee1-em}, but for the 
even-even final nuclei of the considered $\tnbb$ decays.}
\label{fig:ee2-em}
\end{center}
\end{figure}

\begin{table*}
\caption{\label{tab:doo-em}
Comparison of calculated and available experimental data for 
the electric quadrupole $Q(I)$ (in $e$b) and magnetic 
dipole $\mu(I)$ (in $\mu_N$) moments, and the $B(E2)$ 
and $B(M1)$ transition strengths (in W.u.) 
of the intermediate 
odd-odd nuclei. The relativistic functional DD-PC1 
is used. 
The experimental values 
are taken from Refs.~\cite{data,stone2005}.
}
 \begin{center}
 \begin{ruledtabular}
  \begin{tabular}{lccc}
Nucleus 
& Property & Theory & Experiment \\
\hline
$^{48}$Sc
& $\mu(6^+_{1})$ & $+3.098$ & $+3.737\pm0.012$ \\
$^{76}$As
& $\mu(1^+_{1})$ & $+0.388$ & $+0.559\pm0.005$ \\
$^{96}$Nb
& $\mu(6^+_{1})$ & $+4.547$ & $4.976\pm0.004$ \\
& $B(M1;5^+_{1}\to4^+_1)$ & $+1.23$ & $>0.021$ \\
& $B(M1;2^+_{1}\to3^+_1)$ & $+0.24$ & $>0.00017$ \\
$^{110}$Ag
& $\mu(1^+_{1})$ & $+1.900$ & $2.7271\pm0.0008$ \\
& $Q(1^+_{1})$ & $+0.087$ & $0.24\pm0.12$ \\
& $\mu(6^+_{1})$ & $+3.881$ & $3.589\pm0.004$ \\
& $Q(6^+_{1})$ & $+0.016$ & $+1.44\pm0.10$ \\
& $\mu(3^+_{1})$ & $+3.438$ & $3.79\pm0.06$ \\
& $B(M1;2^+_{1}\to1^+_1)$ & $0.014$ & $>0.033$ \\ 
& $B(M1;1^+_{2}\;{\text{or}}\;2^+_2\to1^+_1)$ & $0.27\;{\text{or}}\;0.066$ & $>0.0019$ \\ 
$^{116}$In
& $\mu(1^+_{1})$ & $+2.478$ & $2.7876\pm0.0006$ \\
& $Q(1^+_{1})$ & $+0.213$ & $0.11$ \\
& $\mu(5^+_{1})$ & $+0.177$ & $4.435\pm0.015$ \\
& $Q(5^+_{1})$ & $-0.813$ & $+0.802\pm0.012$ \\
& $B(M1;4^+_{1}\to5^+_1)$ & $0.0073$ & $>0.18$ \\
& $B(M1;2^+_{1}\to1^+_1)$ & $0.23$ & $>0.016$ \\
& $B(M1;4^+_{2}\;{\text{or}}\;5^+_2\to4^+_1)$ & $0.0060\;{\text{or}}\;0.062$ & $0.00013\pm0.00006$ \\
& $B(M1;4^+_{2}\;{\text{or}}\;5^+_2\to5^+_1)$ & $0.00019\;{\text{or}}\;0.0.028$ & $0.00013\pm0.00006$ \\
& $B(M1;3^+_{1}\to4^+_1)$ & $0.0090$ & $>0.0080$ \\
& $B(M1;3^+_{1}\to2^+_1)$ & $0.11$ & $>0.0066$ \\
$^{124}$Sb
& $\mu(2^+_{1})$ & $+0.849$ & $-0.3\pm0.2$ \\
& $Q(2^+_{1})$ & $-0.367$ & $0.0\pm0.2$ \\
& $B(E2;3^+_{1}\;{\text{or}}\;4^+_1\to5^+_1)$ & $3.48\;{\text{or}}\;0.34$ & $0.64\pm0.12$ \\
$^{130}$I
& $\mu(5^+_{1})$ & $+3.900$ & $3.349\pm0.007$ \\
& $B(M1;3^+_{2}\to3^+_1)$ & $0.0068$ & $>0.0017$ \\
& $B(M1;3^+_{2}\to2^+_1)$ & $0.0011$ & $>0.011$ \\
& $B(M1;1^+_{2}\;{\text{or}}\;2^+_2\to2^+_1)$ & $0.0022\;{\text{or}}\;0.022$ & $>0.0026$ \\
& $B(M1;1^+_{2}\;{\text{or}}\;2^+_2\to1^+_1)$ & $0.00045\;{\text{or}}\;0.00058$ & $>0.0046$ \\
& $B(M1;3^+_{3}\to2^+_2)$ & $0.0041$ & $>0.0095$ \\
& $B(M1;3^+_{3}\to3^+_1)$ & $0.0087$ & $>0.00011$ \\
& $B(M1;3^+_{3}\to2^+_1)$ & $0.0021$ & $>0.00051$ \\
& $B(M1;4^+_{2}\to3^+_1)$ & $0.054$ & $>0.0027$ \\
& $B(M1;4^+_{2}\to3^+_2)$ & $0.073$ & $>0.0019$ \\
& $B(M1;4^+_{2}\to4^+_1)$ & $0.0045$ & $>0.00050$ \\
$^{136}$Cs
& $\mu(5^+_{1})$ & $+3.570$ & $+3.711\pm0.005$ \\
& $Q(5^+_{1})$ & $+0.267$ & $+0.225\pm0.010$ \\
$^{198}$Au
& $\mu(5^+_{1})$ & $+6.078$ & $-1.11\pm0.02$ \\
  \end{tabular}
 \end{ruledtabular}
 \end{center}
\end{table*}

Figures~\ref{fig:ee1-em} and \ref{fig:ee2-em} show the 
calculated $B(E2)$ rates between low-lying states, 
$B(E2;2^+_1\to0^+_1)$, $B(E2;4^+_1\to2^+_1)$, 
$B(E2;0^+_2\to2^+_1)$, and $B(E2;2^+_2\to2^+_1)$, 
in comparison with the experimental data \cite{data}. 
The $B(E2;2^+_1\to0^+_1)$ and $B(E2;4^+_1\to2^+_1)$ 
values are generally calculated to be large, 
typically $>25$ W.u. Especially for open-shell nuclei such 
as $^{150}$Nd and $^{150}$Sm, the large $B(E2)$ values 
confirm the pronounced quadrupole collectivity. 
The calculated interband transitions $B(E2;0^+_2\to2^+_1)$ for 
both the initial and final nuclei are systematically small 
[Figs.~\ref{fig:ee1-em}(c) and \ref{fig:ee2-em}(c)]. 
This reflects the fact that the present IBM-2 calculation 
generally yields a more or less rotational energy spectrum 
[see Fig.~\ref{fig:ee}], in which this transition 
is rather weak. Another explanation is that the considered 
IBM-2 model space does not include the intruder excitation 
and the subsequent configuration mixing, which would 
have a strong influence on the nature of the $0^+_2$ 
state. Note that the experimental $B(E2;0^+_2\to2^+_1)$ 
values are quite large for some nuclei, e.g., 
$^{100}$Mo and $^{76}$Se. For the nuclei around these mass 
regions, competition between several intrinsic shapes 
is indeed suggested to play an important role in determining 
the low-lying states \cite{heyde2011,thomas2013}. 
In many of the even-even nuclei considered, the $2^+_2$ 
state is the bandhead of the $\gamma$-vibrational band, and thus 
the $2^+_2\to2^+_1$ $E2$ transition rates are a signature 
of $\gamma$ softness or triaxiality. 
The mapped IBM-2 accounts well for the observed systematics of the 
$B(E2;2^+_2\to2^+_1)$ rates of the initial nuclei 
[Fig.~\ref{fig:ee1-em}(d)]. 
The predicted $B(E2;2^+_2\to2^+_1)$ values 
for the final nuclei, especially $^{100}$Ru and $^{110}$Cd, 
are much more at variance with the data [Fig.~\ref{fig:ee2-em}(d)], 
probably because the triaxiality is not sufficiently taken 
into account for these nuclei. 

Table~\ref{tab:doo-em} lists the calculated and experimental 
$Q(I)$ and $\mu(I)$ moments, and the $B(E2)$ and $B(M1)$ transition 
strengths for the intermediate odd-odd nuclei. 
The calculated $Q(I)$ and $\mu(I)$ moments are, in most cases, in 
a reasonable agreement with the data. 
However, particularly for $Q(5^+_1)$ and $\mu(5^+_1)$ 
of $^{116}$In, $\mu(2^+_1)$ of $^{124}$Sb, and 
$\mu(5^+_1)$ of $^{198}$Au, 
the IBFFM-2 values differ considerably from the observed ones in 
both sign and magnitude. The deviation could be explained 
by the structure of the relevant IBFFM-2 functions, which also 
points to the deficiency of the considered model space. 
For $^{198}$Au, for instance, the $5^+_1$ ground state is mostly 
made of the $[\nu p_{3/2}\otimes\pi h_{11/2}]^{(J=7^+)}$ (34 \%)
and $[\nu f_{5/2}\otimes\pi h_{11/2}]^{(J=3^+)}$ (16 \%) 
pair configurations. 
On the other hand, empirical studies for the 
low-lying structure in this mass region suggested that the $5^+_1$ 
state is composed mainly of the 
$[\nu i_{13/2}\otimes\pi d_{3/2}]^{(J=5^+)}$ configuration, 
which gives rise to the correct sign of the $\mu(5^+_1)$ 
moment \cite{gao2012}, 
but which does not make any contribution in the present 
RHB+IBFFM-2 result. On the other hand, 
in the calculation based on the Gogny-D1M EDF \cite{nomura2019dodd} 
considered in Sec.~\ref{sec:comp-edf}, the same quantity 
is calculated to be $\mu(5^+_1)=-1.06$ $\mu_N$, which agrees 
with the experimental data, $-1.11\pm0.02$ $\mu_N$. 
The $5^+_1$ wave function in this case is decomposed into the 
$[\nu i_{13/2}\otimes\pi s_{1/2}]^{(J=7^+)}$ (27 \%), 
$[\nu i_{13/2}\otimes\pi d_{3/2}]^{(J=5^+)}$ (32 \%), and 
$[\nu i_{13/2}\otimes\pi d_{3/2}]^{(J=7^+)}$ (13 \%) 
pair components, and many other small ones. 
Such a difference in the structure the $5^+_1$ ground-state 
wave function could have arisen mainly from the SPEs used 
in the DD-PC1 and D1M EDFs (see Table~\ref{tab:comp-edf-spe}). 

For completeness, in Table~\ref{tab:doo-em} a comparison 
is made between the calculated and experimental \cite{data} 
$B(E2)$ and $B(M1)$ values of 
the odd-odd nuclei. In most cases, however, 
only the lower limit is known or the spin of the initial or final 
state of the transition is not firmly established, 
which makes it rather hard to make a meaningful comparison 
between theory and experiment. 

\bibliography{refs}

\begin{thebibliography}{86}%
\makeatletter
\providecommand \@ifxundefined [1]{%
 \@ifx{#1\undefined}
}%
\providecommand \@ifnum [1]{%
 \ifnum #1\expandafter \@firstoftwo
 \else \expandafter \@secondoftwo
 \fi
}%
\providecommand \@ifx [1]{%
 \ifx #1\expandafter \@firstoftwo
 \else \expandafter \@secondoftwo
 \fi
}%
\providecommand \natexlab [1]{#1}%
\providecommand \enquote  [1]{``#1''}%
\providecommand \bibnamefont  [1]{#1}%
\providecommand \bibfnamefont [1]{#1}%
\providecommand \citenamefont [1]{#1}%
\providecommand \href@noop [0]{\@secondoftwo}%
\providecommand \href [0]{\begingroup \@sanitize@url \@href}%
\providecommand \@href[1]{\@@startlink{#1}\@@href}%
\providecommand \@@href[1]{\endgroup#1\@@endlink}%
\providecommand \@sanitize@url [0]{\catcode `\\12\catcode `\$12\catcode
  `\&12\catcode `\#12\catcode `\^12\catcode `\_12\catcode `\%12\relax}%
\providecommand \@@startlink[1]{}%
\providecommand \@@endlink[0]{}%
\providecommand \url  [0]{\begingroup\@sanitize@url \@url }%
\providecommand \@url [1]{\endgroup\@href {#1}{\urlprefix }}%
\providecommand \urlprefix  [0]{URL }%
\providecommand \Eprint [0]{\href }%
\providecommand \doibase [0]{https://doi.org/}%
\providecommand \selectlanguage [0]{\@gobble}%
\providecommand \bibinfo  [0]{\@secondoftwo}%
\providecommand \bibfield  [0]{\@secondoftwo}%
\providecommand \translation [1]{[#1]}%
\providecommand \BibitemOpen [0]{}%
\providecommand \bibitemStop [0]{}%
\providecommand \bibitemNoStop [0]{.\EOS\space}%
\providecommand \EOS [0]{\spacefactor3000\relax}%
\providecommand \BibitemShut  [1]{\csname bibitem#1\endcsname}%
\let\auto@bib@innerbib\@empty
\bibitem [{\citenamefont {Avignone}\ \emph {et~al.}(2008)\citenamefont
  {Avignone}, \citenamefont {Elliott},\ and\ \citenamefont
  {Engel}}]{avignone2008}%
  \BibitemOpen
  \bibfield  {author} {\bibinfo {author} {\bibfnamefont {F.~T.}\ \bibnamefont
  {Avignone}}, \bibinfo {author} {\bibfnamefont {S.~R.}\ \bibnamefont
  {Elliott}},\ and\ \bibinfo {author} {\bibfnamefont {J.}~\bibnamefont
  {Engel}},\ }\href {https://doi.org/10.1103/RevModPhys.80.481} {\bibfield
  {journal} {\bibinfo  {journal} {Rev. Mod. Phys.}\ }\textbf {\bibinfo {volume}
  {80}},\ \bibinfo {pages} {481} (\bibinfo {year} {2008})}\BibitemShut
  {NoStop}%
\bibitem [{\citenamefont {Ejiri}\ \emph {et~al.}(2019)\citenamefont {Ejiri},
  \citenamefont {Suhonen},\ and\ \citenamefont {Zuber}}]{ejiri2019}%
  \BibitemOpen
  \bibfield  {author} {\bibinfo {author} {\bibfnamefont {H.}~\bibnamefont
  {Ejiri}}, \bibinfo {author} {\bibfnamefont {J.}~\bibnamefont {Suhonen}},\
  and\ \bibinfo {author} {\bibfnamefont {K.}~\bibnamefont {Zuber}},\ }\href
  {https://doi.org/https://doi.org/10.1016/j.physrep.2018.12.001} {\bibfield
  {journal} {\bibinfo  {journal} {Phys. Rep.}\ }\textbf {\bibinfo {volume}
  {797}},\ \bibinfo {pages} {1} (\bibinfo {year} {2019})}\BibitemShut {NoStop}%
\bibitem [{\citenamefont {Primakoff}\ and\ \citenamefont
  {Rosen}(1959)}]{primakoff1959}%
  \BibitemOpen
  \bibfield  {author} {\bibinfo {author} {\bibfnamefont {H.}~\bibnamefont
  {Primakoff}}\ and\ \bibinfo {author} {\bibfnamefont {S.~P.}\ \bibnamefont
  {Rosen}},\ }\href {https://doi.org/10.1088/0034-4885/22/1/305} {\bibfield
  {journal} {\bibinfo  {journal} {Rep. Prog. Phys.}\ }\textbf {\bibinfo
  {volume} {22}},\ \bibinfo {pages} {121} (\bibinfo {year} {1959})}\BibitemShut
  {NoStop}%
\bibitem [{\citenamefont {Haxton}\ and\ \citenamefont
  {Stephenson}(1984)}]{haxton1984}%
  \BibitemOpen
  \bibfield  {author} {\bibinfo {author} {\bibfnamefont {W.}~\bibnamefont
  {Haxton}}\ and\ \bibinfo {author} {\bibfnamefont {G.}~\bibnamefont
  {Stephenson}},\ }\href
  {https://doi.org/https://doi.org/10.1016/0146-6410(84)90006-1} {\bibfield
  {journal} {\bibinfo  {journal} {Prog. Part. Nucl. Phys.}\ }\textbf {\bibinfo
  {volume} {12}},\ \bibinfo {pages} {409} (\bibinfo {year} {1984})}\BibitemShut
  {NoStop}%
\bibitem [{\citenamefont {Doi}\ \emph {et~al.}(1985)\citenamefont {Doi},
  \citenamefont {Kotani},\ and\ \citenamefont {Takasugi}}]{doi1985}%
  \BibitemOpen
  \bibfield  {author} {\bibinfo {author} {\bibfnamefont {M.}~\bibnamefont
  {Doi}}, \bibinfo {author} {\bibfnamefont {T.}~\bibnamefont {Kotani}},\ and\
  \bibinfo {author} {\bibfnamefont {E.}~\bibnamefont {Takasugi}},\ }\href
  {https://doi.org/10.1143/PTPS.83.1} {\bibfield  {journal} {\bibinfo
  {journal} {Prog. Theor. Phys. Suppl.}\ }\textbf {\bibinfo {volume} {83}},\
  \bibinfo {pages} {1} (\bibinfo {year} {1985})}\BibitemShut {NoStop}%
\bibitem [{\citenamefont {Tomoda}(1991)}]{tomoda1991}%
  \BibitemOpen
  \bibfield  {author} {\bibinfo {author} {\bibfnamefont {T.}~\bibnamefont
  {Tomoda}},\ }\href {https://doi.org/10.1088/0034-4885/54/1/002} {\bibfield
  {journal} {\bibinfo  {journal} {Rep. Prog. Phys.}\ }\textbf {\bibinfo
  {volume} {54}},\ \bibinfo {pages} {53} (\bibinfo {year} {1991})}\BibitemShut
  {NoStop}%
\bibitem [{\citenamefont {Suhonen}\ and\ \citenamefont
  {Civitarese}(1998)}]{suhonen1998}%
  \BibitemOpen
  \bibfield  {author} {\bibinfo {author} {\bibfnamefont {J.}~\bibnamefont
  {Suhonen}}\ and\ \bibinfo {author} {\bibfnamefont {O.}~\bibnamefont
  {Civitarese}},\ }\href
  {https://doi.org/https://doi.org/10.1016/S0370-1573(97)00087-2} {\bibfield
  {journal} {\bibinfo  {journal} {Phys. Rep.}\ }\textbf {\bibinfo {volume}
  {300}},\ \bibinfo {pages} {123} (\bibinfo {year} {1998})}\BibitemShut
  {NoStop}%
\bibitem [{\citenamefont {Faessler}\ and\ \citenamefont
  {Simkovic}(1998)}]{faessler1998}%
  \BibitemOpen
  \bibfield  {author} {\bibinfo {author} {\bibfnamefont {A.}~\bibnamefont
  {Faessler}}\ and\ \bibinfo {author} {\bibfnamefont {F.}~\bibnamefont
  {Simkovic}},\ }\href {https://doi.org/10.1088/0954-3899/24/12/001} {\bibfield
   {journal} {\bibinfo  {journal} {J. Phys. G: Nucl. Part. Phys.}\ }\textbf
  {\bibinfo {volume} {24}},\ \bibinfo {pages} {2139} (\bibinfo {year}
  {1998})}\BibitemShut {NoStop}%
\bibitem [{\citenamefont {Vogel}(2012)}]{vogel2012}%
  \BibitemOpen
  \bibfield  {author} {\bibinfo {author} {\bibfnamefont {P.}~\bibnamefont
  {Vogel}},\ }\href {https://doi.org/10.1088/0954-3899/39/12/124002} {\bibfield
   {journal} {\bibinfo  {journal} {J. Phys. G: Nucl. Part. Phys.}\ }\textbf
  {\bibinfo {volume} {39}},\ \bibinfo {pages} {124002} (\bibinfo {year}
  {2012})}\BibitemShut {NoStop}%
\bibitem [{\citenamefont {Vergados}\ \emph {et~al.}(2012)\citenamefont
  {Vergados}, \citenamefont {Ejiri},\ and\ \citenamefont
  {{\v{S}}imkovic}}]{vergados2012}%
  \BibitemOpen
  \bibfield  {author} {\bibinfo {author} {\bibfnamefont {J.~D.}\ \bibnamefont
  {Vergados}}, \bibinfo {author} {\bibfnamefont {H.}~\bibnamefont {Ejiri}},\
  and\ \bibinfo {author} {\bibfnamefont {F.}~\bibnamefont {{\v{S}}imkovic}},\
  }\href {https://doi.org/10.1088/0034-4885/75/10/106301} {\bibfield  {journal}
  {\bibinfo  {journal} {Rep. Prog. Phys.}\ }\textbf {\bibinfo {volume} {75}},\
  \bibinfo {pages} {106301} (\bibinfo {year} {2012})}\BibitemShut {NoStop}%
\bibitem [{\citenamefont {Engel}\ and\ \citenamefont
  {Men{\'{e}}ndez}(2017)}]{engel2017}%
  \BibitemOpen
  \bibfield  {author} {\bibinfo {author} {\bibfnamefont {J.}~\bibnamefont
  {Engel}}\ and\ \bibinfo {author} {\bibfnamefont {J.}~\bibnamefont
  {Men{\'{e}}ndez}},\ }\href {https://doi.org/10.1088/1361-6633/aa5bc5}
  {\bibfield  {journal} {\bibinfo  {journal} {Rep. Prog. Phys.}\ }\textbf
  {\bibinfo {volume} {80}},\ \bibinfo {pages} {046301} (\bibinfo {year}
  {2017})}\BibitemShut {NoStop}%
\bibitem [{\citenamefont {Barabash}(2020)}]{barabash2020}%
  \BibitemOpen
  \bibfield  {author} {\bibinfo {author} {\bibfnamefont {A.}~\bibnamefont
  {Barabash}},\ }\href {https://www.mdpi.com/2218-1997/6/10/159} {\bibfield
  {journal} {\bibinfo  {journal} {Universe}\ }\textbf {\bibinfo {volume} {6}}
  (\bibinfo {year} {2020})}\BibitemShut {NoStop}%
\bibitem [{\citenamefont {Pirinen}\ and\ \citenamefont
  {Suhonen}(2015)}]{pirinen2015}%
  \BibitemOpen
  \bibfield  {author} {\bibinfo {author} {\bibfnamefont {P.}~\bibnamefont
  {Pirinen}}\ and\ \bibinfo {author} {\bibfnamefont {J.}~\bibnamefont
  {Suhonen}},\ }\href {https://doi.org/10.1103/PhysRevC.91.054309} {\bibfield
  {journal} {\bibinfo  {journal} {Phys. Rev. C}\ }\textbf {\bibinfo {volume}
  {91}},\ \bibinfo {pages} {054309} (\bibinfo {year} {2015})}\BibitemShut
  {NoStop}%
\bibitem [{\citenamefont {\ifmmode~\check{S}\else \v{S}\fi{}imkovic}\ \emph
  {et~al.}(2018)\citenamefont {\ifmmode~\check{S}\else \v{S}\fi{}imkovic},
  \citenamefont {Smetana},\ and\ \citenamefont {Vogel}}]{simkovic2018}%
  \BibitemOpen
  \bibfield  {author} {\bibinfo {author} {\bibfnamefont {F.}~\bibnamefont
  {\ifmmode~\check{S}\else \v{S}\fi{}imkovic}}, \bibinfo {author}
  {\bibfnamefont {A.}~\bibnamefont {Smetana}},\ and\ \bibinfo {author}
  {\bibfnamefont {P.}~\bibnamefont {Vogel}},\ }\href
  {https://doi.org/10.1103/PhysRevC.98.064325} {\bibfield  {journal} {\bibinfo
  {journal} {Phys. Rev. C}\ }\textbf {\bibinfo {volume} {98}},\ \bibinfo
  {pages} {064325} (\bibinfo {year} {2018})}\BibitemShut {NoStop}%
\bibitem [{\citenamefont {Caurier}\ \emph {et~al.}(2007)\citenamefont
  {Caurier}, \citenamefont {Nowacki},\ and\ \citenamefont
  {Poves}}]{caurier2007}%
  \BibitemOpen
  \bibfield  {author} {\bibinfo {author} {\bibfnamefont {E.}~\bibnamefont
  {Caurier}}, \bibinfo {author} {\bibfnamefont {F.}~\bibnamefont {Nowacki}},\
  and\ \bibinfo {author} {\bibfnamefont {A.}~\bibnamefont {Poves}},\ }\href
  {https://doi.org/10.1142/S0218301307005983} {\bibfield  {journal} {\bibinfo
  {journal} {Int. J. Mod. Phys. E}\ }\textbf {\bibinfo {volume} {16}},\
  \bibinfo {pages} {552} (\bibinfo {year} {2007})}\BibitemShut {NoStop}%
\bibitem [{\citenamefont {Yoshinaga}\ \emph {et~al.}(2018)\citenamefont
  {Yoshinaga}, \citenamefont {Yanase}, \citenamefont {Higashiyama},
  \citenamefont {Teruya},\ and\ \citenamefont {Taguchi}}]{yoshinaga2018}%
  \BibitemOpen
  \bibfield  {author} {\bibinfo {author} {\bibfnamefont {N.}~\bibnamefont
  {Yoshinaga}}, \bibinfo {author} {\bibfnamefont {K.}~\bibnamefont {Yanase}},
  \bibinfo {author} {\bibfnamefont {K.}~\bibnamefont {Higashiyama}}, \bibinfo
  {author} {\bibfnamefont {E.}~\bibnamefont {Teruya}},\ and\ \bibinfo {author}
  {\bibfnamefont {D.}~\bibnamefont {Taguchi}},\ }\href
  {https://doi.org/10.1093/ptep/ptx174} {\bibfield  {journal} {\bibinfo
  {journal} {Prog. Theor. Exp. Phys.}\ }\textbf {\bibinfo {volume} {2018}},\
  \bibinfo {pages} {023D02} (\bibinfo {year} {2018})}\BibitemShut {NoStop}%
\bibitem [{\citenamefont {Caurier}\ \emph {et~al.}(1990)\citenamefont
  {Caurier}, \citenamefont {Poves},\ and\ \citenamefont {Zuker}}]{caurier1990}%
  \BibitemOpen
  \bibfield  {author} {\bibinfo {author} {\bibfnamefont {E.}~\bibnamefont
  {Caurier}}, \bibinfo {author} {\bibfnamefont {A.}~\bibnamefont {Poves}},\
  and\ \bibinfo {author} {\bibfnamefont {A.}~\bibnamefont {Zuker}},\ }\href
  {https://doi.org/https://doi.org/10.1016/0370-2693(90)91071-I} {\bibfield
  {journal} {\bibinfo  {journal} {Phys. Lett. B}\ }\textbf {\bibinfo {volume}
  {252}},\ \bibinfo {pages} {13} (\bibinfo {year} {1990})}\BibitemShut
  {NoStop}%
\bibitem [{\citenamefont {Caurier}\ \emph {et~al.}(2012)\citenamefont
  {Caurier}, \citenamefont {Nowacki},\ and\ \citenamefont
  {Poves}}]{caurier2012}%
  \BibitemOpen
  \bibfield  {author} {\bibinfo {author} {\bibfnamefont {E.}~\bibnamefont
  {Caurier}}, \bibinfo {author} {\bibfnamefont {F.}~\bibnamefont {Nowacki}},\
  and\ \bibinfo {author} {\bibfnamefont {A.}~\bibnamefont {Poves}},\ }\href
  {https://doi.org/https://doi.org/10.1016/j.physletb.2012.03.076} {\bibfield
  {journal} {\bibinfo  {journal} {Phys. Lett. B}\ }\textbf {\bibinfo {volume}
  {711}},\ \bibinfo {pages} {62} (\bibinfo {year} {2012})}\BibitemShut
  {NoStop}%
\bibitem [{\citenamefont {Sen'kov}\ and\ \citenamefont
  {Horoi}(2016)}]{senkov2016}%
  \BibitemOpen
  \bibfield  {author} {\bibinfo {author} {\bibfnamefont {R.~A.}\ \bibnamefont
  {Sen'kov}}\ and\ \bibinfo {author} {\bibfnamefont {M.}~\bibnamefont
  {Horoi}},\ }\href {https://doi.org/10.1103/PhysRevC.93.044334} {\bibfield
  {journal} {\bibinfo  {journal} {Phys. Rev. C}\ }\textbf {\bibinfo {volume}
  {93}},\ \bibinfo {pages} {044334} (\bibinfo {year} {2016})}\BibitemShut
  {NoStop}%
\bibitem [{\citenamefont {Coraggio}\ \emph {et~al.}(2019)\citenamefont
  {Coraggio}, \citenamefont {De~Angelis}, \citenamefont {Fukui}, \citenamefont
  {Gargano}, \citenamefont {Itaco},\ and\ \citenamefont
  {Nowacki}}]{coraggio2019}%
  \BibitemOpen
  \bibfield  {author} {\bibinfo {author} {\bibfnamefont {L.}~\bibnamefont
  {Coraggio}}, \bibinfo {author} {\bibfnamefont {L.}~\bibnamefont
  {De~Angelis}}, \bibinfo {author} {\bibfnamefont {T.}~\bibnamefont {Fukui}},
  \bibinfo {author} {\bibfnamefont {A.}~\bibnamefont {Gargano}}, \bibinfo
  {author} {\bibfnamefont {N.}~\bibnamefont {Itaco}},\ and\ \bibinfo {author}
  {\bibfnamefont {F.}~\bibnamefont {Nowacki}},\ }\href
  {https://doi.org/10.1103/PhysRevC.100.014316} {\bibfield  {journal} {\bibinfo
   {journal} {Phys. Rev. C}\ }\textbf {\bibinfo {volume} {100}},\ \bibinfo
  {pages} {014316} (\bibinfo {year} {2019})}\BibitemShut {NoStop}%
\bibitem [{\citenamefont {Yoshida}\ and\ \citenamefont
  {Iachello}(2013)}]{yoshida2013}%
  \BibitemOpen
  \bibfield  {author} {\bibinfo {author} {\bibfnamefont {N.}~\bibnamefont
  {Yoshida}}\ and\ \bibinfo {author} {\bibfnamefont {F.}~\bibnamefont
  {Iachello}},\ }\href {https://doi.org/10.1093/ptep/ptt007} {\bibfield
  {journal} {\bibinfo  {journal} {Prog. Theor. Exp. Phys.}\ }\textbf {\bibinfo
  {volume} {2013}},\ \bibinfo {pages} {043D01} (\bibinfo {year}
  {2013})}\BibitemShut {NoStop}%
\bibitem [{\citenamefont {Ring}\ and\ \citenamefont {Schuck}(1980)}]{RS}%
  \BibitemOpen
  \bibfield  {author} {\bibinfo {author} {\bibfnamefont {P.}~\bibnamefont
  {Ring}}\ and\ \bibinfo {author} {\bibfnamefont {P.}~\bibnamefont {Schuck}},\
  }\href@noop {} {\emph {\bibinfo {title} {The nuclear many-body problem}}}\
  (\bibinfo  {publisher} {Springer, Berlin},\ \bibinfo {year}
  {1980})\BibitemShut {NoStop}%
\bibitem [{\citenamefont {Bender}\ \emph {et~al.}(2003)\citenamefont {Bender},
  \citenamefont {Heenen},\ and\ \citenamefont {Reinhard}}]{bender2003}%
  \BibitemOpen
  \bibfield  {author} {\bibinfo {author} {\bibfnamefont {M.}~\bibnamefont
  {Bender}}, \bibinfo {author} {\bibfnamefont {P.-H.}\ \bibnamefont {Heenen}},\
  and\ \bibinfo {author} {\bibfnamefont {P.-G.}\ \bibnamefont {Reinhard}},\
  }\href {https://doi.org/10.1103/RevModPhys.75.121} {\bibfield  {journal}
  {\bibinfo  {journal} {Rev. Mod. Phys.}\ }\textbf {\bibinfo {volume} {75}},\
  \bibinfo {pages} {121} (\bibinfo {year} {2003})}\BibitemShut {NoStop}%
\bibitem [{\citenamefont {Vretenar}\ \emph {et~al.}(2005)\citenamefont
  {Vretenar}, \citenamefont {Afanasjev}, \citenamefont {Lalazissis},\ and\
  \citenamefont {Ring}}]{vretenar2005}%
  \BibitemOpen
  \bibfield  {author} {\bibinfo {author} {\bibfnamefont {D.}~\bibnamefont
  {Vretenar}}, \bibinfo {author} {\bibfnamefont {A.~V.}\ \bibnamefont
  {Afanasjev}}, \bibinfo {author} {\bibfnamefont {G.~A.}\ \bibnamefont
  {Lalazissis}},\ and\ \bibinfo {author} {\bibfnamefont {P.}~\bibnamefont
  {Ring}},\ }\href {https://doi.org/10.1016/j.physrep.2004.10.001} {\bibfield
  {journal} {\bibinfo  {journal} {Phys. Rep.}\ }\textbf {\bibinfo {volume}
  {409}},\ \bibinfo {pages} {101 } (\bibinfo {year} {2005})}\BibitemShut
  {NoStop}%
\bibitem [{\citenamefont {Nik\ifmmode \check{s}\else
  \v{s}\fi{}i\ifmmode~\acute{c}\else \'{c}\fi{}}\ \emph
  {et~al.}(2011)\citenamefont {Nik\ifmmode \check{s}\else
  \v{s}\fi{}i\ifmmode~\acute{c}\else \'{c}\fi{}}, \citenamefont {Vretenar},\
  and\ \citenamefont {Ring}}]{niksic2011}%
  \BibitemOpen
  \bibfield  {author} {\bibinfo {author} {\bibfnamefont {T.}~\bibnamefont
  {Nik\ifmmode \check{s}\else \v{s}\fi{}i\ifmmode~\acute{c}\else \'{c}\fi{}}},
  \bibinfo {author} {\bibfnamefont {D.}~\bibnamefont {Vretenar}},\ and\
  \bibinfo {author} {\bibfnamefont {P.}~\bibnamefont {Ring}},\ }\href
  {https://doi.org/10.1016/j.ppnp.2011.01.055} {\bibfield  {journal} {\bibinfo
  {journal} {Prog. Part. Nucl. Phys.}\ }\textbf {\bibinfo {volume} {66}},\
  \bibinfo {pages} {519} (\bibinfo {year} {2011})}\BibitemShut {NoStop}%
\bibitem [{\citenamefont {Robledo}\ \emph {et~al.}(2019)\citenamefont
  {Robledo}, \citenamefont {Rodríguez},\ and\ \citenamefont
  {Rodríguez-Guzmán}}]{robledo2019}%
  \BibitemOpen
  \bibfield  {author} {\bibinfo {author} {\bibfnamefont {L.~M.}\ \bibnamefont
  {Robledo}}, \bibinfo {author} {\bibfnamefont {T.~R.}\ \bibnamefont
  {Rodríguez}},\ and\ \bibinfo {author} {\bibfnamefont {R.~R.}\ \bibnamefont
  {Rodríguez-Guzmán}},\ }\href
  {http://stacks.iop.org/0954-3899/46/i=1/a=013001} {\bibfield  {journal}
  {\bibinfo  {journal} {J. Phys. G: Nucl. Part. Phys.}\ }\textbf {\bibinfo
  {volume} {46}},\ \bibinfo {pages} {013001} (\bibinfo {year}
  {2019})}\BibitemShut {NoStop}%
\bibitem [{\citenamefont {Nomura}\ \emph {et~al.}(2008)\citenamefont {Nomura},
  \citenamefont {Shimizu},\ and\ \citenamefont {Otsuka}}]{nomura2008}%
  \BibitemOpen
  \bibfield  {author} {\bibinfo {author} {\bibfnamefont {K.}~\bibnamefont
  {Nomura}}, \bibinfo {author} {\bibfnamefont {N.}~\bibnamefont {Shimizu}},\
  and\ \bibinfo {author} {\bibfnamefont {T.}~\bibnamefont {Otsuka}},\ }\href
  {https://doi.org/10.1103/PhysRevLett.101.142501} {\bibfield  {journal}
  {\bibinfo  {journal} {Phys. Rev. Lett.}\ }\textbf {\bibinfo {volume} {101}},\
  \bibinfo {pages} {142501} (\bibinfo {year} {2008})}\BibitemShut {NoStop}%
\bibitem [{\citenamefont {Brant}\ \emph {et~al.}(1984)\citenamefont {Brant},
  \citenamefont {Paar},\ and\ \citenamefont {Vretenar}}]{brant1984}%
  \BibitemOpen
  \bibfield  {author} {\bibinfo {author} {\bibfnamefont {S.}~\bibnamefont
  {Brant}}, \bibinfo {author} {\bibfnamefont {V.}~\bibnamefont {Paar}},\ and\
  \bibinfo {author} {\bibfnamefont {D.}~\bibnamefont {Vretenar}},\ }\href
  {https://doi.org/10.1007/BF01412551} {\bibfield  {journal} {\bibinfo
  {journal} {Z. Phys. A}\ }\textbf {\bibinfo {volume} {319}},\ \bibinfo {pages}
  {355} (\bibinfo {year} {1984})}\BibitemShut {NoStop}%
\bibitem [{\citenamefont {Iachello}\ and\ \citenamefont {{Van
  Isacker}}(1991)}]{IBFM}%
  \BibitemOpen
  \bibfield  {author} {\bibinfo {author} {\bibfnamefont {F.}~\bibnamefont
  {Iachello}}\ and\ \bibinfo {author} {\bibfnamefont {P.}~\bibnamefont {{Van
  Isacker}}},\ }\href@noop {} {\emph {\bibinfo {title} {The interacting
  boson-fermion model}}}\ (\bibinfo  {publisher} {Cambridge University Press,
  Cambridge},\ \bibinfo {year} {1991})\BibitemShut {NoStop}%
\bibitem [{\citenamefont {Nomura}\ \emph
  {et~al.}(2016{\natexlab{a}})\citenamefont {Nomura}, \citenamefont
  {Nik\ifmmode \check{s}\else \v{s}\fi{}i\ifmmode~\acute{c}\else \'{c}\fi{}},\
  and\ \citenamefont {Vretenar}}]{nomura2016odd}%
  \BibitemOpen
  \bibfield  {author} {\bibinfo {author} {\bibfnamefont {K.}~\bibnamefont
  {Nomura}}, \bibinfo {author} {\bibfnamefont {T.}~\bibnamefont {Nik\ifmmode
  \check{s}\else \v{s}\fi{}i\ifmmode~\acute{c}\else \'{c}\fi{}}},\ and\
  \bibinfo {author} {\bibfnamefont {D.}~\bibnamefont {Vretenar}},\ }\href
  {https://doi.org/10.1103/PhysRevC.93.054305} {\bibfield  {journal} {\bibinfo
  {journal} {Phys. Rev. C}\ }\textbf {\bibinfo {volume} {93}},\ \bibinfo
  {pages} {054305} (\bibinfo {year} {2016}{\natexlab{a}})}\BibitemShut
  {NoStop}%
\bibitem [{\citenamefont {Nomura}\ \emph {et~al.}(2019)\citenamefont {Nomura},
  \citenamefont {Rodr\'{\i}guez-Guzm\'an},\ and\ \citenamefont
  {Robledo}}]{nomura2019dodd}%
  \BibitemOpen
  \bibfield  {author} {\bibinfo {author} {\bibfnamefont {K.}~\bibnamefont
  {Nomura}}, \bibinfo {author} {\bibfnamefont {R.}~\bibnamefont
  {Rodr\'{\i}guez-Guzm\'an}},\ and\ \bibinfo {author} {\bibfnamefont {L.~M.}\
  \bibnamefont {Robledo}},\ }\href {https://doi.org/10.1103/PhysRevC.99.034308}
  {\bibfield  {journal} {\bibinfo  {journal} {Phys. Rev. C}\ }\textbf {\bibinfo
  {volume} {99}},\ \bibinfo {pages} {034308} (\bibinfo {year}
  {2019})}\BibitemShut {NoStop}%
\bibitem [{\citenamefont {Nomura}\ \emph
  {et~al.}(2012{\natexlab{a}})\citenamefont {Nomura}, \citenamefont
  {Rodr\'{\i}guez-Guzm\'an}, \citenamefont {Robledo},\ and\ \citenamefont
  {Shimizu}}]{nomura2012sc}%
  \BibitemOpen
  \bibfield  {author} {\bibinfo {author} {\bibfnamefont {K.}~\bibnamefont
  {Nomura}}, \bibinfo {author} {\bibfnamefont {R.}~\bibnamefont
  {Rodr\'{\i}guez-Guzm\'an}}, \bibinfo {author} {\bibfnamefont {L.~M.}\
  \bibnamefont {Robledo}},\ and\ \bibinfo {author} {\bibfnamefont
  {N.}~\bibnamefont {Shimizu}},\ }\href
  {https://doi.org/10.1103/PhysRevC.86.034322} {\bibfield  {journal} {\bibinfo
  {journal} {Phys. Rev. C}\ }\textbf {\bibinfo {volume} {86}},\ \bibinfo
  {pages} {034322} (\bibinfo {year} {2012}{\natexlab{a}})}\BibitemShut
  {NoStop}%
\bibitem [{\citenamefont {Nomura}\ \emph
  {et~al.}(2016{\natexlab{b}})\citenamefont {Nomura}, \citenamefont {Otsuka},\
  and\ \citenamefont {{Van Isacker}}}]{nomura2016sc}%
  \BibitemOpen
  \bibfield  {author} {\bibinfo {author} {\bibfnamefont {K.}~\bibnamefont
  {Nomura}}, \bibinfo {author} {\bibfnamefont {T.}~\bibnamefont {Otsuka}},\
  and\ \bibinfo {author} {\bibfnamefont {P.}~\bibnamefont {{Van Isacker}}},\
  }\href {https://doi.org/10.1088/0954-3899/43/2/024008} {\bibfield  {journal}
  {\bibinfo  {journal} {J. Phys. G: Nucl. Part. Phys.}\ }\textbf {\bibinfo
  {volume} {43}},\ \bibinfo {pages} {024008} (\bibinfo {year}
  {2016}{\natexlab{b}})}\BibitemShut {NoStop}%
\bibitem [{\citenamefont {Nomura}\ \emph
  {et~al.}(2016{\natexlab{c}})\citenamefont {Nomura}, \citenamefont
  {Rodr\'{\i}guez-Guzm\'an},\ and\ \citenamefont {Robledo}}]{nomura2016zr}%
  \BibitemOpen
  \bibfield  {author} {\bibinfo {author} {\bibfnamefont {K.}~\bibnamefont
  {Nomura}}, \bibinfo {author} {\bibfnamefont {R.}~\bibnamefont
  {Rodr\'{\i}guez-Guzm\'an}},\ and\ \bibinfo {author} {\bibfnamefont {L.~M.}\
  \bibnamefont {Robledo}},\ }\href {https://doi.org/10.1103/PhysRevC.94.044314}
  {\bibfield  {journal} {\bibinfo  {journal} {Phys. Rev. C}\ }\textbf {\bibinfo
  {volume} {94}},\ \bibinfo {pages} {044314} (\bibinfo {year}
  {2016}{\natexlab{c}})}\BibitemShut {NoStop}%
\bibitem [{\citenamefont {Nomura}\ \emph {et~al.}(2013)\citenamefont {Nomura},
  \citenamefont {Vretenar},\ and\ \citenamefont {Lu}}]{nomura2013oct}%
  \BibitemOpen
  \bibfield  {author} {\bibinfo {author} {\bibfnamefont {K.}~\bibnamefont
  {Nomura}}, \bibinfo {author} {\bibfnamefont {D.}~\bibnamefont {Vretenar}},\
  and\ \bibinfo {author} {\bibfnamefont {B.-N.}\ \bibnamefont {Lu}},\ }\href
  {https://doi.org/10.1103/PhysRevC.88.021303} {\bibfield  {journal} {\bibinfo
  {journal} {Phys. Rev. C}\ }\textbf {\bibinfo {volume} {88}},\ \bibinfo
  {pages} {021303} (\bibinfo {year} {2013})}\BibitemShut {NoStop}%
\bibitem [{\citenamefont {Nomura}\ \emph {et~al.}(2014)\citenamefont {Nomura},
  \citenamefont {Vretenar}, \citenamefont {Nik\ifmmode \check{s}\else
  \v{s}\fi{}i\ifmmode~\acute{c}\else \'{c}\fi{}},\ and\ \citenamefont
  {Lu}}]{nomura2014}%
  \BibitemOpen
  \bibfield  {author} {\bibinfo {author} {\bibfnamefont {K.}~\bibnamefont
  {Nomura}}, \bibinfo {author} {\bibfnamefont {D.}~\bibnamefont {Vretenar}},
  \bibinfo {author} {\bibfnamefont {T.}~\bibnamefont {Nik\ifmmode
  \check{s}\else \v{s}\fi{}i\ifmmode~\acute{c}\else \'{c}\fi{}}},\ and\
  \bibinfo {author} {\bibfnamefont {B.-N.}\ \bibnamefont {Lu}},\ }\href
  {https://doi.org/10.1103/PhysRevC.89.024312} {\bibfield  {journal} {\bibinfo
  {journal} {Phys. Rev. C}\ }\textbf {\bibinfo {volume} {89}},\ \bibinfo
  {pages} {024312} (\bibinfo {year} {2014})}\BibitemShut {NoStop}%
\bibitem [{\citenamefont {Nomura}\ \emph
  {et~al.}(2016{\natexlab{d}})\citenamefont {Nomura}, \citenamefont
  {Nik\ifmmode \check{s}\else \v{s}\fi{}i\ifmmode~\acute{c}\else \'{c}\fi{}},\
  and\ \citenamefont {Vretenar}}]{nomura2016qpt}%
  \BibitemOpen
  \bibfield  {author} {\bibinfo {author} {\bibfnamefont {K.}~\bibnamefont
  {Nomura}}, \bibinfo {author} {\bibfnamefont {T.}~\bibnamefont {Nik\ifmmode
  \check{s}\else \v{s}\fi{}i\ifmmode~\acute{c}\else \'{c}\fi{}}},\ and\
  \bibinfo {author} {\bibfnamefont {D.}~\bibnamefont {Vretenar}},\ }\href
  {https://doi.org/10.1103/PhysRevC.94.064310} {\bibfield  {journal} {\bibinfo
  {journal} {Phys. Rev. C}\ }\textbf {\bibinfo {volume} {94}},\ \bibinfo
  {pages} {064310} (\bibinfo {year} {2016}{\natexlab{d}})}\BibitemShut
  {NoStop}%
\bibitem [{\citenamefont {Nomura}\ \emph {et~al.}(2017)\citenamefont {Nomura},
  \citenamefont {Nik\ifmmode \check{s}\else \v{s}\fi{}i\ifmmode~\acute{c}\else
  \'{c}\fi{}},\ and\ \citenamefont {Vretenar}}]{nomura2017odd-1}%
  \BibitemOpen
  \bibfield  {author} {\bibinfo {author} {\bibfnamefont {K.}~\bibnamefont
  {Nomura}}, \bibinfo {author} {\bibfnamefont {T.}~\bibnamefont {Nik\ifmmode
  \check{s}\else \v{s}\fi{}i\ifmmode~\acute{c}\else \'{c}\fi{}}},\ and\
  \bibinfo {author} {\bibfnamefont {D.}~\bibnamefont {Vretenar}},\ }\href
  {https://doi.org/10.1103/PhysRevC.96.014304} {\bibfield  {journal} {\bibinfo
  {journal} {Phys. Rev. C}\ }\textbf {\bibinfo {volume} {96}},\ \bibinfo
  {pages} {014304} (\bibinfo {year} {2017})}\BibitemShut {NoStop}%
\bibitem [{\citenamefont {Nomura}\ \emph
  {et~al.}(2020{\natexlab{a}})\citenamefont {Nomura}, \citenamefont
  {Rodr\'{\i}guez-Guzm\'an},\ and\ \citenamefont {Robledo}}]{nomura2020beta-1}%
  \BibitemOpen
  \bibfield  {author} {\bibinfo {author} {\bibfnamefont {K.}~\bibnamefont
  {Nomura}}, \bibinfo {author} {\bibfnamefont {R.}~\bibnamefont
  {Rodr\'{\i}guez-Guzm\'an}},\ and\ \bibinfo {author} {\bibfnamefont {L.~M.}\
  \bibnamefont {Robledo}},\ }\href
  {https://doi.org/10.1103/PhysRevC.101.024311} {\bibfield  {journal} {\bibinfo
   {journal} {Phys. Rev. C}\ }\textbf {\bibinfo {volume} {101}},\ \bibinfo
  {pages} {024311} (\bibinfo {year} {2020}{\natexlab{a}})}\BibitemShut
  {NoStop}%
\bibitem [{\citenamefont {Nomura}\ \emph
  {et~al.}(2020{\natexlab{b}})\citenamefont {Nomura}, \citenamefont
  {Rodr\'{\i}guez-Guzm\'an},\ and\ \citenamefont {Robledo}}]{nomura2020beta-2}%
  \BibitemOpen
  \bibfield  {author} {\bibinfo {author} {\bibfnamefont {K.}~\bibnamefont
  {Nomura}}, \bibinfo {author} {\bibfnamefont {R.}~\bibnamefont
  {Rodr\'{\i}guez-Guzm\'an}},\ and\ \bibinfo {author} {\bibfnamefont {L.~M.}\
  \bibnamefont {Robledo}},\ }\href
  {https://doi.org/10.1103/PhysRevC.101.044318} {\bibfield  {journal} {\bibinfo
   {journal} {Phys. Rev. C}\ }\textbf {\bibinfo {volume} {101}},\ \bibinfo
  {pages} {044318} (\bibinfo {year} {2020}{\natexlab{b}})}\BibitemShut
  {NoStop}%
\bibitem [{\citenamefont {Barea}\ and\ \citenamefont
  {Iachello}(2009)}]{barea2009}%
  \BibitemOpen
  \bibfield  {author} {\bibinfo {author} {\bibfnamefont {J.}~\bibnamefont
  {Barea}}\ and\ \bibinfo {author} {\bibfnamefont {F.}~\bibnamefont
  {Iachello}},\ }\href {https://doi.org/10.1103/PhysRevC.79.044301} {\bibfield
  {journal} {\bibinfo  {journal} {Phys. Rev. C}\ }\textbf {\bibinfo {volume}
  {79}},\ \bibinfo {pages} {044301} (\bibinfo {year} {2009})}\BibitemShut
  {NoStop}%
\bibitem [{\citenamefont {Barea}\ \emph {et~al.}(2013)\citenamefont {Barea},
  \citenamefont {Kotila},\ and\ \citenamefont {Iachello}}]{barea2013}%
  \BibitemOpen
  \bibfield  {author} {\bibinfo {author} {\bibfnamefont {J.}~\bibnamefont
  {Barea}}, \bibinfo {author} {\bibfnamefont {J.}~\bibnamefont {Kotila}},\ and\
  \bibinfo {author} {\bibfnamefont {F.}~\bibnamefont {Iachello}},\ }\href
  {https://doi.org/10.1103/PhysRevC.87.014315} {\bibfield  {journal} {\bibinfo
  {journal} {Phys. Rev. C}\ }\textbf {\bibinfo {volume} {87}},\ \bibinfo
  {pages} {014315} (\bibinfo {year} {2013})}\BibitemShut {NoStop}%
\bibitem [{\citenamefont {Barea}\ \emph
  {et~al.}(2015{\natexlab{a}})\citenamefont {Barea}, \citenamefont {Kotila},\
  and\ \citenamefont {Iachello}}]{barea2015}%
  \BibitemOpen
  \bibfield  {author} {\bibinfo {author} {\bibfnamefont {J.}~\bibnamefont
  {Barea}}, \bibinfo {author} {\bibfnamefont {J.}~\bibnamefont {Kotila}},\ and\
  \bibinfo {author} {\bibfnamefont {F.}~\bibnamefont {Iachello}},\ }\href
  {https://doi.org/10.1103/PhysRevC.91.034304} {\bibfield  {journal} {\bibinfo
  {journal} {Phys. Rev. C}\ }\textbf {\bibinfo {volume} {91}},\ \bibinfo
  {pages} {034304} (\bibinfo {year} {2015}{\natexlab{a}})}\BibitemShut
  {NoStop}%
\bibitem [{\citenamefont {Barea}\ \emph
  {et~al.}(2015{\natexlab{b}})\citenamefont {Barea}, \citenamefont {Kotila},\
  and\ \citenamefont {Iachello}}]{barea2015b}%
  \BibitemOpen
  \bibfield  {author} {\bibinfo {author} {\bibfnamefont {J.}~\bibnamefont
  {Barea}}, \bibinfo {author} {\bibfnamefont {J.}~\bibnamefont {Kotila}},\ and\
  \bibinfo {author} {\bibfnamefont {F.}~\bibnamefont {Iachello}},\ }\href
  {https://doi.org/10.1103/PhysRevD.92.093001} {\bibfield  {journal} {\bibinfo
  {journal} {Phys. Rev. D}\ }\textbf {\bibinfo {volume} {92}},\ \bibinfo
  {pages} {093001} (\bibinfo {year} {2015}{\natexlab{b}})}\BibitemShut
  {NoStop}%
\bibitem [{\citenamefont {Deppisch}\ \emph {et~al.}(2020)\citenamefont
  {Deppisch}, \citenamefont {Graf}, \citenamefont {Iachello},\ and\
  \citenamefont {Kotila}}]{deppisch2020}%
  \BibitemOpen
  \bibfield  {author} {\bibinfo {author} {\bibfnamefont {F.~F.}\ \bibnamefont
  {Deppisch}}, \bibinfo {author} {\bibfnamefont {L.}~\bibnamefont {Graf}},
  \bibinfo {author} {\bibfnamefont {F.}~\bibnamefont {Iachello}},\ and\
  \bibinfo {author} {\bibfnamefont {J.}~\bibnamefont {Kotila}},\ }\href
  {https://doi.org/10.1103/PhysRevD.102.095016} {\bibfield  {journal} {\bibinfo
   {journal} {Phys. Rev. D}\ }\textbf {\bibinfo {volume} {102}},\ \bibinfo
  {pages} {095016} (\bibinfo {year} {2020})}\BibitemShut {NoStop}%
\bibitem [{\citenamefont {Kotila}\ and\ \citenamefont
  {Iachello}(2021)}]{kotila2021}%
  \BibitemOpen
  \bibfield  {author} {\bibinfo {author} {\bibfnamefont {J.}~\bibnamefont
  {Kotila}}\ and\ \bibinfo {author} {\bibfnamefont {F.}~\bibnamefont
  {Iachello}},\ }\href {https://doi.org/10.1103/PhysRevC.103.044302} {\bibfield
   {journal} {\bibinfo  {journal} {Phys. Rev. C}\ }\textbf {\bibinfo {volume}
  {103}},\ \bibinfo {pages} {044302} (\bibinfo {year} {2021})}\BibitemShut
  {NoStop}%
\bibitem [{\citenamefont {Otsuka}\ \emph
  {et~al.}(1978{\natexlab{a}})\citenamefont {Otsuka}, \citenamefont {Arima},\
  and\ \citenamefont {Iachello}}]{OAI}%
  \BibitemOpen
  \bibfield  {author} {\bibinfo {author} {\bibfnamefont {T.}~\bibnamefont
  {Otsuka}}, \bibinfo {author} {\bibfnamefont {A.}~\bibnamefont {Arima}},\ and\
  \bibinfo {author} {\bibfnamefont {F.}~\bibnamefont {Iachello}},\ }\href
  {https://doi.org/10.1016/0375-9474(78)90532-8} {\bibfield  {journal}
  {\bibinfo  {journal} {Nucl. Phys. A}\ }\textbf {\bibinfo {volume} {309}},\
  \bibinfo {pages} {1} (\bibinfo {year} {1978}{\natexlab{a}})}\BibitemShut
  {NoStop}%
\bibitem [{\citenamefont {Van~Isacker}\ \emph {et~al.}(2017)\citenamefont
  {Van~Isacker}, \citenamefont {Engel},\ and\ \citenamefont
  {Nomura}}]{vanisacker2017}%
  \BibitemOpen
  \bibfield  {author} {\bibinfo {author} {\bibfnamefont {P.}~\bibnamefont
  {Van~Isacker}}, \bibinfo {author} {\bibfnamefont {J.}~\bibnamefont {Engel}},\
  and\ \bibinfo {author} {\bibfnamefont {K.}~\bibnamefont {Nomura}},\ }\href
  {https://doi.org/10.1103/PhysRevC.96.064305} {\bibfield  {journal} {\bibinfo
  {journal} {Phys. Rev. C}\ }\textbf {\bibinfo {volume} {96}},\ \bibinfo
  {pages} {064305} (\bibinfo {year} {2017})}\BibitemShut {NoStop}%
\bibitem [{\citenamefont {Nik\v{s}i\'c}\ \emph {et~al.}(2014)\citenamefont
  {Nik\v{s}i\'c}, \citenamefont {Paar}, \citenamefont {Vretenar},\ and\
  \citenamefont {Ring}}]{DIRHB}%
  \BibitemOpen
  \bibfield  {author} {\bibinfo {author} {\bibfnamefont {T.}~\bibnamefont
  {Nik\v{s}i\'c}}, \bibinfo {author} {\bibfnamefont {N.}~\bibnamefont {Paar}},
  \bibinfo {author} {\bibfnamefont {D.}~\bibnamefont {Vretenar}},\ and\
  \bibinfo {author} {\bibfnamefont {P.}~\bibnamefont {Ring}},\ }\href
  {https://doi.org/https://doi.org/10.1016/j.cpc.2014.02.027} {\bibfield
  {journal} {\bibinfo  {journal} {Comput. Phys. Commun.}\ }\textbf {\bibinfo
  {volume} {185}},\ \bibinfo {pages} {1808} (\bibinfo {year}
  {2014})}\BibitemShut {NoStop}%
\bibitem [{\citenamefont {Nik\ifmmode \check{s}\else
  \v{s}\fi{}i\ifmmode~\acute{c}\else \'{c}\fi{}}\ \emph
  {et~al.}(2008)\citenamefont {Nik\ifmmode \check{s}\else
  \v{s}\fi{}i\ifmmode~\acute{c}\else \'{c}\fi{}}, \citenamefont {Vretenar},\
  and\ \citenamefont {Ring}}]{DDPC1}%
  \BibitemOpen
  \bibfield  {author} {\bibinfo {author} {\bibfnamefont {T.}~\bibnamefont
  {Nik\ifmmode \check{s}\else \v{s}\fi{}i\ifmmode~\acute{c}\else \'{c}\fi{}}},
  \bibinfo {author} {\bibfnamefont {D.}~\bibnamefont {Vretenar}},\ and\
  \bibinfo {author} {\bibfnamefont {P.}~\bibnamefont {Ring}},\ }\href
  {https://doi.org/10.1103/PhysRevC.78.034318} {\bibfield  {journal} {\bibinfo
  {journal} {Phys. Rev. C}\ }\textbf {\bibinfo {volume} {78}},\ \bibinfo
  {pages} {034318} (\bibinfo {year} {2008})}\BibitemShut {NoStop}%
\bibitem [{\citenamefont {Tian}\ \emph {et~al.}(2009)\citenamefont {Tian},
  \citenamefont {Ma},\ and\ \citenamefont {Ring}}]{tian2009}%
  \BibitemOpen
  \bibfield  {author} {\bibinfo {author} {\bibfnamefont {Y.}~\bibnamefont
  {Tian}}, \bibinfo {author} {\bibfnamefont {Z.~Y.}\ \bibnamefont {Ma}},\ and\
  \bibinfo {author} {\bibfnamefont {P.}~\bibnamefont {Ring}},\ }\href
  {https://doi.org/10.1016/j.physletb.2009.04.067} {\bibfield  {journal}
  {\bibinfo  {journal} {Phys. Lett. B}\ }\textbf {\bibinfo {volume} {676}},\
  \bibinfo {pages} {44 } (\bibinfo {year} {2009})}\BibitemShut {NoStop}%
\bibitem [{\citenamefont {Bohr}\ and\ \citenamefont {Mottelson}(1975)}]{BM}%
  \BibitemOpen
  \bibfield  {author} {\bibinfo {author} {\bibfnamefont {A.}~\bibnamefont
  {Bohr}}\ and\ \bibinfo {author} {\bibfnamefont {B.~R.}\ \bibnamefont
  {Mottelson}},\ }\href@noop {} {\emph {\bibinfo {title} {Nuclear Structure}}}\
  (\bibinfo  {publisher} {Benjamin, New York},\ \bibinfo {year}
  {1975})\BibitemShut {NoStop}%
\bibitem [{\citenamefont {Lalazissis}\ \emph {et~al.}(2005)\citenamefont
  {Lalazissis}, \citenamefont {Nik\ifmmode \check{s}\else
  \v{s}\fi{}i\ifmmode~\acute{c}\else \'{c}\fi{}}, \citenamefont {Vretenar},\
  and\ \citenamefont {Ring}}]{DDME2}%
  \BibitemOpen
  \bibfield  {author} {\bibinfo {author} {\bibfnamefont {G.~A.}\ \bibnamefont
  {Lalazissis}}, \bibinfo {author} {\bibfnamefont {T.}~\bibnamefont
  {Nik\ifmmode \check{s}\else \v{s}\fi{}i\ifmmode~\acute{c}\else \'{c}\fi{}}},
  \bibinfo {author} {\bibfnamefont {D.}~\bibnamefont {Vretenar}},\ and\
  \bibinfo {author} {\bibfnamefont {P.}~\bibnamefont {Ring}},\ }\href
  {https://doi.org/10.1103/PhysRevC.71.024312} {\bibfield  {journal} {\bibinfo
  {journal} {Phys. Rev. C}\ }\textbf {\bibinfo {volume} {71}},\ \bibinfo
  {pages} {024312} (\bibinfo {year} {2005})}\BibitemShut {NoStop}%
\bibitem [{\citenamefont {Otsuka}\ \emph
  {et~al.}(1978{\natexlab{b}})\citenamefont {Otsuka}, \citenamefont {Arima},
  \citenamefont {Iachello},\ and\ \citenamefont {Talmi}}]{OAIT}%
  \BibitemOpen
  \bibfield  {author} {\bibinfo {author} {\bibfnamefont {T.}~\bibnamefont
  {Otsuka}}, \bibinfo {author} {\bibfnamefont {A.}~\bibnamefont {Arima}},
  \bibinfo {author} {\bibfnamefont {F.}~\bibnamefont {Iachello}},\ and\
  \bibinfo {author} {\bibfnamefont {I.}~\bibnamefont {Talmi}},\ }\href
  {https://doi.org/10.1016/0370-2693(78)90260-5} {\bibfield  {journal}
  {\bibinfo  {journal} {Phys. Lett. B}\ }\textbf {\bibinfo {volume} {76}},\
  \bibinfo {pages} {139 } (\bibinfo {year} {1978}{\natexlab{b}})}\BibitemShut
  {NoStop}%
\bibitem [{\citenamefont {Dieperink}\ \emph {et~al.}(1980)\citenamefont
  {Dieperink}, \citenamefont {Scholten},\ and\ \citenamefont
  {Iachello}}]{dieperink1980}%
  \BibitemOpen
  \bibfield  {author} {\bibinfo {author} {\bibfnamefont {A.~E.~L.}\
  \bibnamefont {Dieperink}}, \bibinfo {author} {\bibfnamefont {O.}~\bibnamefont
  {Scholten}},\ and\ \bibinfo {author} {\bibfnamefont {F.}~\bibnamefont
  {Iachello}},\ }\href {https://doi.org/10.1103/PhysRevLett.44.1747} {\bibfield
   {journal} {\bibinfo  {journal} {Phys. Rev. Lett.}\ }\textbf {\bibinfo
  {volume} {44}},\ \bibinfo {pages} {1747} (\bibinfo {year}
  {1980})}\BibitemShut {NoStop}%
\bibitem [{\citenamefont {Ginocchio}\ and\ \citenamefont
  {Kirson}(1980)}]{ginocchio1980}%
  \BibitemOpen
  \bibfield  {author} {\bibinfo {author} {\bibfnamefont {J.~N.}\ \bibnamefont
  {Ginocchio}}\ and\ \bibinfo {author} {\bibfnamefont {M.~W.}\ \bibnamefont
  {Kirson}},\ }\href {https://doi.org/10.1016/0375-9474(80)90387-5} {\bibfield
  {journal} {\bibinfo  {journal} {Nucl. Phys. A}\ }\textbf {\bibinfo {volume}
  {350}},\ \bibinfo {pages} {31} (\bibinfo {year} {1980})}\BibitemShut
  {NoStop}%
\bibitem [{\citenamefont {Nomura}\ \emph
  {et~al.}(2011{\natexlab{a}})\citenamefont {Nomura}, \citenamefont {Otsuka},
  \citenamefont {Shimizu},\ and\ \citenamefont {Guo}}]{nomura2011rot}%
  \BibitemOpen
  \bibfield  {author} {\bibinfo {author} {\bibfnamefont {K.}~\bibnamefont
  {Nomura}}, \bibinfo {author} {\bibfnamefont {T.}~\bibnamefont {Otsuka}},
  \bibinfo {author} {\bibfnamefont {N.}~\bibnamefont {Shimizu}},\ and\ \bibinfo
  {author} {\bibfnamefont {L.}~\bibnamefont {Guo}},\ }\href
  {https://doi.org/10.1103/PhysRevC.83.041302} {\bibfield  {journal} {\bibinfo
  {journal} {Phys. Rev. C}\ }\textbf {\bibinfo {volume} {83}},\ \bibinfo
  {pages} {041302} (\bibinfo {year} {2011}{\natexlab{a}})}\BibitemShut
  {NoStop}%
\bibitem [{\citenamefont {Inglis}(1956)}]{inglis1956}%
  \BibitemOpen
  \bibfield  {author} {\bibinfo {author} {\bibfnamefont {D.~R.}\ \bibnamefont
  {Inglis}},\ }\href {https://doi.org/10.1103/PhysRev.103.1786} {\bibfield
  {journal} {\bibinfo  {journal} {Phys. Rev.}\ }\textbf {\bibinfo {volume}
  {103}},\ \bibinfo {pages} {1786 } (\bibinfo {year} {1956})}\BibitemShut
  {NoStop}%
\bibitem [{\citenamefont {Beliaev}(1961)}]{belyaev1961}%
  \BibitemOpen
  \bibfield  {author} {\bibinfo {author} {\bibfnamefont {S.~T.}\ \bibnamefont
  {Beliaev}},\ }\href {https://doi.org/10.1016/0029-5582(61)90384-4} {\bibfield
   {journal} {\bibinfo  {journal} {Nucl. Phys.}\ }\textbf {\bibinfo {volume}
  {24}},\ \bibinfo {pages} {322 } (\bibinfo {year} {1961})}\BibitemShut
  {NoStop}%
\bibitem [{\citenamefont {Nomura}\ \emph
  {et~al.}(2012{\natexlab{b}})\citenamefont {Nomura}, \citenamefont {Shimizu},
  \citenamefont {Vretenar}, \citenamefont {Nik\ifmmode \check{s}\else
  \v{s}\fi{}i\ifmmode~\acute{c}\else \'{c}\fi{}},\ and\ \citenamefont
  {Otsuka}}]{nomura2012tri}%
  \BibitemOpen
  \bibfield  {author} {\bibinfo {author} {\bibfnamefont {K.}~\bibnamefont
  {Nomura}}, \bibinfo {author} {\bibfnamefont {N.}~\bibnamefont {Shimizu}},
  \bibinfo {author} {\bibfnamefont {D.}~\bibnamefont {Vretenar}}, \bibinfo
  {author} {\bibfnamefont {T.}~\bibnamefont {Nik\ifmmode \check{s}\else
  \v{s}\fi{}i\ifmmode~\acute{c}\else \'{c}\fi{}}},\ and\ \bibinfo {author}
  {\bibfnamefont {T.}~\bibnamefont {Otsuka}},\ }\href
  {https://doi.org/10.1103/PhysRevLett.108.132501} {\bibfield  {journal}
  {\bibinfo  {journal} {Phys. Rev. Lett.}\ }\textbf {\bibinfo {volume} {108}},\
  \bibinfo {pages} {132501} (\bibinfo {year} {2012}{\natexlab{b}})}\BibitemShut
  {NoStop}%
\bibitem [{\citenamefont {Scholten}(1985)}]{scholten1985}%
  \BibitemOpen
  \bibfield  {author} {\bibinfo {author} {\bibfnamefont {O.}~\bibnamefont
  {Scholten}},\ }\href
  {https://doi.org/https://doi.org/10.1016/0146-6410(85)90054-7} {\bibfield
  {journal} {\bibinfo  {journal} {Prog. Part. Nucl. Phys.}\ }\textbf {\bibinfo
  {volume} {14}},\ \bibinfo {pages} {189} (\bibinfo {year} {1985})}\BibitemShut
  {NoStop}%
\bibitem [{\citenamefont {Brant}\ and\ \citenamefont {Paar}(1988)}]{brant1988}%
  \BibitemOpen
  \bibfield  {author} {\bibinfo {author} {\bibfnamefont {S.}~\bibnamefont
  {Brant}}\ and\ \bibinfo {author} {\bibfnamefont {V.}~\bibnamefont {Paar}},\
  }\href {https://doi.org/10.1007/BF01283770} {\bibfield  {journal} {\bibinfo
  {journal} {Z. Phys. A}\ }\textbf {\bibinfo {volume} {329}},\ \bibinfo {pages}
  {151} (\bibinfo {year} {1988})}\BibitemShut {NoStop}%
\bibitem [{\citenamefont {Iachello}\ and\ \citenamefont
  {Scholten}(1979)}]{iachello1979}%
  \BibitemOpen
  \bibfield  {author} {\bibinfo {author} {\bibfnamefont {F.}~\bibnamefont
  {Iachello}}\ and\ \bibinfo {author} {\bibfnamefont {O.}~\bibnamefont
  {Scholten}},\ }\href {https://doi.org/10.1103/PhysRevLett.43.679} {\bibfield
  {journal} {\bibinfo  {journal} {Phys. Rev. Lett.}\ }\textbf {\bibinfo
  {volume} {43}},\ \bibinfo {pages} {679} (\bibinfo {year} {1979})}\BibitemShut
  {NoStop}%
\bibitem [{\citenamefont {Yoshida}()}]{TWBOS}%
  \BibitemOpen
  \bibfield  {author} {\bibinfo {author} {\bibfnamefont {N.}~\bibnamefont
  {Yoshida}},\ }\bibinfo {note} {private communication (2018)}\BibitemShut
  {NoStop}%
\bibitem [{\citenamefont {{Brookhaven National Nuclear Data Center}}()}]{data}%
  \BibitemOpen
  \bibfield  {author} {\bibinfo {author} {\bibnamefont {{Brookhaven National
  Nuclear Data Center}}},\ }\href@noop {} {}\bibinfo {howpublished}
  {{http://www.nndc.bnl.gov}}\BibitemShut {NoStop}%
\bibitem [{\citenamefont {Otsuka}\ and\ \citenamefont {Yoshida}()}]{NPBOS}%
  \BibitemOpen
  \bibfield  {author} {\bibinfo {author} {\bibfnamefont {T.}~\bibnamefont
  {Otsuka}}\ and\ \bibinfo {author} {\bibfnamefont {N.}~\bibnamefont
  {Yoshida}},\ }\href@noop {} {}\bibinfo {note} {{}JAERI-M (Japan At. Ener.
  Res. Inst.) Report No. 85 (1985)}\BibitemShut {NoStop}%
\bibitem [{\citenamefont {Nomura}\ \emph {et~al.}(2010)\citenamefont {Nomura},
  \citenamefont {Shimizu},\ and\ \citenamefont {Otsuka}}]{nomura2010}%
  \BibitemOpen
  \bibfield  {author} {\bibinfo {author} {\bibfnamefont {K.}~\bibnamefont
  {Nomura}}, \bibinfo {author} {\bibfnamefont {N.}~\bibnamefont {Shimizu}},\
  and\ \bibinfo {author} {\bibfnamefont {T.}~\bibnamefont {Otsuka}},\ }\href
  {https://doi.org/10.1103/PhysRevC.81.044307} {\bibfield  {journal} {\bibinfo
  {journal} {Phys. Rev. C}\ }\textbf {\bibinfo {volume} {81}},\ \bibinfo
  {pages} {044307} (\bibinfo {year} {2010})}\BibitemShut {NoStop}%
\bibitem [{\citenamefont {Nomura}\ \emph
  {et~al.}(2011{\natexlab{b}})\citenamefont {Nomura}, \citenamefont {Otsuka},
  \citenamefont {Rodr\'{\i}guez-Guzm\'an}, \citenamefont {Robledo},\ and\
  \citenamefont {Sarriguren}}]{nomura2011sys}%
  \BibitemOpen
  \bibfield  {author} {\bibinfo {author} {\bibfnamefont {K.}~\bibnamefont
  {Nomura}}, \bibinfo {author} {\bibfnamefont {T.}~\bibnamefont {Otsuka}},
  \bibinfo {author} {\bibfnamefont {R.}~\bibnamefont
  {Rodr\'{\i}guez-Guzm\'an}}, \bibinfo {author} {\bibfnamefont {L.~M.}\
  \bibnamefont {Robledo}},\ and\ \bibinfo {author} {\bibfnamefont
  {P.}~\bibnamefont {Sarriguren}},\ }\href
  {https://doi.org/10.1103/PhysRevC.84.054316} {\bibfield  {journal} {\bibinfo
  {journal} {Phys. Rev. C}\ }\textbf {\bibinfo {volume} {84}},\ \bibinfo
  {pages} {054316} (\bibinfo {year} {2011}{\natexlab{b}})}\BibitemShut
  {NoStop}%
\bibitem [{\citenamefont {Duval}\ and\ \citenamefont
  {Barrett}(1981)}]{duval1981}%
  \BibitemOpen
  \bibfield  {author} {\bibinfo {author} {\bibfnamefont {P.~D.}\ \bibnamefont
  {Duval}}\ and\ \bibinfo {author} {\bibfnamefont {B.~R.}\ \bibnamefont
  {Barrett}},\ }\href {https://doi.org/10.1016/0370-2693(81)90321-X} {\bibfield
   {journal} {\bibinfo  {journal} {Phys. Lett. B}\ }\textbf {\bibinfo {volume}
  {100}},\ \bibinfo {pages} {223} (\bibinfo {year} {1981})}\BibitemShut
  {NoStop}%
\bibitem [{\citenamefont {Kremer}\ \emph {et~al.}(2016)\citenamefont {Kremer},
  \citenamefont {Aslanidou}, \citenamefont {Bassauer}, \citenamefont {Hilcker},
  \citenamefont {Krugmann}, \citenamefont {von Neumann-Cosel}, \citenamefont
  {Otsuka}, \citenamefont {Pietralla}, \citenamefont {Ponomarev}, \citenamefont
  {Shimizu}, \citenamefont {Singer}, \citenamefont {Steinhilber}, \citenamefont
  {Togashi}, \citenamefont {Tsunoda}, \citenamefont {Werner},\ and\
  \citenamefont {Zweidinger}}]{kremer2016}%
  \BibitemOpen
  \bibfield  {author} {\bibinfo {author} {\bibfnamefont {C.}~\bibnamefont
  {Kremer}}, \bibinfo {author} {\bibfnamefont {S.}~\bibnamefont {Aslanidou}},
  \bibinfo {author} {\bibfnamefont {S.}~\bibnamefont {Bassauer}}, \bibinfo
  {author} {\bibfnamefont {M.}~\bibnamefont {Hilcker}}, \bibinfo {author}
  {\bibfnamefont {A.}~\bibnamefont {Krugmann}}, \bibinfo {author}
  {\bibfnamefont {P.}~\bibnamefont {von Neumann-Cosel}}, \bibinfo {author}
  {\bibfnamefont {T.}~\bibnamefont {Otsuka}}, \bibinfo {author} {\bibfnamefont
  {N.}~\bibnamefont {Pietralla}}, \bibinfo {author} {\bibfnamefont {V.~Y.}\
  \bibnamefont {Ponomarev}}, \bibinfo {author} {\bibfnamefont {N.}~\bibnamefont
  {Shimizu}}, \bibinfo {author} {\bibfnamefont {M.}~\bibnamefont {Singer}},
  \bibinfo {author} {\bibfnamefont {G.}~\bibnamefont {Steinhilber}}, \bibinfo
  {author} {\bibfnamefont {T.}~\bibnamefont {Togashi}}, \bibinfo {author}
  {\bibfnamefont {Y.}~\bibnamefont {Tsunoda}}, \bibinfo {author} {\bibfnamefont
  {V.}~\bibnamefont {Werner}},\ and\ \bibinfo {author} {\bibfnamefont
  {M.}~\bibnamefont {Zweidinger}},\ }\href
  {https://doi.org/10.1103/PhysRevLett.117.172503} {\bibfield  {journal}
  {\bibinfo  {journal} {Phys. Rev. Lett.}\ }\textbf {\bibinfo {volume} {117}},\
  \bibinfo {pages} {172503} (\bibinfo {year} {2016})}\BibitemShut {NoStop}%
\bibitem [{\citenamefont {Bucurescu}\ \emph {et~al.}(2012)\citenamefont
  {Bucurescu}, \citenamefont {Dr\ifmmode~\u{a}\else \u{a}\fi{}gulescu},
  \citenamefont {Pascu}, \citenamefont {Wirth}, \citenamefont {Filipescu},
  \citenamefont {C\ifmmode \u{a}\else~\u{a}\fi{}ta Danil}, \citenamefont
  {C\ifmmode \u{a}\else~\u{a}\fi{}ta Danil}, \citenamefont {Deleanu},
  \citenamefont {Eppinger}, \citenamefont {Faestermann}, \citenamefont
  {Ghi\ifmmode \mbox{\c{t}}\else \c{t}\fi{}\ifmmode~\u{a}\else \u{a}\fi{}},
  \citenamefont {Glodariu}, \citenamefont {Hertenberger}, \citenamefont
  {Iva\ifmmode~\mbox{\c{s}}\else \c{s}\fi{}cu}, \citenamefont {Kr\"ucken},
  \citenamefont {M\ifmmode~\u{a}\else \u{a}\fi{}rginean}, \citenamefont
  {M\ifmmode~\u{a}\else \u{a}\fi{}rginean}, \citenamefont {Mihai},
  \citenamefont {Negret}, \citenamefont {Sava}, \citenamefont {Stroe},
  \citenamefont {Wimmer},\ and\ \citenamefont {Zamfir}}]{bucurescu2012}%
  \BibitemOpen
  \bibfield  {author} {\bibinfo {author} {\bibfnamefont {D.}~\bibnamefont
  {Bucurescu}}, \bibinfo {author} {\bibfnamefont {E.}~\bibnamefont
  {Dr\ifmmode~\u{a}\else \u{a}\fi{}gulescu}}, \bibinfo {author} {\bibfnamefont
  {S.}~\bibnamefont {Pascu}}, \bibinfo {author} {\bibfnamefont {H.-F.}\
  \bibnamefont {Wirth}}, \bibinfo {author} {\bibfnamefont {D.}~\bibnamefont
  {Filipescu}}, \bibinfo {author} {\bibfnamefont {G.}~\bibnamefont {C\ifmmode
  \u{a}\else~\u{a}\fi{}ta Danil}}, \bibinfo {author} {\bibfnamefont
  {I.}~\bibnamefont {C\ifmmode \u{a}\else~\u{a}\fi{}ta Danil}}, \bibinfo
  {author} {\bibfnamefont {D.}~\bibnamefont {Deleanu}}, \bibinfo {author}
  {\bibfnamefont {K.}~\bibnamefont {Eppinger}}, \bibinfo {author}
  {\bibfnamefont {T.}~\bibnamefont {Faestermann}}, \bibinfo {author}
  {\bibfnamefont {D.~G.}\ \bibnamefont {Ghi\ifmmode \mbox{\c{t}}\else
  \c{t}\fi{}\ifmmode~\u{a}\else \u{a}\fi{}}}, \bibinfo {author} {\bibfnamefont
  {T.}~\bibnamefont {Glodariu}}, \bibinfo {author} {\bibfnamefont
  {R.}~\bibnamefont {Hertenberger}}, \bibinfo {author} {\bibfnamefont
  {M.}~\bibnamefont {Iva\ifmmode~\mbox{\c{s}}\else \c{s}\fi{}cu}}, \bibinfo
  {author} {\bibfnamefont {R.}~\bibnamefont {Kr\"ucken}}, \bibinfo {author}
  {\bibfnamefont {N.}~\bibnamefont {M\ifmmode~\u{a}\else \u{a}\fi{}rginean}},
  \bibinfo {author} {\bibfnamefont {R.}~\bibnamefont {M\ifmmode~\u{a}\else
  \u{a}\fi{}rginean}}, \bibinfo {author} {\bibfnamefont {C.}~\bibnamefont
  {Mihai}}, \bibinfo {author} {\bibfnamefont {A.}~\bibnamefont {Negret}},
  \bibinfo {author} {\bibfnamefont {T.}~\bibnamefont {Sava}}, \bibinfo {author}
  {\bibfnamefont {L.}~\bibnamefont {Stroe}}, \bibinfo {author} {\bibfnamefont
  {K.}~\bibnamefont {Wimmer}},\ and\ \bibinfo {author} {\bibfnamefont {N.~V.}\
  \bibnamefont {Zamfir}},\ }\href {https://doi.org/10.1103/PhysRevC.85.017304}
  {\bibfield  {journal} {\bibinfo  {journal} {Phys. Rev. C}\ }\textbf {\bibinfo
  {volume} {85}},\ \bibinfo {pages} {017304} (\bibinfo {year}
  {2012})}\BibitemShut {NoStop}%
\bibitem [{\citenamefont {Iachello}\ and\ \citenamefont {Arima}(1987)}]{IBM}%
  \BibitemOpen
  \bibfield  {author} {\bibinfo {author} {\bibfnamefont {F.}~\bibnamefont
  {Iachello}}\ and\ \bibinfo {author} {\bibfnamefont {A.}~\bibnamefont
  {Arima}},\ }\href@noop {} {\emph {\bibinfo {title} {The interacting boson
  model}}}\ (\bibinfo  {publisher} {Cambridge University Press, Cambridge},\
  \bibinfo {year} {1987})\BibitemShut {NoStop}%
\bibitem [{\citenamefont {\'Alvarez-Rodr\'{\i}guez}\ \emph
  {et~al.}(2004)\citenamefont {\'Alvarez-Rodr\'{\i}guez}, \citenamefont
  {Sarriguren}, \citenamefont {de~Guerra}, \citenamefont {Pacearescu},
  \citenamefont {Faessler},\ and\ \citenamefont {\ifmmode~\check{S}\else
  \v{S}\fi{}imkovic}}]{alvarez2004}%
  \BibitemOpen
  \bibfield  {author} {\bibinfo {author} {\bibfnamefont {R.}~\bibnamefont
  {\'Alvarez-Rodr\'{\i}guez}}, \bibinfo {author} {\bibfnamefont
  {P.}~\bibnamefont {Sarriguren}}, \bibinfo {author} {\bibfnamefont {E.~M.}\
  \bibnamefont {de~Guerra}}, \bibinfo {author} {\bibfnamefont {L.}~\bibnamefont
  {Pacearescu}}, \bibinfo {author} {\bibfnamefont {A.}~\bibnamefont
  {Faessler}},\ and\ \bibinfo {author} {\bibfnamefont {F.}~\bibnamefont
  {\ifmmode~\check{S}\else \v{S}\fi{}imkovic}},\ }\href
  {https://doi.org/10.1103/PhysRevC.70.064309} {\bibfield  {journal} {\bibinfo
  {journal} {Phys. Rev. C}\ }\textbf {\bibinfo {volume} {70}},\ \bibinfo
  {pages} {064309} (\bibinfo {year} {2004})}\BibitemShut {NoStop}%
\bibitem [{\citenamefont {Griffiths}\ and\ \citenamefont
  {Vogel}(1992)}]{griffiths1992}%
  \BibitemOpen
  \bibfield  {author} {\bibinfo {author} {\bibfnamefont {A.}~\bibnamefont
  {Griffiths}}\ and\ \bibinfo {author} {\bibfnamefont {P.}~\bibnamefont
  {Vogel}},\ }\href {https://doi.org/10.1103/PhysRevC.46.181} {\bibfield
  {journal} {\bibinfo  {journal} {Phys. Rev. C}\ }\textbf {\bibinfo {volume}
  {46}},\ \bibinfo {pages} {181} (\bibinfo {year} {1992})}\BibitemShut
  {NoStop}%
\bibitem [{\citenamefont {Civitarese}\ and\ \citenamefont
  {Suhonen}(1998)}]{civitarese1998}%
  \BibitemOpen
  \bibfield  {author} {\bibinfo {author} {\bibfnamefont {O.}~\bibnamefont
  {Civitarese}}\ and\ \bibinfo {author} {\bibfnamefont {J.}~\bibnamefont
  {Suhonen}},\ }\href {https://doi.org/10.1103/PhysRevC.58.1535} {\bibfield
  {journal} {\bibinfo  {journal} {Phys. Rev. C}\ }\textbf {\bibinfo {volume}
  {58}},\ \bibinfo {pages} {1535} (\bibinfo {year} {1998})}\BibitemShut
  {NoStop}%
\bibitem [{\citenamefont {Moreno}\ \emph {et~al.}(2008)\citenamefont {Moreno},
  \citenamefont {{\'{A}}lvarez-Rodr{\'{\i}}guez}, \citenamefont {Sarriguren},
  \citenamefont {de~Guerra}, \citenamefont {{\v{S}}imkovic},\ and\
  \citenamefont {Faessler}}]{moreno2008}%
  \BibitemOpen
  \bibfield  {author} {\bibinfo {author} {\bibfnamefont {O.}~\bibnamefont
  {Moreno}}, \bibinfo {author} {\bibfnamefont {R.}~\bibnamefont
  {{\'{A}}lvarez-Rodr{\'{\i}}guez}}, \bibinfo {author} {\bibfnamefont
  {P.}~\bibnamefont {Sarriguren}}, \bibinfo {author} {\bibfnamefont {E.~M.}\
  \bibnamefont {de~Guerra}}, \bibinfo {author} {\bibfnamefont {F.}~\bibnamefont
  {{\v{S}}imkovic}},\ and\ \bibinfo {author} {\bibfnamefont {A.}~\bibnamefont
  {Faessler}},\ }\href {https://doi.org/10.1088/0954-3899/36/1/015106}
  {\bibfield  {journal} {\bibinfo  {journal} {J. Phys. G: Nucl. Part. Phys.}\
  }\textbf {\bibinfo {volume} {36}},\ \bibinfo {pages} {015106} (\bibinfo
  {year} {2008})}\BibitemShut {NoStop}%
\bibitem [{\citenamefont {Kotila}\ and\ \citenamefont
  {Iachello}(2012)}]{kotila2012}%
  \BibitemOpen
  \bibfield  {author} {\bibinfo {author} {\bibfnamefont {J.}~\bibnamefont
  {Kotila}}\ and\ \bibinfo {author} {\bibfnamefont {F.}~\bibnamefont
  {Iachello}},\ }\href {https://doi.org/10.1103/PhysRevC.85.034316} {\bibfield
  {journal} {\bibinfo  {journal} {Phys. Rev. C}\ }\textbf {\bibinfo {volume}
  {85}},\ \bibinfo {pages} {034316} (\bibinfo {year} {2012})}\BibitemShut
  {NoStop}%
\bibitem [{\citenamefont {Honma}\ \emph {et~al.}(2009)\citenamefont {Honma},
  \citenamefont {Otsuka}, \citenamefont {Mizusaki},\ and\ \citenamefont
  {Hjorth-Jensen}}]{honma2009}%
  \BibitemOpen
  \bibfield  {author} {\bibinfo {author} {\bibfnamefont {M.}~\bibnamefont
  {Honma}}, \bibinfo {author} {\bibfnamefont {T.}~\bibnamefont {Otsuka}},
  \bibinfo {author} {\bibfnamefont {T.}~\bibnamefont {Mizusaki}},\ and\
  \bibinfo {author} {\bibfnamefont {M.}~\bibnamefont {Hjorth-Jensen}},\ }\href
  {https://doi.org/10.1103/PhysRevC.80.064323} {\bibfield  {journal} {\bibinfo
  {journal} {Phys. Rev. C}\ }\textbf {\bibinfo {volume} {80}},\ \bibinfo
  {pages} {064323} (\bibinfo {year} {2009})}\BibitemShut {NoStop}%
\bibitem [{\citenamefont {Goriely}\ \emph {et~al.}(2009)\citenamefont
  {Goriely}, \citenamefont {Hilaire}, \citenamefont {Girod},\ and\
  \citenamefont {P\'eru}}]{D1M}%
  \BibitemOpen
  \bibfield  {author} {\bibinfo {author} {\bibfnamefont {S.}~\bibnamefont
  {Goriely}}, \bibinfo {author} {\bibfnamefont {S.}~\bibnamefont {Hilaire}},
  \bibinfo {author} {\bibfnamefont {M.}~\bibnamefont {Girod}},\ and\ \bibinfo
  {author} {\bibfnamefont {S.}~\bibnamefont {P\'eru}},\ }\href
  {https://doi.org/10.1103/PhysRevLett.102.242501} {\bibfield  {journal}
  {\bibinfo  {journal} {Phys. Rev. Lett.}\ }\textbf {\bibinfo {volume} {102}},\
  \bibinfo {pages} {242501} (\bibinfo {year} {2009})}\BibitemShut {NoStop}%
\bibitem [{\citenamefont {Stone}(2005)}]{stone2005}%
  \BibitemOpen
  \bibfield  {author} {\bibinfo {author} {\bibfnamefont {N.}~\bibnamefont
  {Stone}},\ }\href@noop {} {\bibfield  {journal} {\bibinfo  {journal} {At.
  Data Nucl. Data Tables}\ }\textbf {\bibinfo {volume} {90}},\ \bibinfo {pages}
  {75} (\bibinfo {year} {2005})}\BibitemShut {NoStop}%
\bibitem [{\citenamefont {Dellagiacoma}(1988)}]{dellagiacoma1988phdthesis}%
  \BibitemOpen
  \bibfield  {author} {\bibinfo {author} {\bibfnamefont {F.}~\bibnamefont
  {Dellagiacoma}},\ }\emph {\bibinfo {title} {Beta decay of odd mass nuclei in
  the interacting boson-fermion model}},\ \href@noop {} {Ph.D. thesis},\
  \bibinfo  {school} {Yale University} (\bibinfo {year} {1988})\BibitemShut
  {NoStop}%
\bibitem [{\citenamefont {Dellagiacoma}\ and\ \citenamefont
  {Iachello}(1989)}]{DELLAGIACOMA1989}%
  \BibitemOpen
  \bibfield  {author} {\bibinfo {author} {\bibfnamefont {F.}~\bibnamefont
  {Dellagiacoma}}\ and\ \bibinfo {author} {\bibfnamefont {F.}~\bibnamefont
  {Iachello}},\ }\href
  {https://doi.org/https://doi.org/10.1016/0370-2693(89)91434-2} {\bibfield
  {journal} {\bibinfo  {journal} {Phys. Lett. B}\ }\textbf {\bibinfo {volume}
  {218}},\ \bibinfo {pages} {399 } (\bibinfo {year} {1989})}\BibitemShut
  {NoStop}%
\bibitem [{\citenamefont {Nomura}\ \emph
  {et~al.}(2020{\natexlab{c}})\citenamefont {Nomura}, \citenamefont
  {Rodr\'{\i}guez-Guzm\'an},\ and\ \citenamefont {Robledo}}]{nomura2020cs}%
  \BibitemOpen
  \bibfield  {author} {\bibinfo {author} {\bibfnamefont {K.}~\bibnamefont
  {Nomura}}, \bibinfo {author} {\bibfnamefont {R.}~\bibnamefont
  {Rodr\'{\i}guez-Guzm\'an}},\ and\ \bibinfo {author} {\bibfnamefont {L.~M.}\
  \bibnamefont {Robledo}},\ }\href
  {https://doi.org/10.1103/PhysRevC.101.014306} {\bibfield  {journal} {\bibinfo
   {journal} {Phys. Rev. C}\ }\textbf {\bibinfo {volume} {101}},\ \bibinfo
  {pages} {014306} (\bibinfo {year} {2020}{\natexlab{c}})}\BibitemShut
  {NoStop}%
\bibitem [{\citenamefont {Heyde}\ and\ \citenamefont {Wood}(2011)}]{heyde2011}%
  \BibitemOpen
  \bibfield  {author} {\bibinfo {author} {\bibfnamefont {K.}~\bibnamefont
  {Heyde}}\ and\ \bibinfo {author} {\bibfnamefont {J.~L.}\ \bibnamefont
  {Wood}},\ }\href {https://doi.org/10.1103/RevModPhys.83.1467} {\bibfield
  {journal} {\bibinfo  {journal} {Rev. Mod. Phys.}\ }\textbf {\bibinfo {volume}
  {83}},\ \bibinfo {pages} {1467} (\bibinfo {year} {2011})}\BibitemShut
  {NoStop}%
\bibitem [{\citenamefont {Thomas}\ \emph {et~al.}(2013)\citenamefont {Thomas},
  \citenamefont {Nomura}, \citenamefont {Werner}, \citenamefont {Ahn},
  \citenamefont {Cooper}, \citenamefont {Duckwitz}, \citenamefont {Hinton},
  \citenamefont {Ilie}, \citenamefont {Jolie}, \citenamefont {Petkov},\ and\
  \citenamefont {Radeck}}]{thomas2013}%
  \BibitemOpen
  \bibfield  {author} {\bibinfo {author} {\bibfnamefont {T.}~\bibnamefont
  {Thomas}}, \bibinfo {author} {\bibfnamefont {K.}~\bibnamefont {Nomura}},
  \bibinfo {author} {\bibfnamefont {V.}~\bibnamefont {Werner}}, \bibinfo
  {author} {\bibfnamefont {T.}~\bibnamefont {Ahn}}, \bibinfo {author}
  {\bibfnamefont {N.}~\bibnamefont {Cooper}}, \bibinfo {author} {\bibfnamefont
  {H.}~\bibnamefont {Duckwitz}}, \bibinfo {author} {\bibfnamefont
  {M.}~\bibnamefont {Hinton}}, \bibinfo {author} {\bibfnamefont
  {G.}~\bibnamefont {Ilie}}, \bibinfo {author} {\bibfnamefont {J.}~\bibnamefont
  {Jolie}}, \bibinfo {author} {\bibfnamefont {P.}~\bibnamefont {Petkov}},\ and\
  \bibinfo {author} {\bibfnamefont {D.}~\bibnamefont {Radeck}},\ }\href
  {https://doi.org/10.1103/PhysRevC.88.044305} {\bibfield  {journal} {\bibinfo
  {journal} {Phys. Rev. C}\ }\textbf {\bibinfo {volume} {88}},\ \bibinfo
  {pages} {044305} (\bibinfo {year} {2013})}\BibitemShut {NoStop}%
\bibitem [{\citenamefont {Gao}\ \emph {et~al.}(2012)\citenamefont {Gao},
  \citenamefont {Zhou}, \citenamefont {Fang}, \citenamefont {Zhang},
  \citenamefont {Wang}, \citenamefont {Zhang}, \citenamefont {Liu},
  \citenamefont {Wang}, \citenamefont {Ma}, \citenamefont {Guo}, \citenamefont
  {Wu}, \citenamefont {He}, \citenamefont {Zheng}, \citenamefont {Wang},
  \citenamefont {Li}, \citenamefont {Yan}, \citenamefont {He}, \citenamefont
  {Wang}, \citenamefont {Zheng}, \citenamefont {Fang},\ and\ \citenamefont
  {Chen}}]{gao2012}%
  \BibitemOpen
  \bibfield  {author} {\bibinfo {author} {\bibfnamefont {B.~S.}\ \bibnamefont
  {Gao}}, \bibinfo {author} {\bibfnamefont {X.~H.}\ \bibnamefont {Zhou}},
  \bibinfo {author} {\bibfnamefont {Y.~D.}\ \bibnamefont {Fang}}, \bibinfo
  {author} {\bibfnamefont {Y.~H.}\ \bibnamefont {Zhang}}, \bibinfo {author}
  {\bibfnamefont {S.~C.}\ \bibnamefont {Wang}}, \bibinfo {author}
  {\bibfnamefont {N.~T.}\ \bibnamefont {Zhang}}, \bibinfo {author}
  {\bibfnamefont {M.~L.}\ \bibnamefont {Liu}}, \bibinfo {author} {\bibfnamefont
  {J.~G.}\ \bibnamefont {Wang}}, \bibinfo {author} {\bibfnamefont
  {F.}~\bibnamefont {Ma}}, \bibinfo {author} {\bibfnamefont {Y.~X.}\
  \bibnamefont {Guo}}, \bibinfo {author} {\bibfnamefont {X.~G.}\ \bibnamefont
  {Wu}}, \bibinfo {author} {\bibfnamefont {C.~Y.}\ \bibnamefont {He}}, \bibinfo
  {author} {\bibfnamefont {Y.}~\bibnamefont {Zheng}}, \bibinfo {author}
  {\bibfnamefont {Z.~M.}\ \bibnamefont {Wang}}, \bibinfo {author}
  {\bibfnamefont {S.~C.}\ \bibnamefont {Li}}, \bibinfo {author} {\bibfnamefont
  {X.~L.}\ \bibnamefont {Yan}}, \bibinfo {author} {\bibfnamefont
  {L.}~\bibnamefont {He}}, \bibinfo {author} {\bibfnamefont {Z.~G.}\
  \bibnamefont {Wang}}, \bibinfo {author} {\bibfnamefont {Y.}~\bibnamefont
  {Zheng}}, \bibinfo {author} {\bibfnamefont {F.}~\bibnamefont {Fang}},\ and\
  \bibinfo {author} {\bibfnamefont {X.~M.}\ \bibnamefont {Chen}},\ }\href
  {https://doi.org/10.1103/PhysRevC.86.054310} {\bibfield  {journal} {\bibinfo
  {journal} {Phys. Rev. C}\ }\textbf {\bibinfo {volume} {86}},\ \bibinfo
  {pages} {054310} (\bibinfo {year} {2012})}\BibitemShut {NoStop}%
\end{thebibliography}%

\end{document}